\documentclass[12pt]{article}

\usepackage{graphics}
\usepackage{epsfig}
\usepackage{cite}
\input epsf

\title{Prompt charmonia production and polarization at LHC in the NRQCD with $k_T$-factorization. \\Part III: $J/\psi$ meson}
\author{S.P.~Baranov$^{1}$, A.V.~Lipatov$^{2,\,3}$}

\begin{document}

\maketitle

\begin{center}

{\it $^1$P.N.~Lebedev Physics Institute, 119991 Moscow, Russia}\\
{\it $^2$Skobeltsyn Institute of Nuclear Physics, Lomonosov Moscow State University, 119991 Moscow, Russia}\\
{\it $^3$Joint Institute for Nuclear Research, Dubna 141980, Moscow region, Russia}\\

\end{center}

\vspace{5mm}

\begin{center}

{\bf Abstract }

\end{center}

In the framework of $k_T$-factorization approach, 
the production and polarization of prompt $J/\psi$ mesons 
at the LHC energies is studied. Our consideration is based on the non-relativistic
QCD formalism for bound states and off-shell amplitudes for hard partonic 
subprocesses.
Both the direct production mechanism and feed-down 
contributions from $\chi_c$ and $\psi(2S)$ decays
are taken into account. 
The transverse momentum dependent (or unintegrated) gluon densities in a proton were derived from
Ciafaloni-Catani-Fiorani-Marchesini evolution equation or, alternatively, were
chosen in accordance with Kimber-Martin-Ryskin prescription.
The non-perturbative color-octet matrix elements were first deduced from the 
fits to the latest CMS data on $J/\psi$ transverse momentum distributions
and then applied to describe the ATLAS and LHCb data on $J/\psi$ production 
and polarization at $\sqrt s = 7$, $8$ and $13$~TeV.
We perform an estimation of polarization parameters $\lambda_\theta$, 
$\lambda_\phi$ and $\lambda_{\theta \phi}$ which determine $J/\psi$ 
spin density matrix and demonstrate that treating the soft gluon emission 
as a series of explicit color-electric dipole transitions within NRQCD 
leads to unpolarized $J/\psi$ production at high
transverse momenta, that is in qualitative agreement with the LHC data.

\vspace{1.0cm}

\noindent
PACS number(s): 12.38.-t, 13.20.Gd, 14.40.Pq

\newpage

\section{Introduction} \indent

Since it was first observed, inclusive
$J/\psi$ meson production in hadronic collisions is a subject of considerable
theoretical and experimental interest.
It serves as a complex probe of the
hadron structure, perturbative QCD and
the formation mechanism of charmed quark bound states.
Indeed, the production
of unbound $c\bar{c}$ pairs in hard scattering (described by perturbative
QCD) is followed by the formation of bound states that is essentially
a non-perturbative process.
The latter seems to be the most tricky ingredient in the
theory.

Two theoretical approaches describing this
non-perturbative step are known in the literature under the names of
color-singlet (CS)\cite{1} and color-octet(CO)\cite{2} models.
In general, the charmed quark pair is produced in a state $^{2S+1}L_J^{(a)}$
with spin $S$, orbital angular momentum $L$, total angular momentum $J$ and
color $a$, which can be either identical to the final charmonium quantum numbers
(as accepted in the CS model) or different from those.
In the latter case, the produced $c\bar c$ pair transforms into physical
charmonium state by means of soft (non-perturbative) gluon radiation, as
considered in the formalism of non-relativistic QCD (NRQCD)\cite{3,4}.
The probability to form a given bound state is determined by the respective
non-perturbative long-distance matrix elements (NMEs), which are assumed to be
universal (process-independent), not depending on the charmonium momentum and
obeying certain hierarchy in powers of the relative charmed quarks velocity $v$.

As we have explained in our previous papers\cite{5,6}, none of the existing 
theoretical approaches is able to describe all of the data in
their integrity. The CS predictions obtained
at the dominant tree-level next-to-next-to-leading order (NNLO$^*$)\cite{7,8}
underestimate the measured $J/\psi$ and $\psi(2S)$ cross sections by a factor
of $5$, that is well off the theoretical and experimental uncertainty band\cite{9}.
In the NRQCD formalism, a reasonably good description of the transverse momentum
distributions can be achieved at the next-to-leading order (NLO) by adjusting the
NMEs values, which play the role of free parameters\cite{10,11,12,13,14,15}. However,
these NMEs dramatically depend on the minimal charmonium transverse
momentum used in the fits\cite{16} and are incompatible
with each other (both in size and even in sign) when obtained from fitting the
different data sets.
Moreover, the worst problem of the calculations 
is connected with $J/\psi$ spin alignment. If, as predicted,
the dominant contrubution comes from the gluon fragmentation into an octet
$c\bar{c}$ pair, the outgoing $J/\psi$ must have strong transverse polarization.
The latter disagrees with the Tevatron\cite{17,18} and LHC measurements\cite{19,20,21}
which point to unpolarized or even longitudinally polarized particles.  
The overall situation is known as ``quarkonium
polarization puzzle'' and understood as a deep crisis. 

Recently, a new solution to this polarization puzzle has been proposed\cite{22}
in the framework of a model that interprets the soft final state gluon
radiation (transforming an unbound $c\bar c$ pair 
into physical charmonium state) as a series of color-electric dipole transitions.
In contrast with the usual NRQCD calculations where the emitted 
gluons are not presented explicitly in the final state, it was proposed
to represent the long-distance NMEs in an explicit form ispired by 
classical multipole radiation theory, so that the spin structure
of the transition amplitudes is specified. 
This model differs in its predictions from the conventional 
LO and NLO calculations (both CS and CO), as
the final-state soft gluons are now emitted 
by the entire $c\bar c$ system and not by individual quarks.
This scenario  
results in the unpolarized (or only weakly polarized)
heavy quarkonia, providing us with an easy and
natural solution of the long-standing puzzle.
Moreover, it was already succesfully applied\cite{5} to describe the recent 
LHC data on the production and polarization of 
$\psi(2S)$ mesons.

In the present note we consider the
production and polarization of $J/\psi$ mesons at the LHC conditions
using the approach\cite{22}.
This study is a continuation of the works\cite{5,6},
where the $\psi(2S)$ and $\chi_c$ meson production at the LHC has been considered.
We give a systematic analysis of ATLAS\cite{23}, CMS\cite{21,24}
and LHCb\cite{20,25,26} data collected at $\sqrt s = 7$, $8$ and $13$~TeV
regarding the transverse momentum distributions and polarization parameters
$\lambda_\theta$, $\lambda_\phi$ and $\lambda_{\theta \phi}$
which determine the spin density matrix of the produced $J/\psi$ mesons.
To describe the perturbative production of $c\bar c$ pair in
a short-distance gluon-gluon fusion subprocess we employ 
the $k_T$-factorization approach\cite{27,28}.
The latter is based on the Balitsky-Fadin-Kuraev-Lipatov (BFKL)\cite{29} or 
Ciafaloni-Catani-Fiorani-Marchesini (CCFM)\cite{30} gluon
evolution equations.
We see certain advantages in the fact that, 
even with the leading-order (LO) partonic amplitudes, 
one then includes a large piece of high-order corrections
(namely, part of NLO + NNLO + ... terms containing the leading
logarithms of the type
$\log 1/x$ due to real parton emissions in initial state)
taking them into
account in the form of transverse momentum dependent (TMD) parton 
densities\footnote{A detailed description of the $k_T$-factorization
approach can be found, for example, in reviews\cite{31}.}.
Besides that, the latter absorb the effects of soft gluon resummation, that
regularises the infrared divergences and makes our predictions applicable even
at low transverse momenta.
Two sources of $J/\psi$ production are taken into account: direct 
$J/\psi$ production and feed-down
from radiative decays of heavier charmonium states like $\chi_{c1}$, 
$\chi_{c2}$ or $\psi(2S)$, that is in full agreement
with the experimental setup\cite{20,21,23,24,25,26}. In the literature, these two joint 
sources are referred to as prompt $J/\psi$ production\footnote{In $pp$ collisions,
$J/\psi$ mesons can be also produced via decays of $b$-flavored hadrons. This 
process which is usually referred to as non-prompt $J/\psi$ production or "$J/\psi$-from-$b$" is
out of our present consideration.}.

The outline of our paper is the following. In Section~2 we briefly recall the 
NRQCD formalism and the $k_T$-factorization approach. In Section~3 we perform 
a numerical fit to the latest CMS data and extract the color-octet NMEs for $J/\psi$ mesons 
using three different sets of TMD gluon distributions. 
Later in this section we check the compatibility of the extracted 
papameters with ATLAS and LHCb
data on $J/\psi$ production and polarization. The comparison
is followed by a discussion. Our conclusions are collected in Section~4.

\section{Theoretical framework} \indent

Our consideration is based on the following leading-order off-shell gluon-gluon fusion 
subprocesses:
\begin{equation}
  g^*(k_1) + g^*(k_2) \to c \bar c \left[^3S_1^{(1)}\right](p) + g(k),
\end{equation}
\begin{equation}
  g^*(k_1) + g^*(k_2) \to c \bar c \left[^1S_0^{(8)}, \, ^3S_1^{(8)}, \, ^3P_J^{(8)}\right](p),
\end{equation}
\noindent 
for $S$-wave charmonia, i.e., $J/\psi$ or $\psi(2S)$ mesons, followed by
non-perturbative transitions $c\bar c \to J/\psi$,~$\psi(2S) + X$,
and 
\begin{equation}
  g^*(k_1)+g^*(k_2)\to c \bar c \left[^3P_J^{(1)}, \, ^3S_1^{(8)}\right](p),
\end{equation}

\noindent
followed by non-perturbative transitions $c\bar c \to \chi_{cJ} + X$
for $P$-wave states, where $J = 0, 1$ or $2$ and the four-momenta of all particles are indicated 
in the parentheses. The corresponding production amplitudes can be obtained from the 
one for an unspecified $c\bar c$ state by applying the appropriate projection 
operators, which guarantee the proper quantum numbers of the $c\bar c$ state 
under consideration.
The details of calculations can be found in our previous papers\cite{5,6}.
Here we briefly discuss only the transition of unbound $c\bar c$ pair to $J/\psi$ meson
applied in our calculations.

As it was already mentioned above, it is usually assumed that the emitted soft gluons bring away the unwanted
color and change other quantum numbers of the $c\bar c$ system 
but do not carry any energy, thus keeping the kinematics intact.
However, this is in obvious contradiction with confinement which 
prohibits the emission of infinitely soft colored  quanta.
In order that the quantum numbers get changed, one needs to radiate
a real gluon with some energy $E \sim \Lambda_{\rm QCD}$, giving us the confidence 
that we do not enter into the confinement or perturbative domains\cite{33}. 
This issue is not the matter of only kinematic corrections, because 
we cannot organize a
transition amplitude with correct spin properties
without having a non-zero energy-momentum transfer.
Following\cite{22},
we describe this step in terms of electric dipole ($E1$) transitions
that dominante the multipole expansion.
In the case of $^3P_J^{(8)}$ states, a single $E1$ transition is 
needed to transform them into $^3S_1$ mesons. The corresponding 
amplitudes can be written as\cite{34}:
\begin{equation}
  {\cal A}(^3P_0^{(8)} \to \psi + g) \sim k_\mu^{(g)} p^{{\rm (CO)}\mu} \epsilon_\nu^{(\psi)} \epsilon^{{(g)}\nu},
\end{equation}
\begin{equation}
  {\cal A}(^3P_1^{(8)} \to \psi + g) \sim e^{\mu \nu \alpha \beta} k_\mu^{(g)} \epsilon_\nu^{\rm (CO)} \epsilon_\alpha^{(\psi)} \epsilon_\beta^{(g)},
\end{equation}
\begin{equation}
  {\cal A}(^3P_2^{(8)} \to \psi + g) \sim p^{{\rm (CO)}\mu} \epsilon^{{\rm (CO)}\alpha \beta} \epsilon_\alpha^{(\psi)} \left[ k_\mu^{(g)} \epsilon_\beta^{(g)} - k_\beta^{(g)} \epsilon_\mu^{(g)} \right],
\end{equation}

\noindent
where $p_\mu^{\rm (CO)}$, $k_\mu^{(g)}$, $\epsilon_\mu^{(\psi)}$, $\epsilon_\mu^{(g)}$,
$\epsilon_\mu^{\rm (CO)}$ and $\epsilon_{\mu \nu}^{\rm (CO)}$ are the four-momenta and 
polarization four-vectors (tensor) of the respective particles, and
$e^{\mu \nu \alpha \beta}$ is the fully antisymmetric Levi-Civita tensor.
These amplitudes are practically the same as the ones for radiative decays of $\chi_c$ mesons,
the only difference is in the overall normalization factors.
In the case of $^3S_1^{(8)}$ state, we treat its transformation into real $^3S_1$ meson
as two successive color-electric dipole transitions, $^3S_1^{(8)} \to \, ^3P_J^{(8)} + g$,
$^3P_J^{(8)} \to \psi + g$, proceeding via either of the three intermediate $^3P_J^{(8)}$ states, where 
$J = 0$, $1$ or $2$, and expoit the same effective coupling vertices (9) --- (11).
%(see\cite{22} for more details). 
Now, the polarization of the 
outgoing mesons can also be calculated without any ambiguity.
Below we use the expressions derived in our previous papers\cite{5,6}.

As we did in our previous papers\cite{5,6}, we tried numerically 
several sets of TMD gluon densities in a proton.
Two of them (A0\cite{38} and JH'2013\cite{39}) were obtained from CCFM 
equation where all input parameters were fitted to the proton structure 
function $F_2(x, Q^2)$. Besides that, we used a parametrization obtained with
Kimber-Martin-Ryskin (KMR) prescription\cite{40} which provides a method
to construct the TMD quark and gluon densities out of conventional (collinear) 
distributions. In that case, we used for the input the leading-order 
Martin-Stirling-Thorn-Watt (MSTW'2008) set\cite{41}.

\section{Numerical results} \indent

We are now in a position to present our numerical results. First we describe our
input and kinematic conditions. Having the TMD gluon denisities chosen,
the cross sections~(12) and~(13) depend on the renormalization and factorization 
scales $\mu_R$ and $\mu_F$. We set 
$\mu_R^2 = m^2 + {\mathbf p}_{T}^2$ and
$\mu_F^2 = \hat s + {\mathbf Q}_T^2$, where ${\mathbf Q}_T$ is the 
transverse momentum of the initial off-shell gluon pair.
The choice of $\mu_R$ is the standard one for studying the 
charmonia production, whereas the special choice of $\mu_F$ is connected 
with the CCFM evolution\cite{38,39}.
Following\cite{42}, we set $J/\psi$ mass $m = 3.0969$~GeV, 
branching fraction $B(J/\psi \to l^+ l^-) = 0.05961$ and use the LO formula 
for the coupling constant $\alpha_s(\mu^2)$ with $n_f = 4$ quark flavours
and $\Lambda_{\rm QCD} = 200$~MeV, so that $\alpha_s(M_Z^2) = 0.1232$.
When calculating the feed-down contributions from radiative decays
of $\chi_c$ and $\psi(2S)$ mesons, we used exactly the same parameters and 
NMEs as in our previous papers\cite{5,6}.

First, we determine the whole set of $J/\psi$ meson NMEs.
We have fitted the transverse momentum distributions  of $J/\psi$ mesons 
measured recently by the CMS Collaboration at 
$\sqrt s = 7$~TeV\cite{24}. These measurements were done at central rapidities $|y| < 1.2$ and moderate and high 
transverse momenta $10 < p_T < 100$~GeV, where the NRQCD formalism is believed 
to be most reliable. We performed the fitting 
procedure under requirement that the NMEs be strictly positive.
Further on, instead of taking the color singlet NME from the 
$J/\psi \to l^+ l^-$ partial decay width, we 
treat it as free parameter.
A comparison of such fitted NME with the ones known
from leptonic decay
will provide an independent cross-check of our calculations
and an additional test for the TMD gluon densities in a proton.

In Table~1 we list our results for the NMEs fits obtained for three different
TMD gluon distributions. For comparison, 
we also present here two sets of NMEs\cite{11,15}, obtained within the 
NLO NRQCD by other authors. The main difference between them is in that these fits were based on differently selected data sets.
One can see that our fitting procedure leads to very similar values of the 
CS NME extracted with the considered TMD gluon densities.
On the contrary, the fitted CO NME values strongly depend on 
the choice of TMD gluon distribution. Typically, they are 
smaller than the ones obtained in the NLO NRQCD fits\cite{11,15}, that is
almost consistent with the estimates performed by other authors\cite{43,44}.
We find that the $^3P_J^{(8)}$ contributions are compatible with zero,
that agrees with the early consideration\cite{32}.
Note that, as it was expected, our fitted CS MNEs are close
to the ones determined from the $J/\psi \to l^+ l^-$ decay, 
$\langle {\cal O}^{\psi}\left[ ^3S_1^{(1)}\right] \rangle \sim 1.3 - 1.5$~GeV$^3$,
that demonstrates self-consistency of our calculations and 
good agreement with the previous ones\cite{45}, which were
performed in the CS model alone at relatively low transverse momenta $p_T < 30$~GeV.

\begin{table}
\begin{center}
\begin{tabular}{|c|c|c|c|c|}
\hline
  & & & & \\
    & $\langle {\cal O}^{\psi}\left[ ^3S_1^{(1)}\right] \rangle$/GeV$^3$ & $\langle {\cal O}^{\psi}\left[ ^1S_0^{(8)}\right] \rangle$/GeV$^3$ & $\langle {\cal O}^{\psi}\left[ ^3S_1^{(8)}\right] \rangle$/GeV$^3$ & $\langle {\cal O}^{\psi}\left[ ^3P_0^{(8)}\right] \rangle$/GeV$^5$ \\
  & & & & \\
\hline
  & & & & \\
  A0 & $1.97$ & 0.0 & $9.01 \times 10^{-4}$ & 0.0 \\
  & & & & \\
  JH & $1.62$ & $1.71 \times 10^{-2}$ & $2.83 \times 10^{-4}$ & 0.0 \\
  & & & & \\
  KMR & $1.58$ & $8.35 \times 10^{-3}$ & $2.32 \times 10^{-4}$ & 0.0 \\
  & & & & \\
  $[11]$ & $1.32$ & $3.04 \times 10^{-2}$ & $1.68 \times 10^{-3}$ & $-9.08 \times 10^{-3}$ \\
  & & & & \\
  $[15]$ & $1.16$ & $9.7 \times 10^{-2}$ & $-4.6 \times 10^{-3}$ & $-2.14 \times 10^{-2}$ \\
  & & & & \\
\hline
\end{tabular}
\caption{The NMEs for $J/\psi$ meson derived from the fit of the CMS data\cite{24}. The NMEs
obtained in the NLO NRQCD fits\cite{11,15} are shown for comparison.}
\label{table1}
\end{center}
\end{table}

Now we turn to comparing our predictions with the latest data collected by the ATLAS\cite{23},
CMS\cite{21,24} and LHCb\cite{20,25,26} Collaborations.
Recently, the ATLAS Collaboration presented prompt $J/\psi$ transverse momentum 
distributions at $8 < p_T < 100$~GeV and $\sqrt s = 7$~TeV and 
$8 < p_T < 110$~GeV and $\sqrt s = 8$~TeV for eight subdivisions in $J/\psi$ rapidity $y$.
The CMS Collaboration measured $J/\psi$ transverse momentum spectra 
in the kinematic range $10 < p_T < 100$~GeV and $|y| < 1.2$,
and the LHCb Collaboration measured them in the range $p_T < 14$~GeV and $2 < y < 4.5$ at 
different energies $\sqrt s = 7$, $8$ and $13$~TeV.
Our predictions for the differential cross sections are 
presented in Figs.~1 --- 6 in comparison with the LHC data.
As usual, to estimate the theoretical uncertainty of our calculations coming from the 
hard scales, we vary them by a factor of $2$ around the default values.
One can see that at central rapidities we achieved a reasonably good agreement between our calculations
(with any of the considered TMD gluon densities) and CMS and ATLAS data  
in the whole $p_T$ region within the uncertainties.
However, with increasing rapidity, the overall description of 
ATLAS data becomes a bit worse:
our central predictions tend to slightly underestimate the data at low and moderate $p_T$ 
for both energies $\sqrt s = 7$ and $8$~TeV (see Figs.~2 and 4).
We note that the observed disrepancy is not catastrophic, because some reasonable
variation in the factorization and/or renormalization scales (shown by the shaded bands) 
eliminate visible disagreement.
In the forward rapidity region, covered by the LHCb experiment, 
the difference between the predictions obtained with the different 
TMD gluon densities becomes more clear (see Figs.~3, 5 and 6).
The best description of LHCb data for $\sqrt s = 7$ and $8$~TeV 
is provided by the A0 gluon distribution. However, this gluon density
is unable to describe LHCb data at $\sqrt s = 13$~TeV (see Fig.~6),
that spoils the belief in the universality of the fitted NMEs.
The JH'2013 gluon significantly overestimates these data at low $p_T < 6$~GeV
and practically coincides with A0 predictions at larger transverse momenta.
Therefore, we conclude that, similarly to the collinear factorization,
including the low $p_T$ data into the fitting procedure in the $k_T$-factorization approach
can change the relative weight of different NMEs.
However, as it was already mentioned above, the applicability of the NRQCD 
formulas at low transverse momenta is questionable.
The results of calculations obtained with KMR gluon density are similar to the 
A0 ones at $y < 3$ and tend to overestimate the data at more forward rapidities,
though still agree with the data within the uncertainties.

The ratios of the $\psi(2S)$ to $J/\psi$ differential cross sections were measured
by CMS Collaboration at $|y| < 1.2$ and $\sqrt s = 7$~TeV\cite{24} and 
by ATLAS Collaboration at $\sqrt s = 7$ and $8$~TeV 
in several rapidity subdivisions\cite{23}.
Additionally, the ratio $R_{13/8}$ of the double differential  
$J/\psi$ production cross sections at $\sqrt s = 13$~TeV and $\sqrt s = 8$~TeV was presented 
by LHCb Collaboration\cite{26}. Many of the 
experimental and theoretical uncertainties (as connected, for example, with the NMEs 
or hard scales) cancel out in these ratios, giving us possibility to further test
the production dynamics. Our predictions are shown in Figs.~7 --- 9 and compared to the data.
The cross sections of $\psi(2S)$ production are calculated 
in the same way as in our previous paper\cite{5}.
One can see that at low and moderate transverse momenta 
the $\psi(2S)$ to $J/\psi$ production ratios are reasonably described 
by the considered TMD gluon densities everywhere except two last rapidity subintervals.
The high $p_T$ data (to be precise, at $p_T \geq 20$~GeV) 
can be used to discriminate between the TMD gluons.
The difference becomes even more prononced in the $R_{13/8}$ ratio (see Fig.~9).

Now we turn to the polarization of $J/\psi$ mesons, which is the most 
interesting part of our study.
In general, the spin density matrix of a vector particle decaying into a lepton
pair depends on three angular parameters $\lambda_{\theta}$, $\lambda_\phi$ and 
$\lambda_{\theta \phi}$ which can be measured experimentally. The double 
differential angular distribution of the decay leptons can be written as\cite{46}:
\begin{equation}
  {d\sigma \over d\cos \theta^* d\phi^*} \sim 1 + \lambda_\theta \cos^2 \theta^* + 
    \lambda_\phi \sin^2 \theta^* \cos 2 \phi^* + \lambda_{\theta \phi} \sin 2 \theta^* \cos \phi^*,
\end{equation}

\noindent
where $\theta^*$ and $\phi^*$ are the polar and azimuthal angles of the decay 
lepton measured in the $J/\psi$ rest frame. The case of $(\lambda_{\theta}, 
\lambda_\phi, \lambda_{\theta \phi}) = (0,0,0)$ corresponds to unpolarized state, 
while 
$(\lambda_{\theta}, \lambda_\phi, \lambda_{\theta \phi}) = (1,0,0)$ and
$(\lambda_{\theta}, \lambda_\phi, \lambda_{\theta \phi}) = (-1,0,0)$ refer 
to fully transverse and fully longitudinal polarizations.
CMS\cite{21} and LHCb\cite{20} Collaborations have measured these 
parameters as functions of $J/\psi$ transverse momentum in two complementary 
frames: the Collins-Soper and helicity ones. 
In addition, CMS Collaboration provided measurements in the perpendicular 
helicity frame.
In the Collins-Soper frame the polarization axis $z$ bisects the two beam 
directions whereas the polarization axis in the helicity frame
coincides with the direction of $J/\psi$ momentum in the laboratory frame.
In the perpendicular helicity frame the $z$ axis is orthogonal to that
in the Collins-Soper frame and lies in the plane spanned by the two beam
momenta. Additionally, the frame-independent polarization parameter\cite{21}
$\lambda^* = (\lambda_\theta + 3 \lambda_\phi)/(1 - \lambda_\phi)$
was investigated. Below we estimate the 
polarization parameters $\lambda_\theta$, $\lambda_\phi$, $\lambda_{\theta \phi}$
and $\lambda^*$ for the CMS and LHCb conditions.
Our calculation generally follows the experimental procedure. We collect 
the simulated events in the kinematical region defined by the CMS and LHCb 
experiments, generate the decay lepton angular distributions according
to the production and decay matrix elements, and then apply a three-parametric
fit based on~(14).

In Figs.~10 --- 14 we confront our predictions for
parameters $\lambda_\theta$, $\lambda_\phi$, 
$\lambda_{\theta \phi}$ and $\lambda^*$ with the latest 
CMS\cite{20} and LHCb\cite{21} data.
We find practically zero polarization ($\lambda_\theta \sim 0$)
of the produced $J/\psi$ mesons at moderate and large transverse momenta $p_T \geq 20$~GeV 
(see Figs.~10 --- 12) and weak longitudinal polarization  
at low transverse momenta $p_T \leq 14$~GeV covered by the LHCb experiment.
To be precise, we obtained $\lambda_\theta \sim 0.1$ in the 
Collins-Soper frame and $\lambda_\theta \sim - 0.2$ in the helicity frame, respectively.
Moreover, these results are practically independent of the $J/\psi$ rapidity.
As it was mentioned above, the traditional NLO CS calculations predict large 
longitudinal polarization at high $p_T$, while the NRQCD predicts large transverse polarization, 
and none of these predictions is supported by experimental results.
As one can see, treating the soft gluon emission within the NRQCD
as a series of explicit color-electric dipole transitions\cite{22} 
leads to unpolarized $J/\psi$ production at high
transverse momenta, that is in qualitative agreement with available LHC data.
This remarkable property is the main result of our study, and
the same conclusion was made
previously in the case of prompt $\psi(2S)$ meson production\cite{5}.
Note that our interpretation of gluon radiation is not the same as in the
conventional CS model at high-order pQCD, because the gluons
are emitted by the entire $c \bar c$ system, not by individual quarks.

The absense of strong $J/\psi$ polarization is
not connected with parameter tuning, but
seems to be a natural and rather general feature of the
scenario\cite{22}.
We note, however, that while our predictions for $\lambda_\phi$ and 
$\lambda_{\theta \phi}$ parameters agree with the data,
the description of $\lambda_{\theta}$ and $\lambda^*$ 
is still rather qualitative than quantitative.
Despite the huge experimental uncertainties,
it could be due to the significant theoretical uncertainties 
connected, in particular, with the inclusion of NLO subprocesses and precise definition of NMEs.
The detailed study of these uncertainties is out of our present paper.
Nevertheless, the proposed way, in our opinion, can provide an easy and natural solution 
to a long-standing quarkonia polarization puzzle.

\section{Conclusions} \indent 

We have considered prompt $J/\psi$ production and polarization in $pp$ 
collisions at the LHC in the framework 
of $k_T$-factorization approach. We have used the LO non-relativistic QCD
formalism including both color-singlet and color-octet contributions
and took into account both the direct production mechanism and the feed-down 
contributions from $\chi_{c1}$, $\chi_{c2}$ and $\psi(2S)$ decays.
Using the TMD gluon densities in a proton derived 
from the CCFM equation and from the Kimber-Martin-Ryskin prescription,
we extracted the color-octet NMEs 
$\langle {\cal O}^{\psi}\left[ ^1S_0^{(8)}\right] \rangle$,
$\langle {\cal O}^{\psi}\left[ ^3S_1^{(8)}\right] \rangle$ and
$\langle {\cal O}^{\psi}\left[ ^3P_0^{(8)}\right] \rangle$ for $J/\psi$
mesons from fits to transverse momentum distributions provided by the latest CMS 
measurements at $\sqrt s = 7$~TeV.
Using the fitted NMEs, we have analyzed the data taken by 
the ATLAS, CMS and LHCb Collaborations at different energies 
$\sqrt s = 7$, $8$ and $13$~TeV.
We demonstrated the sensitivity of the different production rates,
in particular, the ratios of the cross sections calculated at different 
energies, to the TMD gluon densities in a proton.
We estimated the polarization parameters $\lambda_\theta$, 
$\lambda_\phi$ and $\lambda_{\theta \phi}$ which determine the $J/\psi$ 
spin density matrix and demonstrated that treating the soft gluon emission 
as a series of explicit color-electric dipole transitions within the NRQCD 
leads to unpolarized $J/\psi$ production at high
transverse momenta, that is in qualitative agreement with the LHC data.

\section{Acknowledgements} \indent 

The authors are grateful to H.~Jung
for very useful discussions and remarks.
This work was supported in part by grant of the President of Russian Federation NS-7989.2016.2 and
by the DESY Directorate in the framework of Moscow-DESY project on Monte-Carlo 
implementations for HERA---LHC.
We are also grateful to Nikolai Zotov, who passed avay in January 2016, for all his enthusiasm and many
discussions on the topic.

\newpage

\begin{figure}
\begin{center}
\epsfig{figure=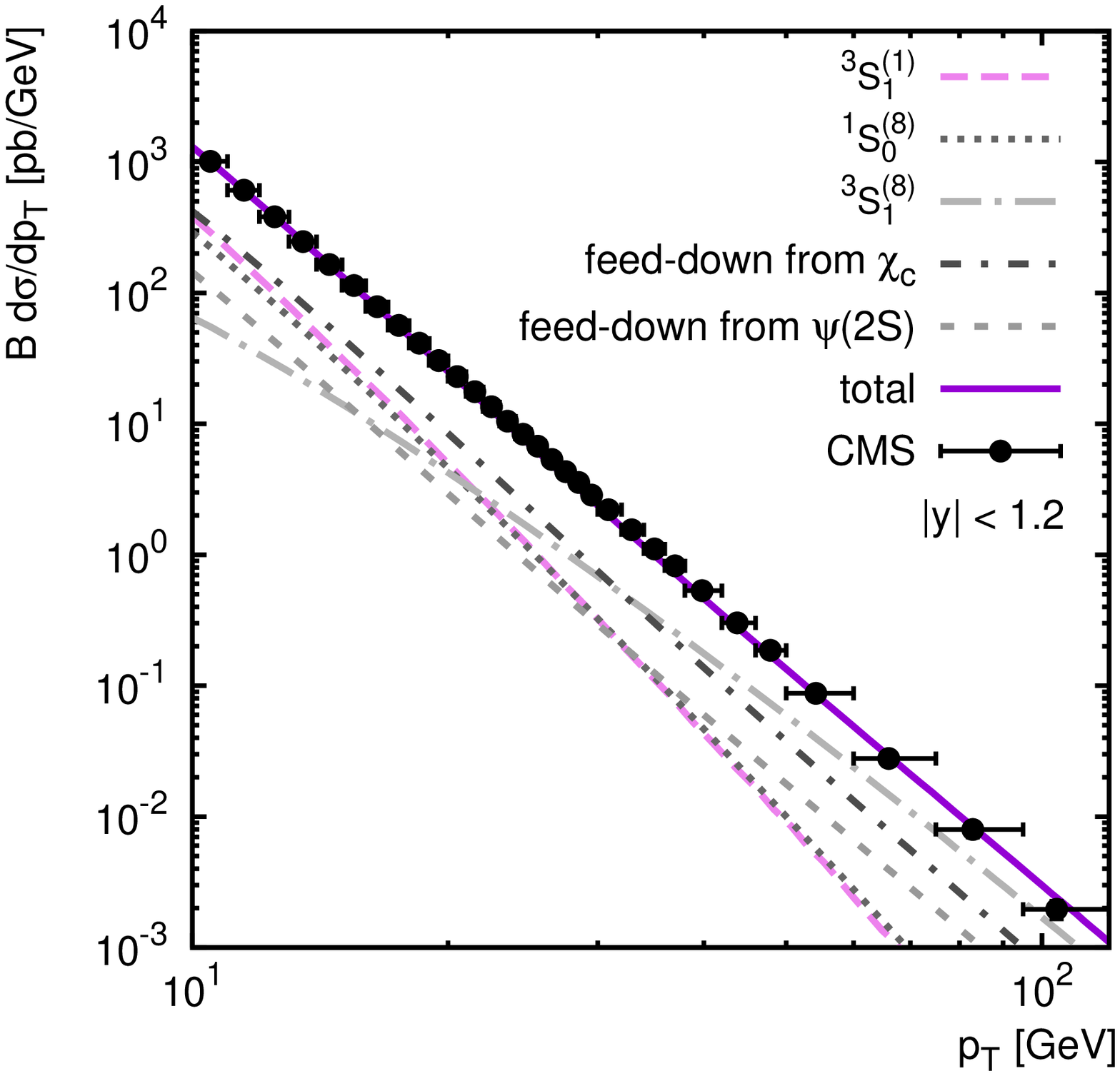, width = 5cm, angle = 270}
\epsfig{figure=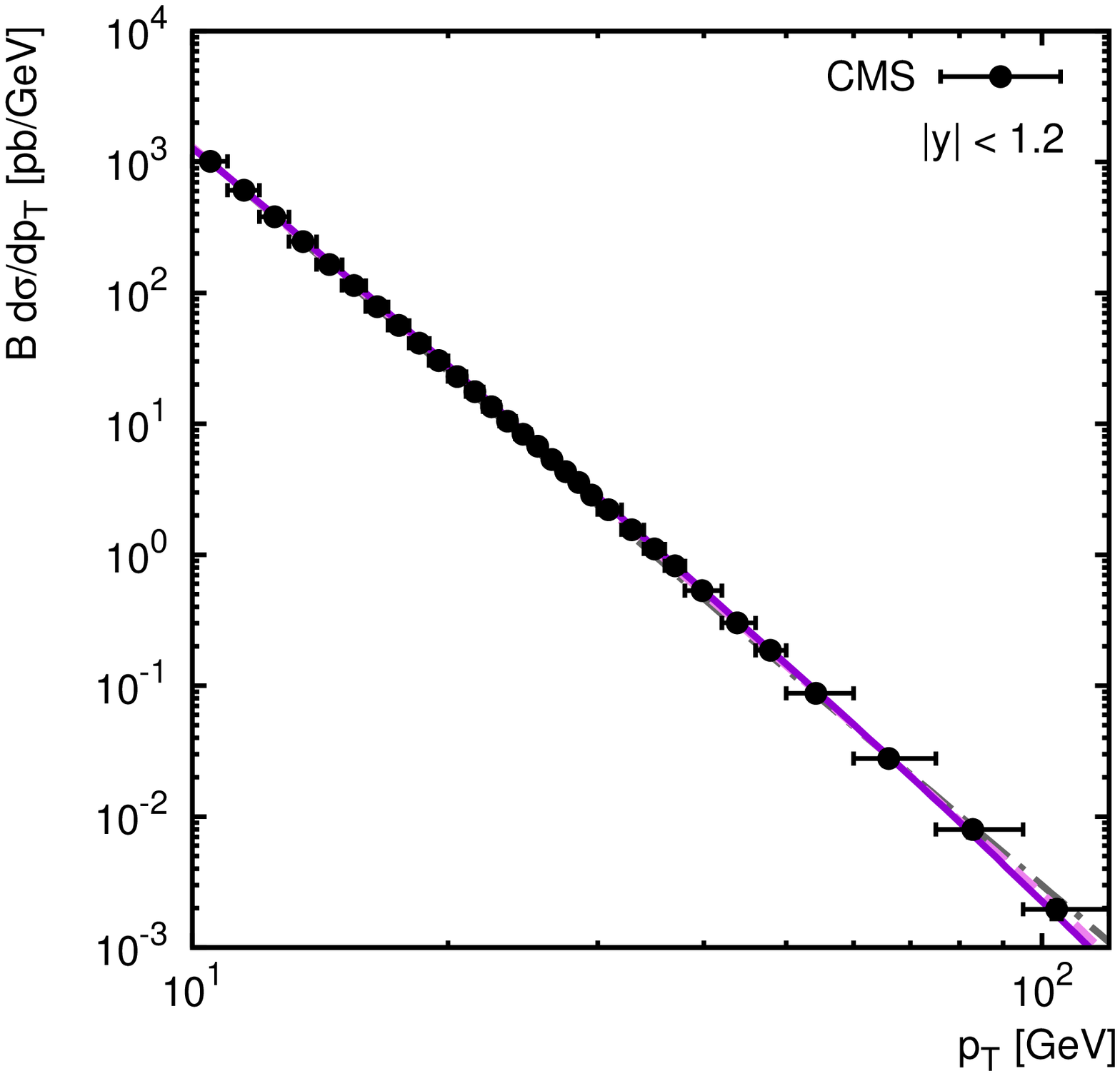, width = 5cm, angle = 270}
\caption{The transverse momentum distribution of prompt $J/\psi$ meson production
in $pp$ collisions at $\sqrt s = 7$~TeV. Left panel: the dashed, dotted and dash-dotted 
curves correspond to the color-singlet $^3S_1^{(1)}$ and color-octet 
$^1S_0^{(8)}$ and $^3S_1^{(8)}$ contributions calculated with the KMR gluon density.
The short dashed and short dash-dotted curves represent the feed-down contributions
from the radiative decays of $\chi_c$ and $\psi(2S)$ mesons.
The solid curve represent the sum of all these terms.
Right panel: the solid, dashed and dash-dotted curves correspond to the 
predictions obtained with the A0, JH and KMR gluon distributions, 
respectively. The experimental data are from CMS\cite{24}.}
\label{fig1}
\end{center}
\end{figure}

\begin{figure}
\begin{center}
\epsfig{figure=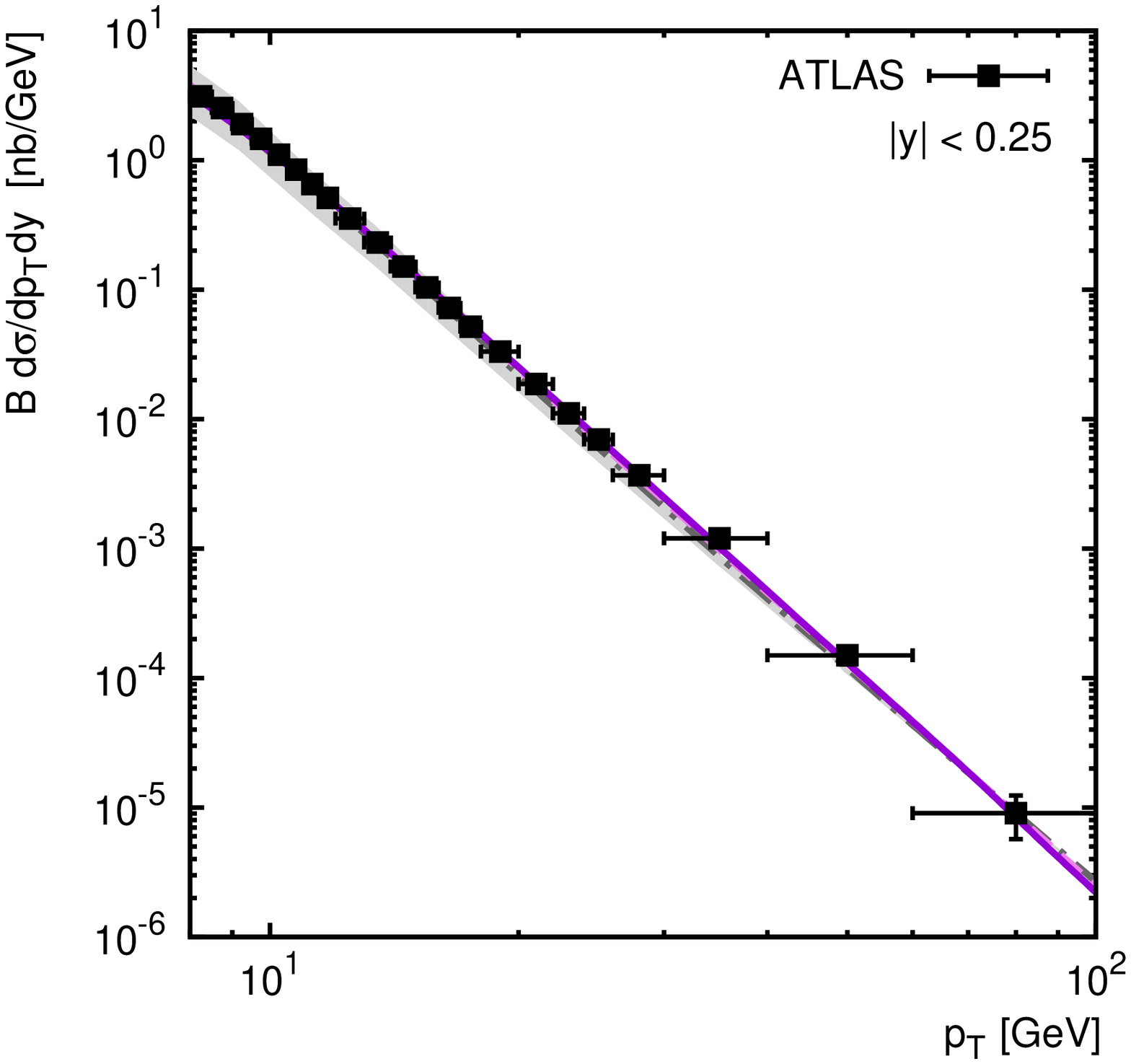, width = 5cm, angle = 270} 
\epsfig{figure=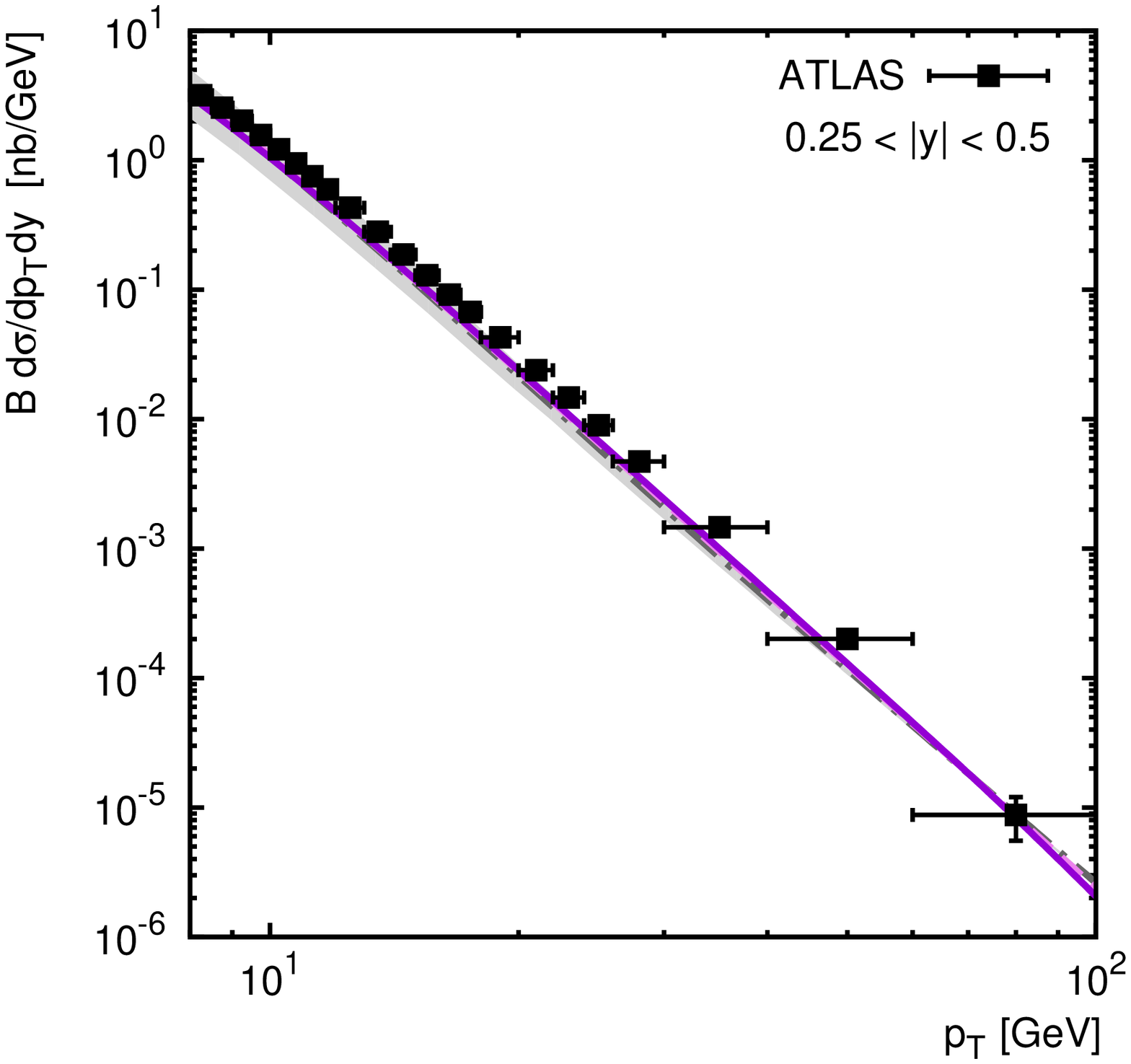, width = 5cm, angle = 270} 
\epsfig{figure=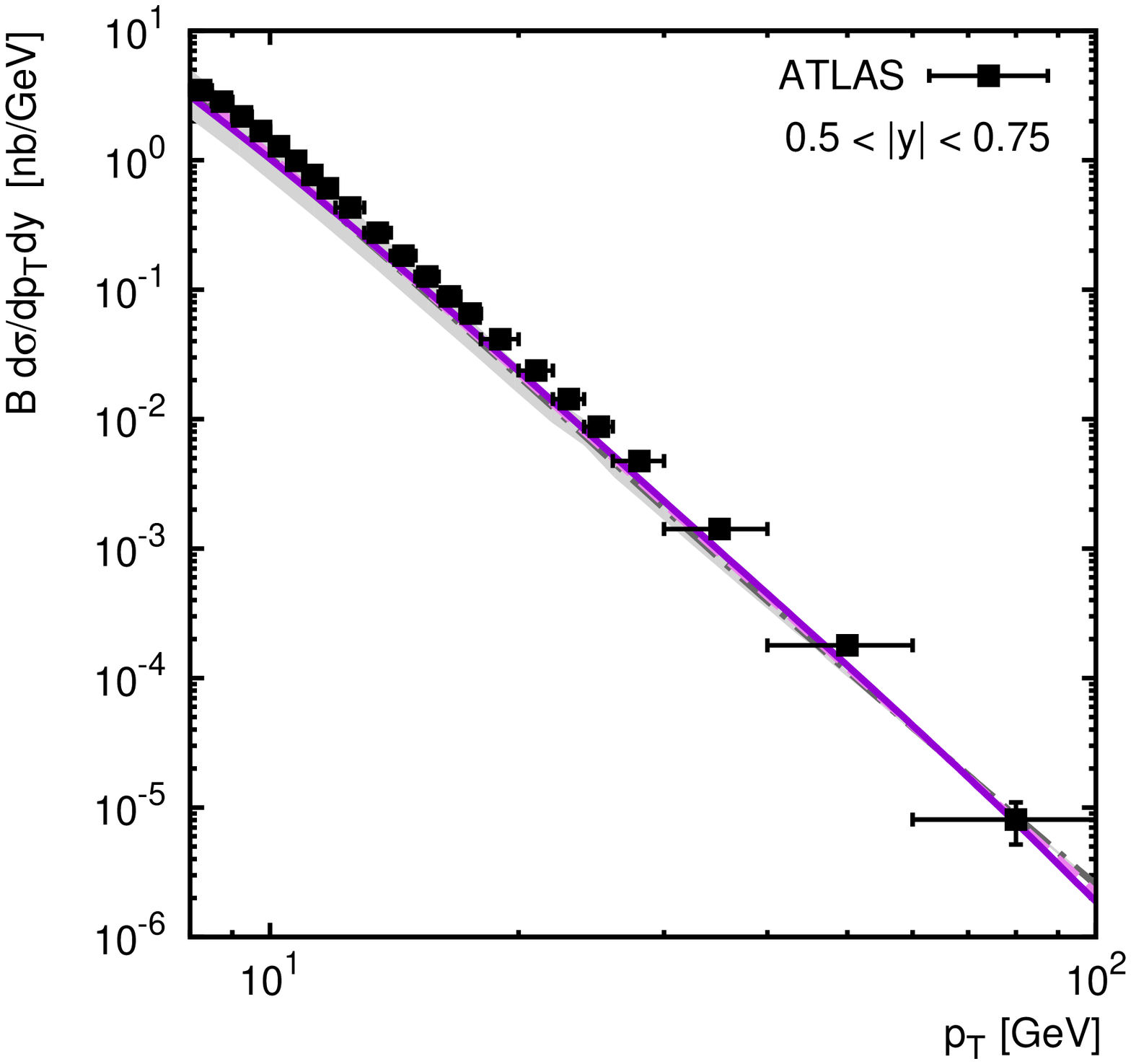, width = 5cm, angle = 270}
\epsfig{figure=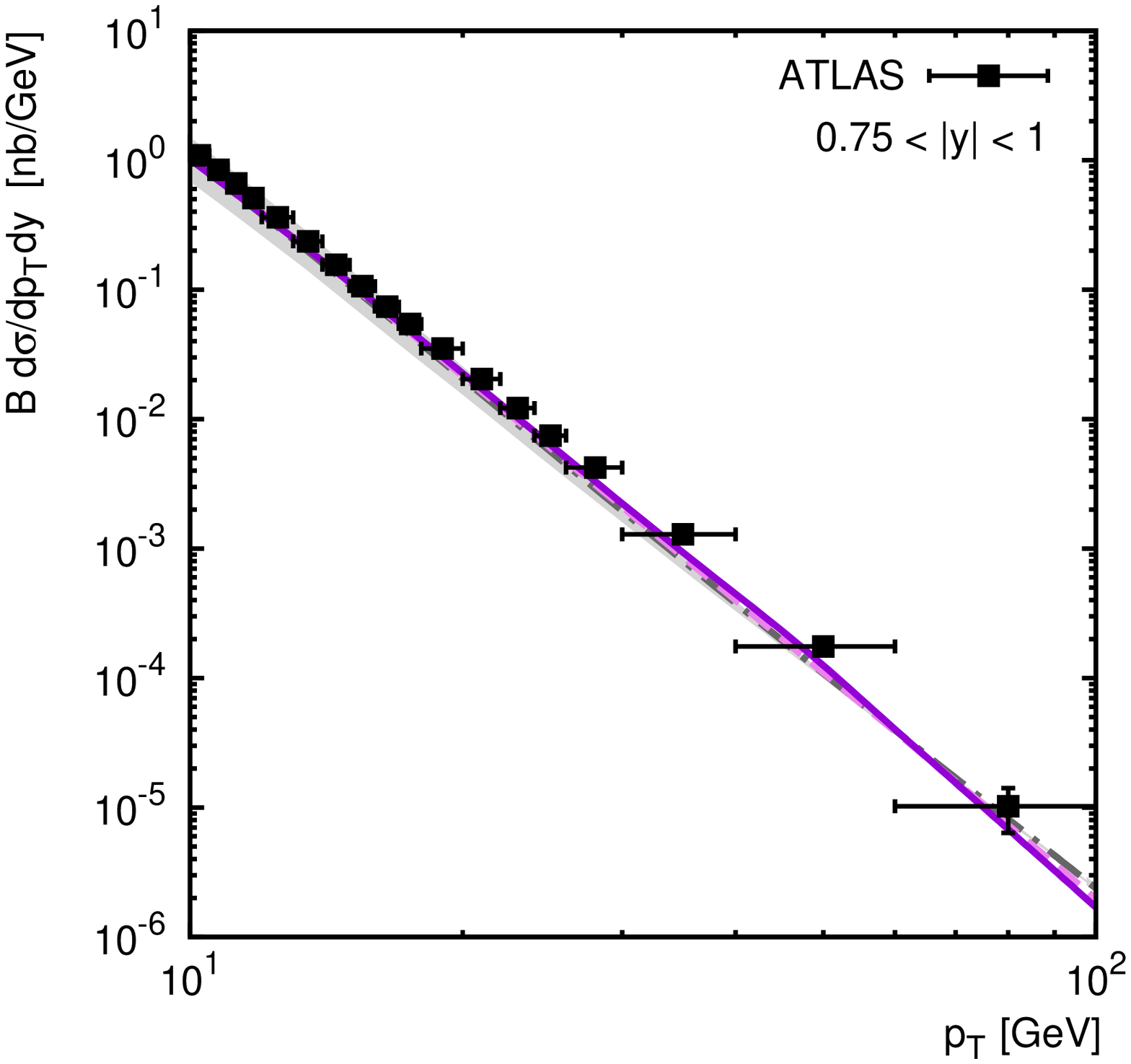, width = 5cm, angle = 270}
\epsfig{figure=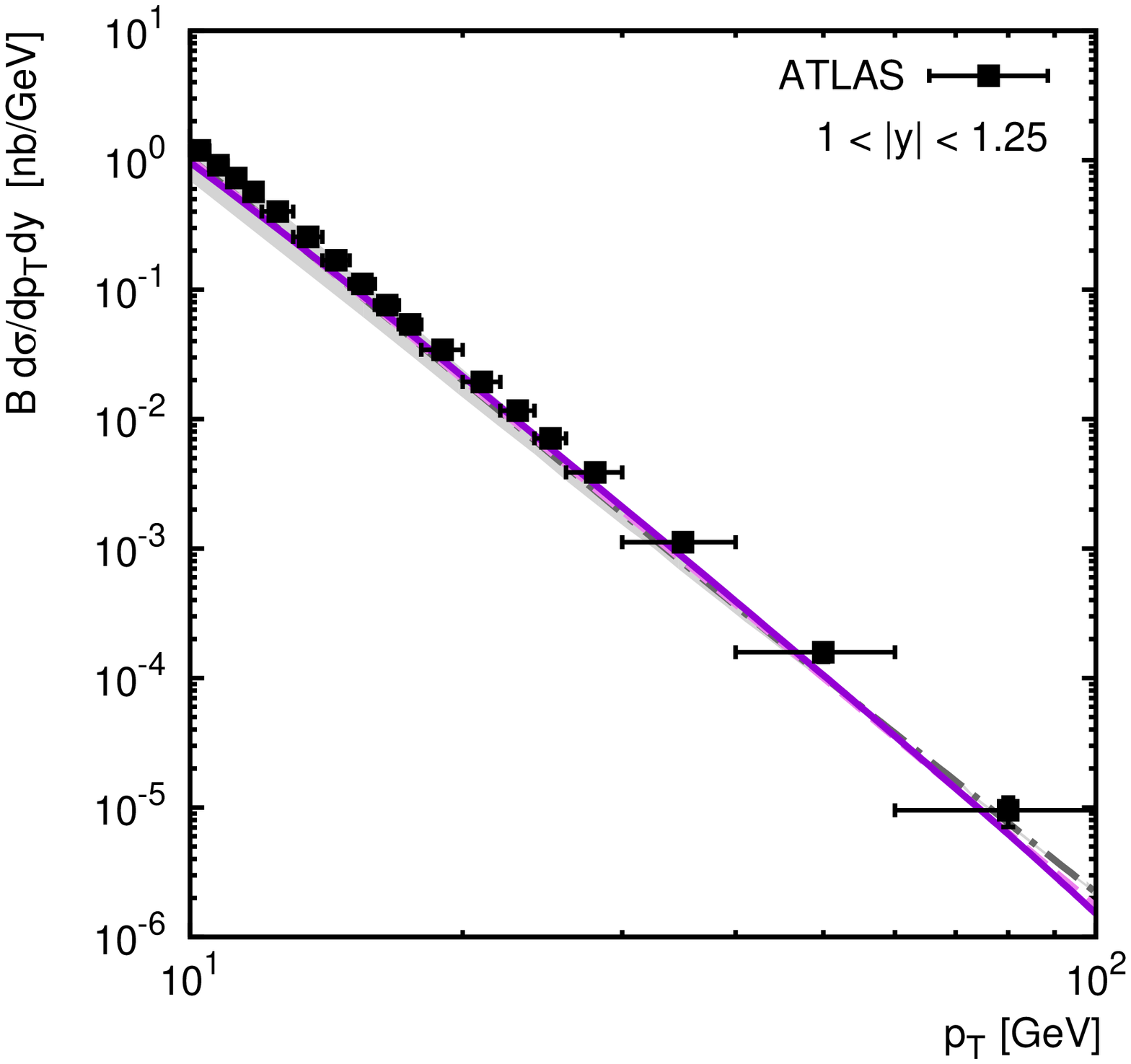, width = 5cm, angle = 270}
\epsfig{figure=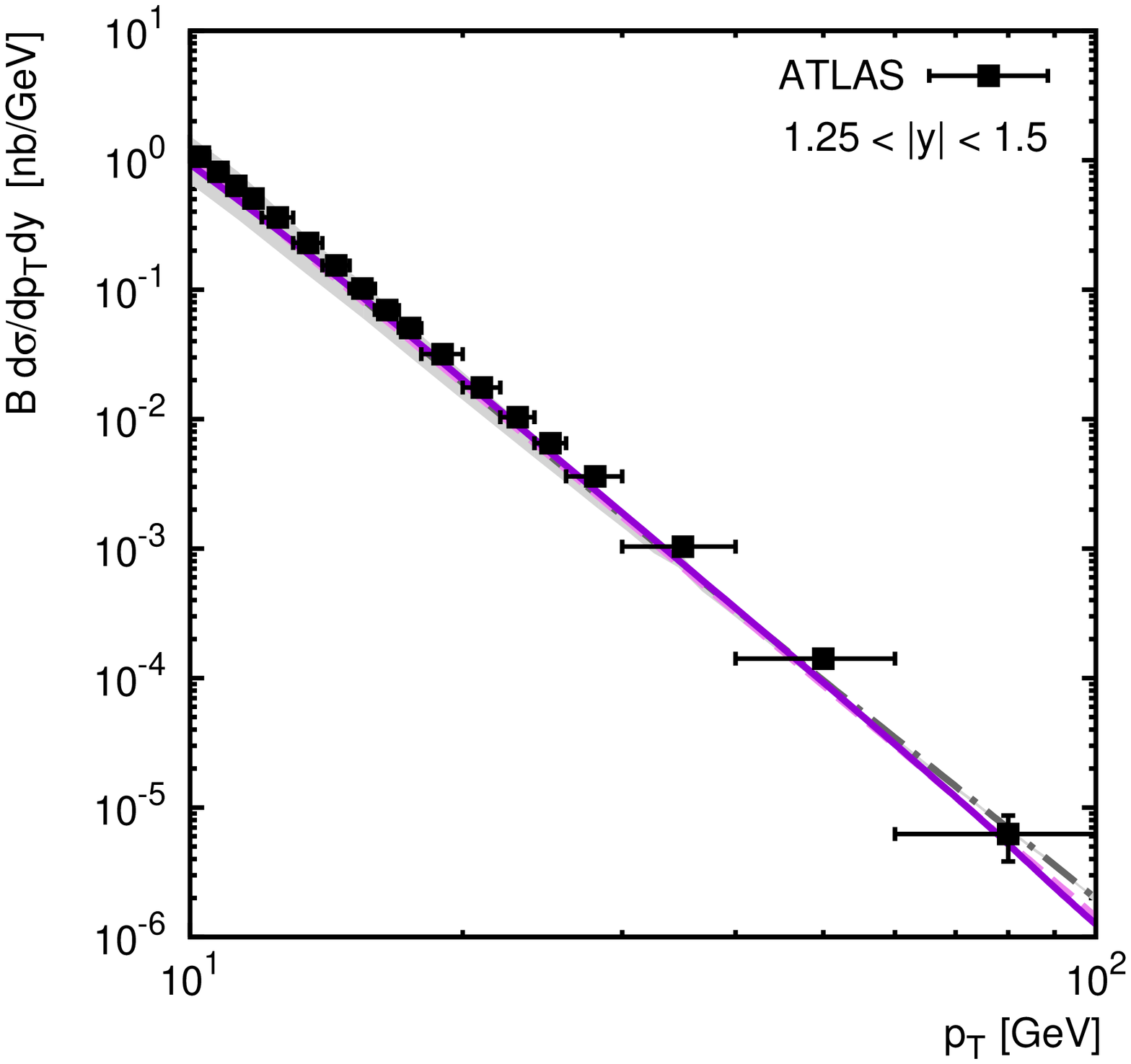, width = 5cm, angle = 270}
\epsfig{figure=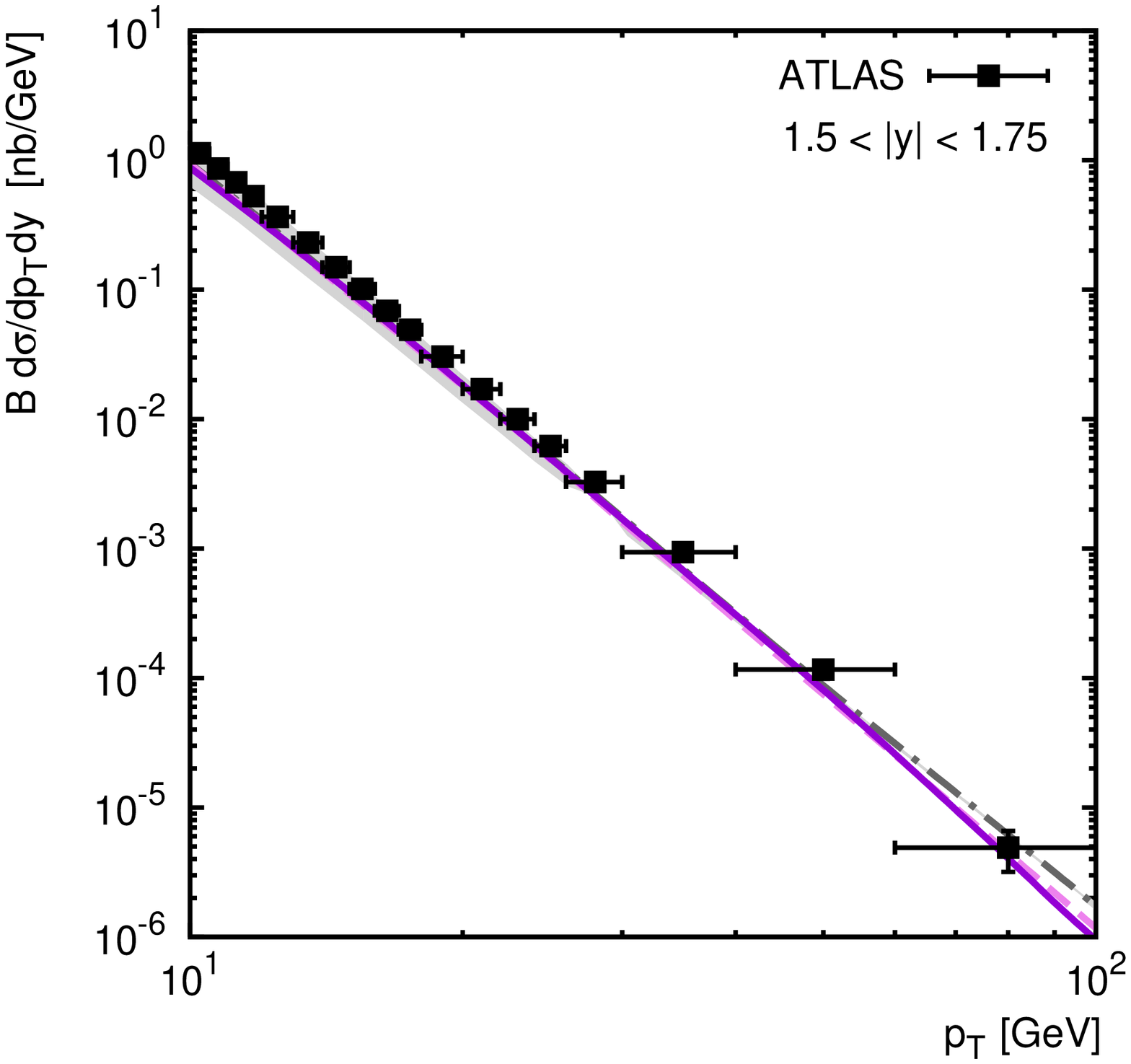, width = 5cm, angle = 270}
\epsfig{figure=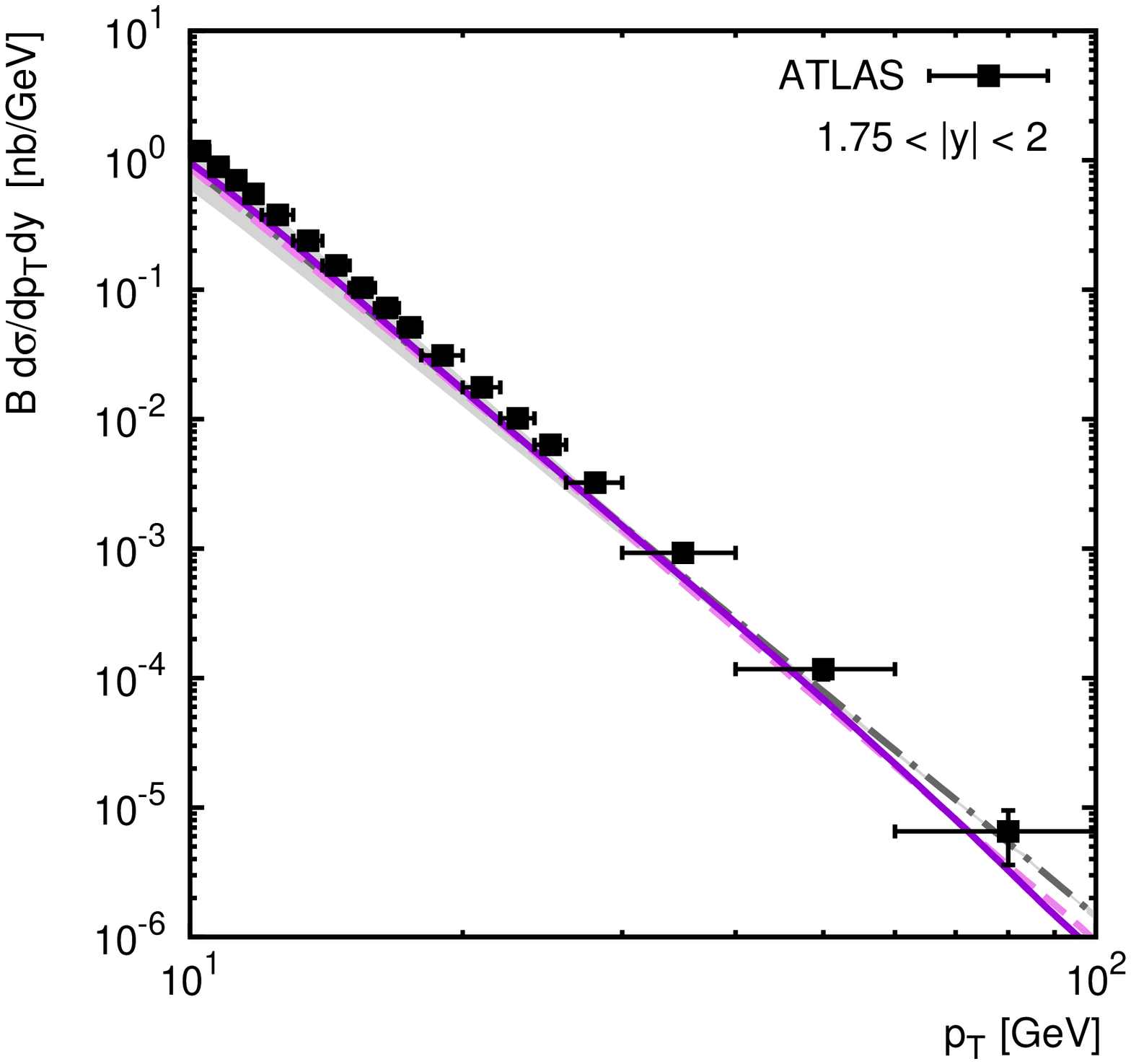, width = 5cm, angle = 270}
\caption{The double differential cross sections of prompt $J/\psi$ meson production
in $pp$ collisions at $\sqrt s = 7$~TeV. The solid, dashed and dash-dotted curves correspond to the 
predictions obtained with the A0, JH and KMR gluon densities, 
respectively. The shaded bands represent the usual scale variations in the 
KMR predictions, as it is described in the text.
The experimental data are from ATLAS\cite{23}.}
\label{fig2}
\end{center}
\end{figure}

\begin{figure}
\begin{center}
\epsfig{figure=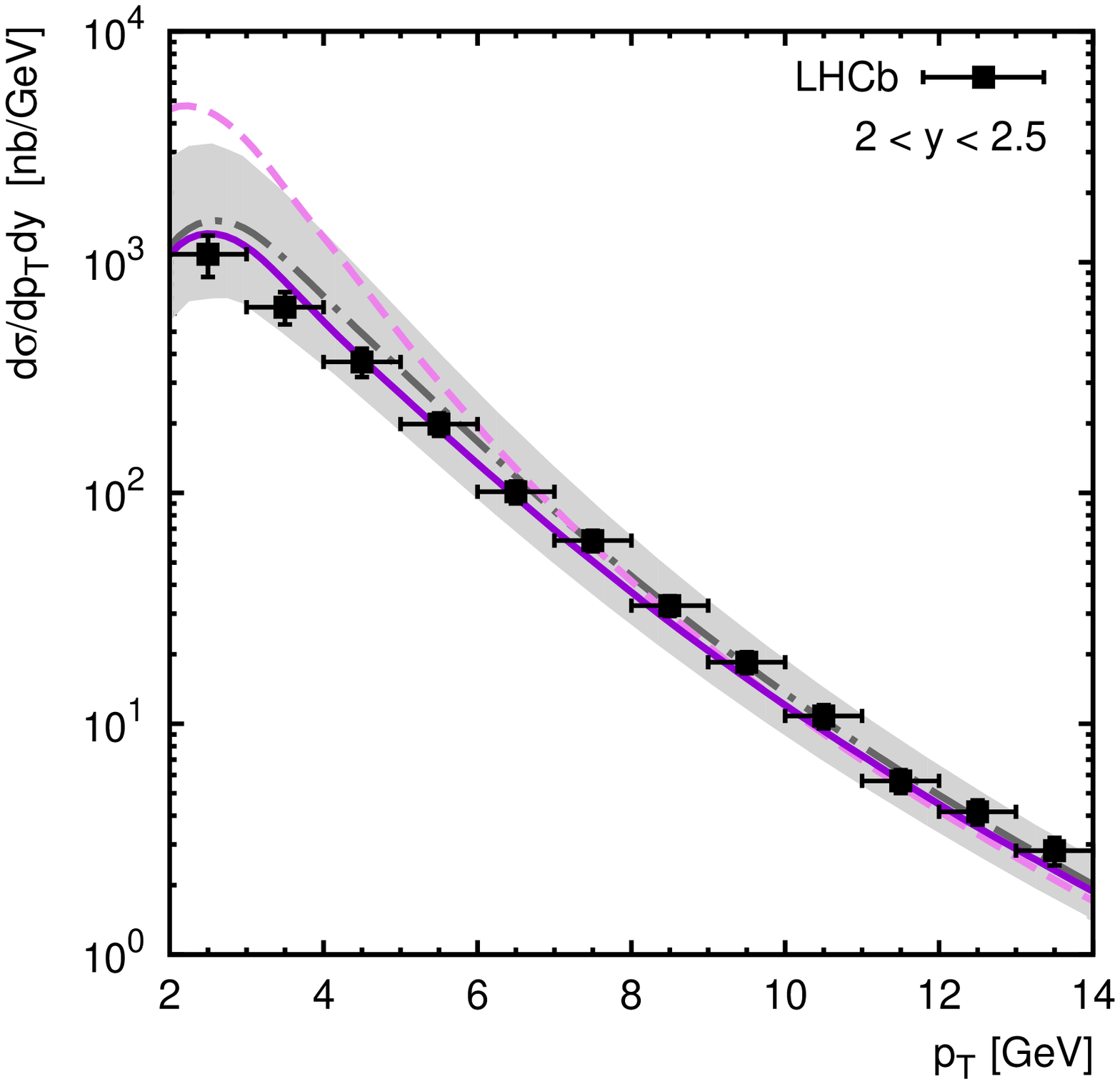, width = 5cm, angle = 270} 
\epsfig{figure=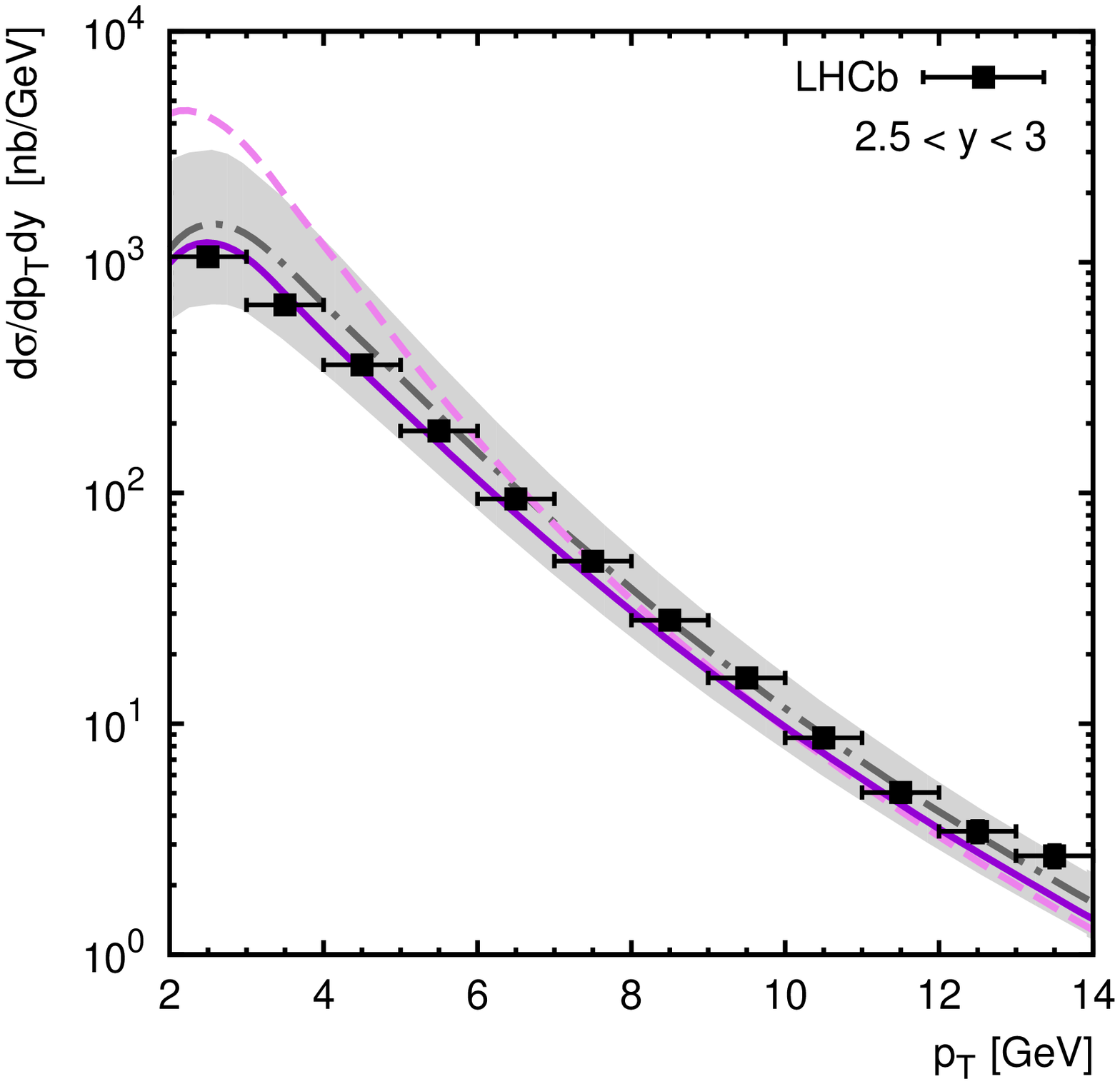, width = 5cm, angle = 270} 
\epsfig{figure=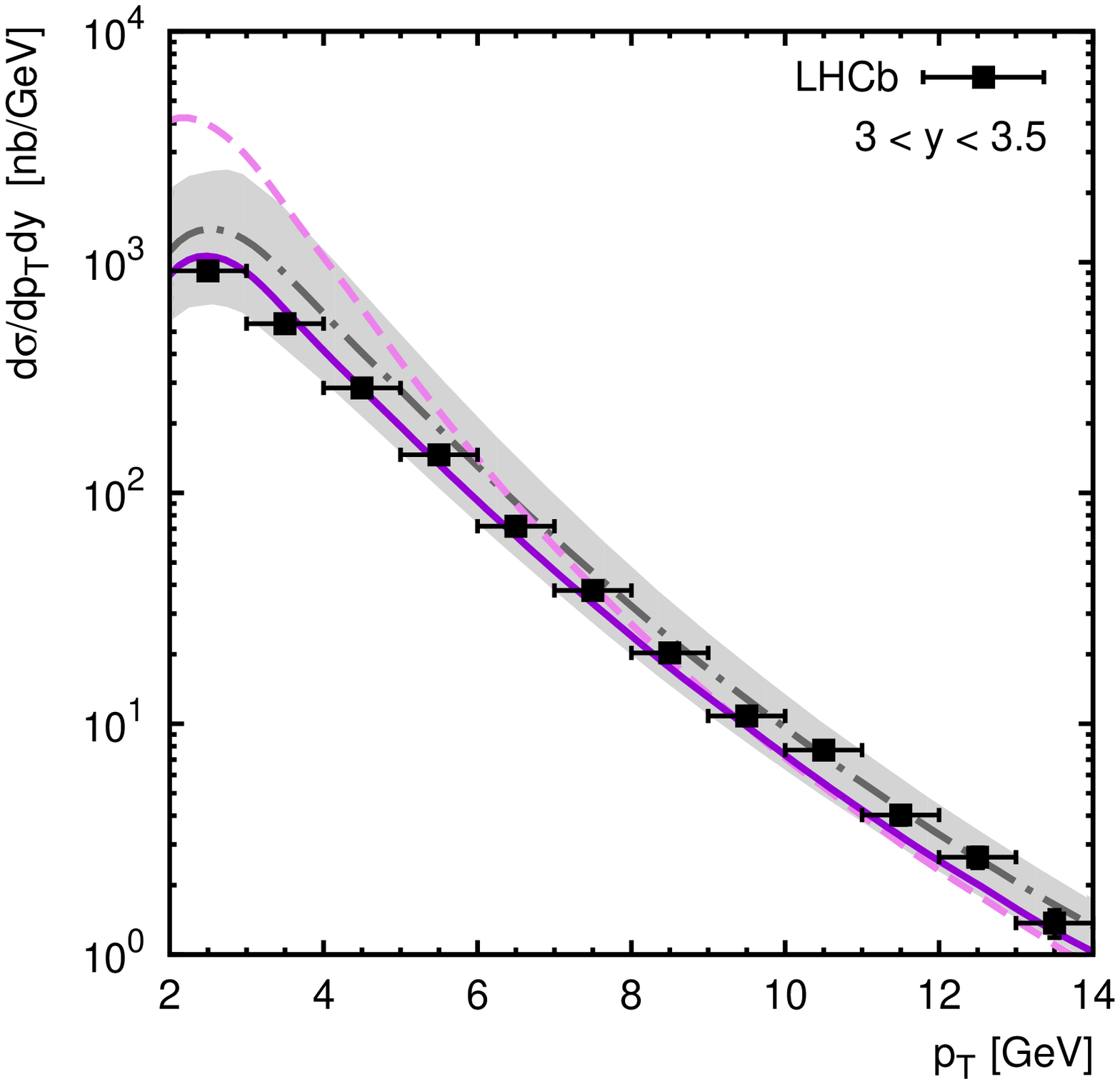, width = 5cm, angle = 270}
\epsfig{figure=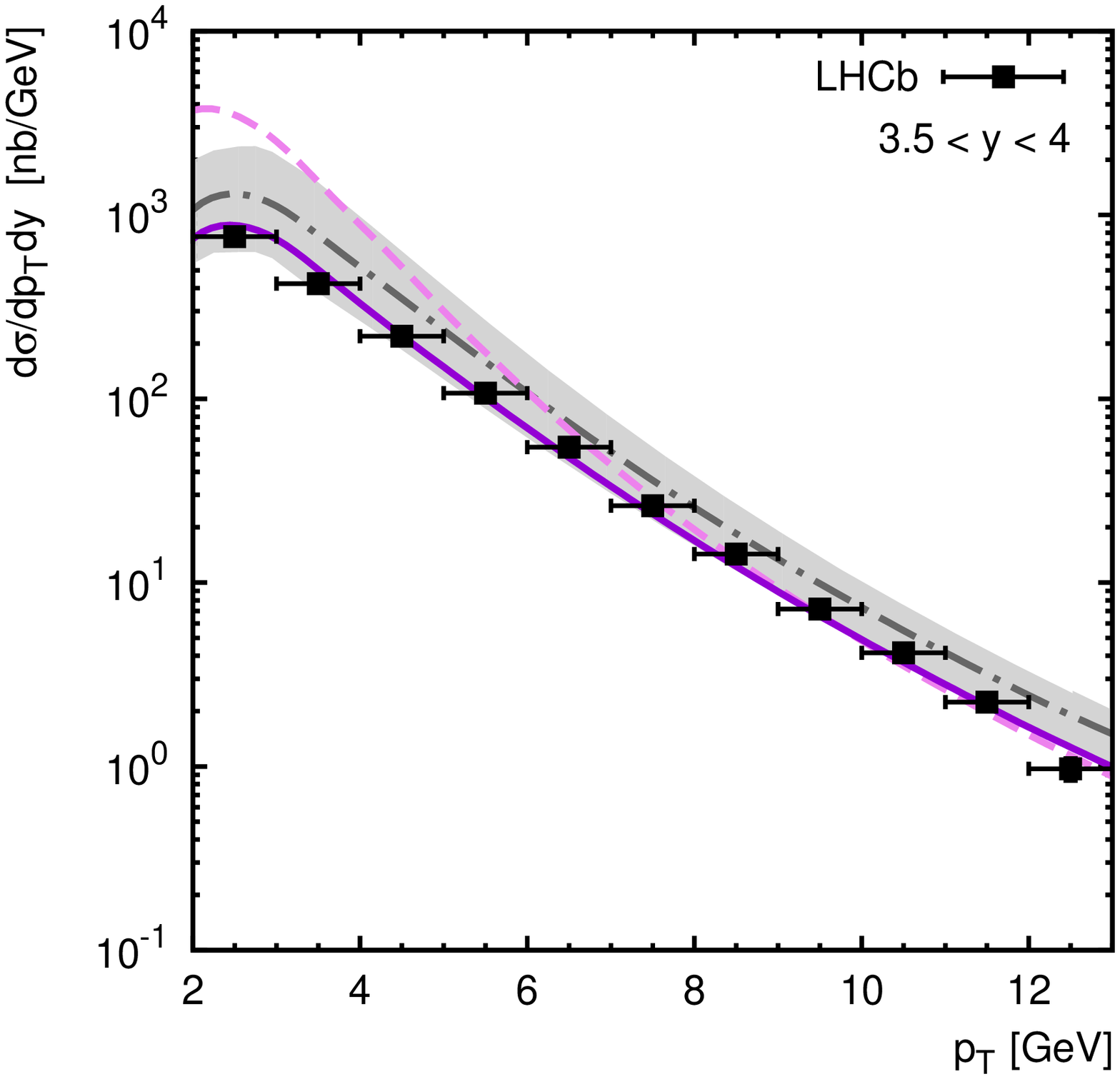, width = 5cm, angle = 270}
\epsfig{figure=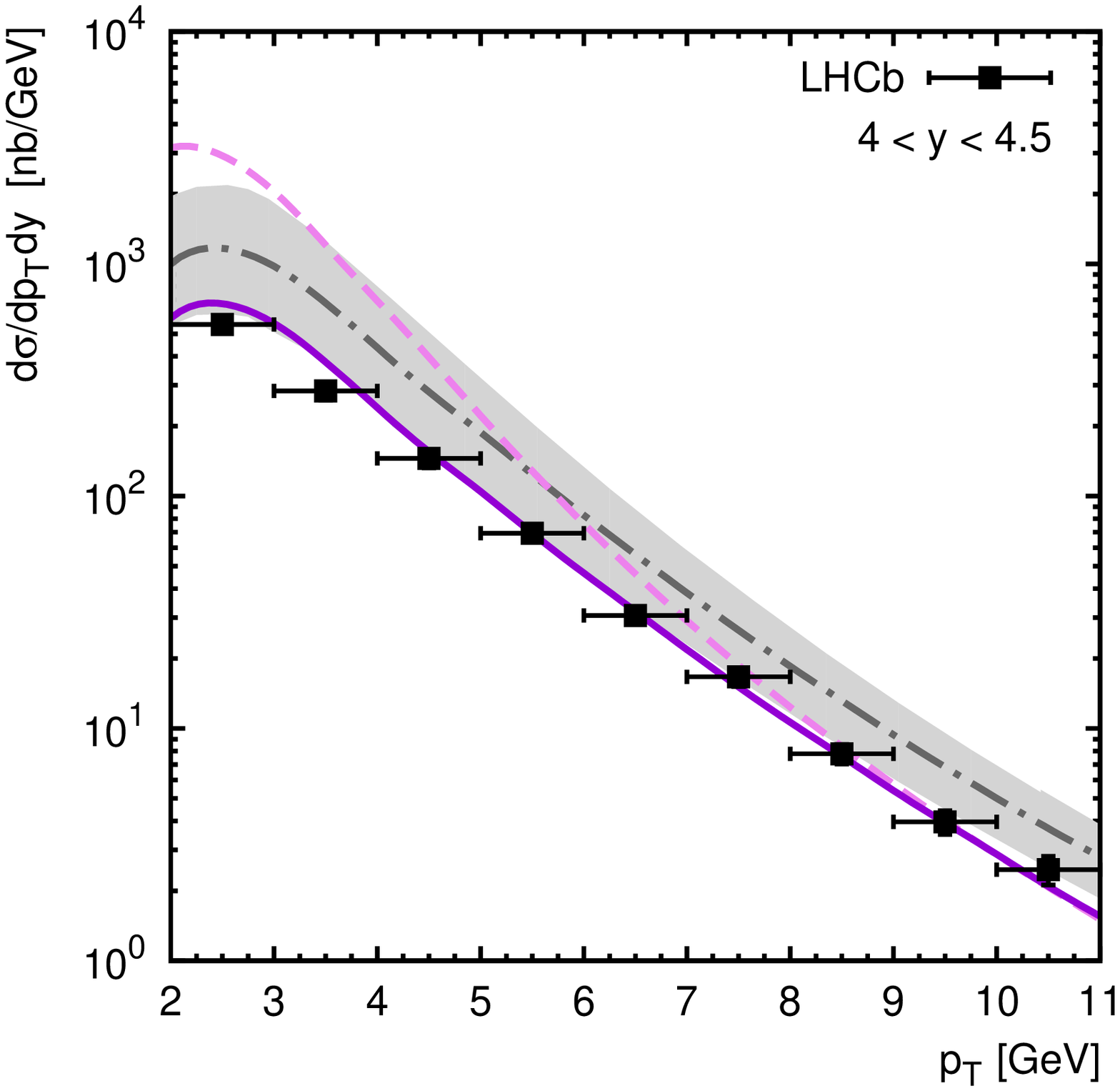, width = 5cm, angle = 270}
\caption{The double differential cross sections of prompt $J/\psi$ meson production in $pp$ collisions
at $\sqrt s = 7$~TeV. Notation of all curves is the same as in Fig.~2.
The experimental data are from LHCb\cite{20}.}
\label{fig3}
\end{center}
\end{figure}

\begin{figure}
\begin{center}
\epsfig{figure=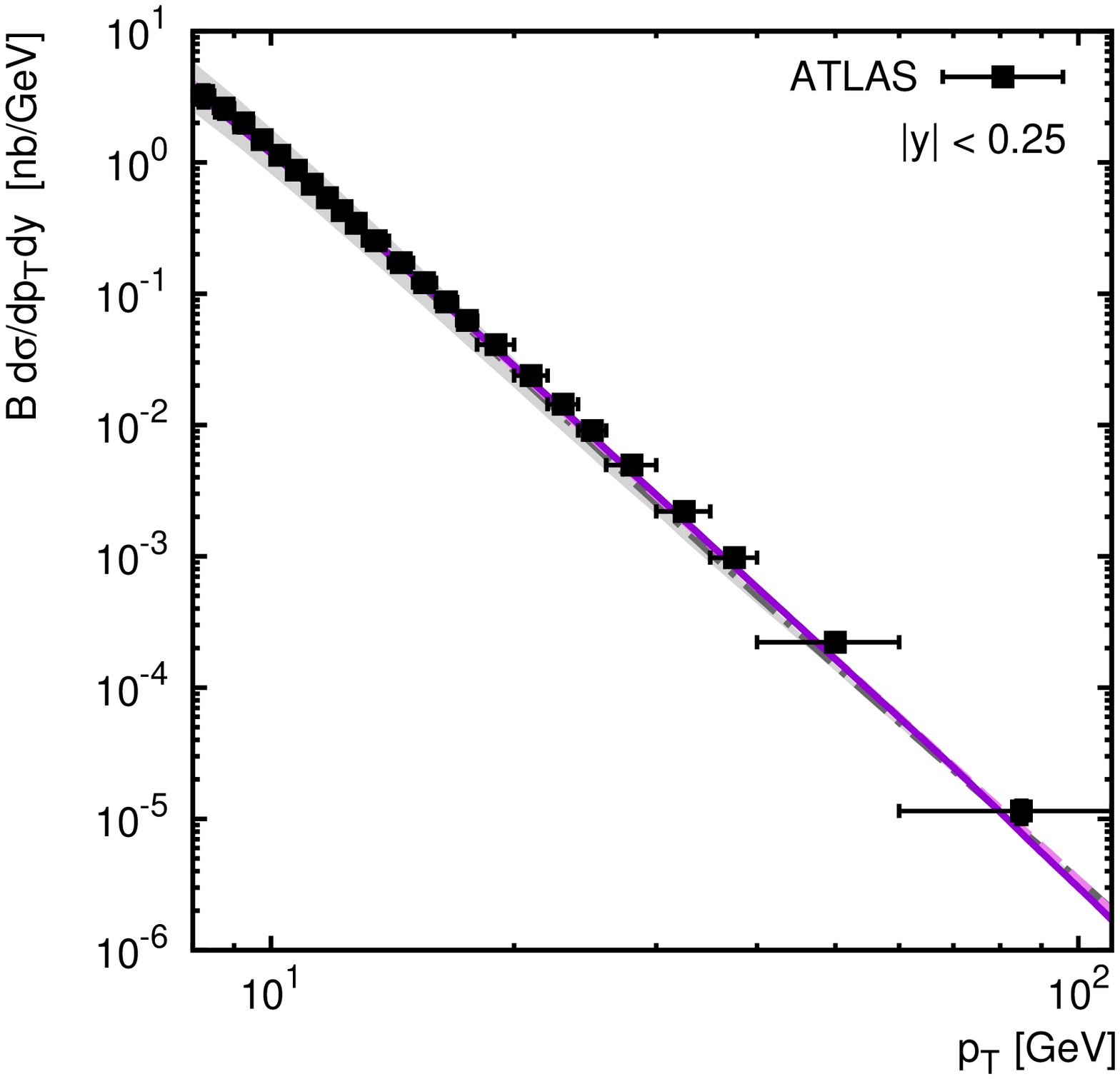, width = 5cm, angle = 270} 
\epsfig{figure=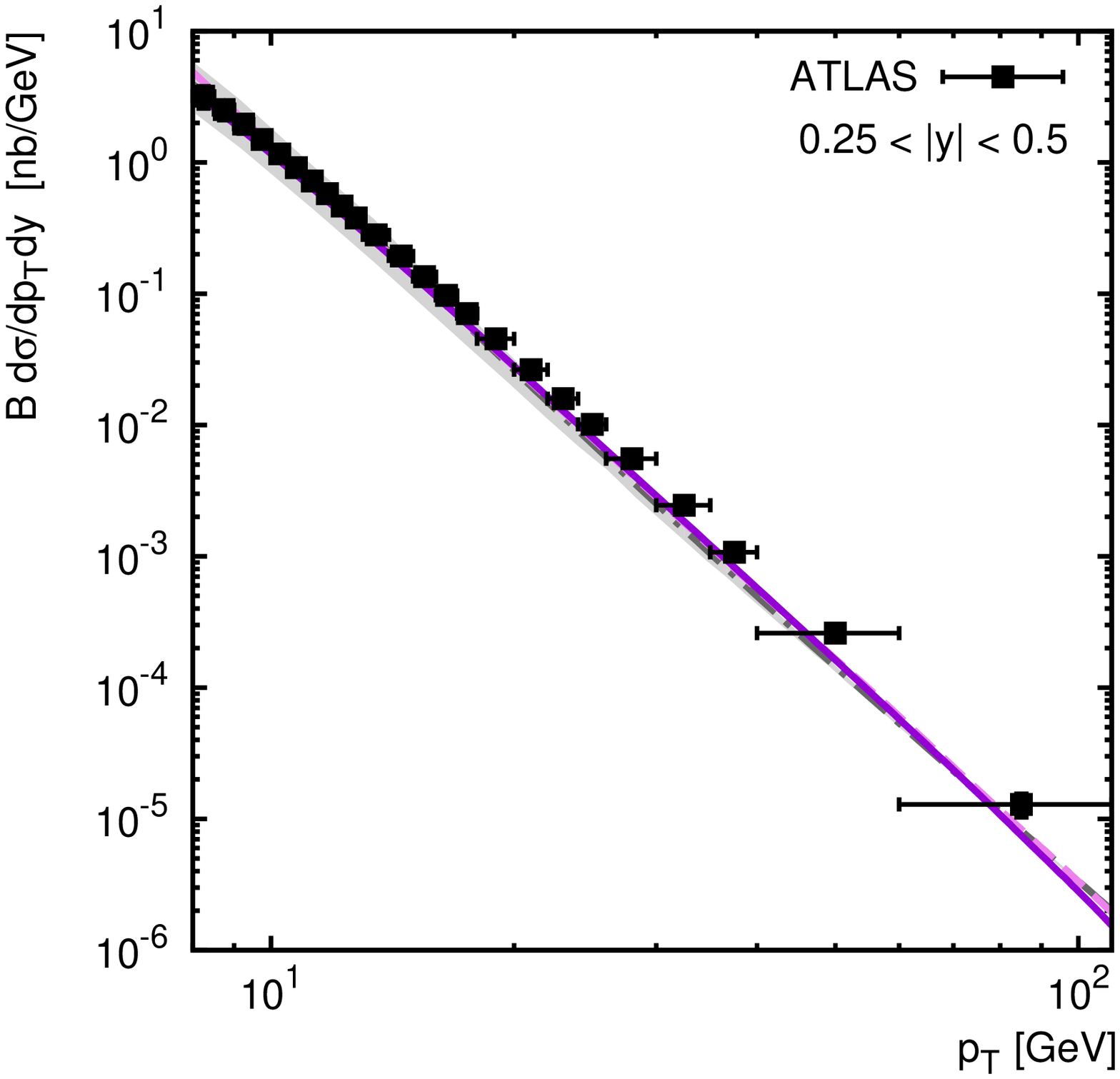, width = 5cm, angle = 270} 
\epsfig{figure=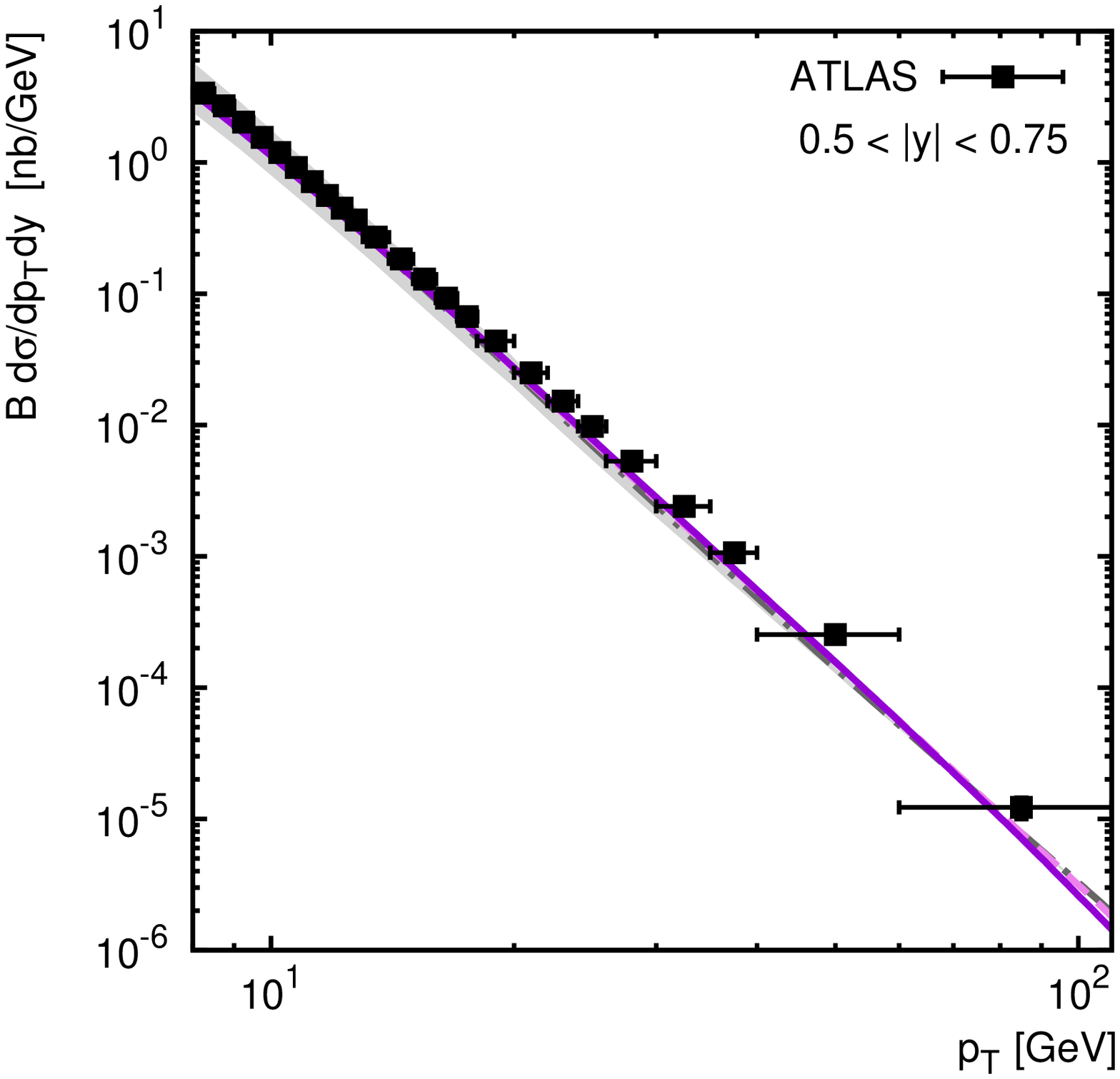, width = 5cm, angle = 270}
\epsfig{figure=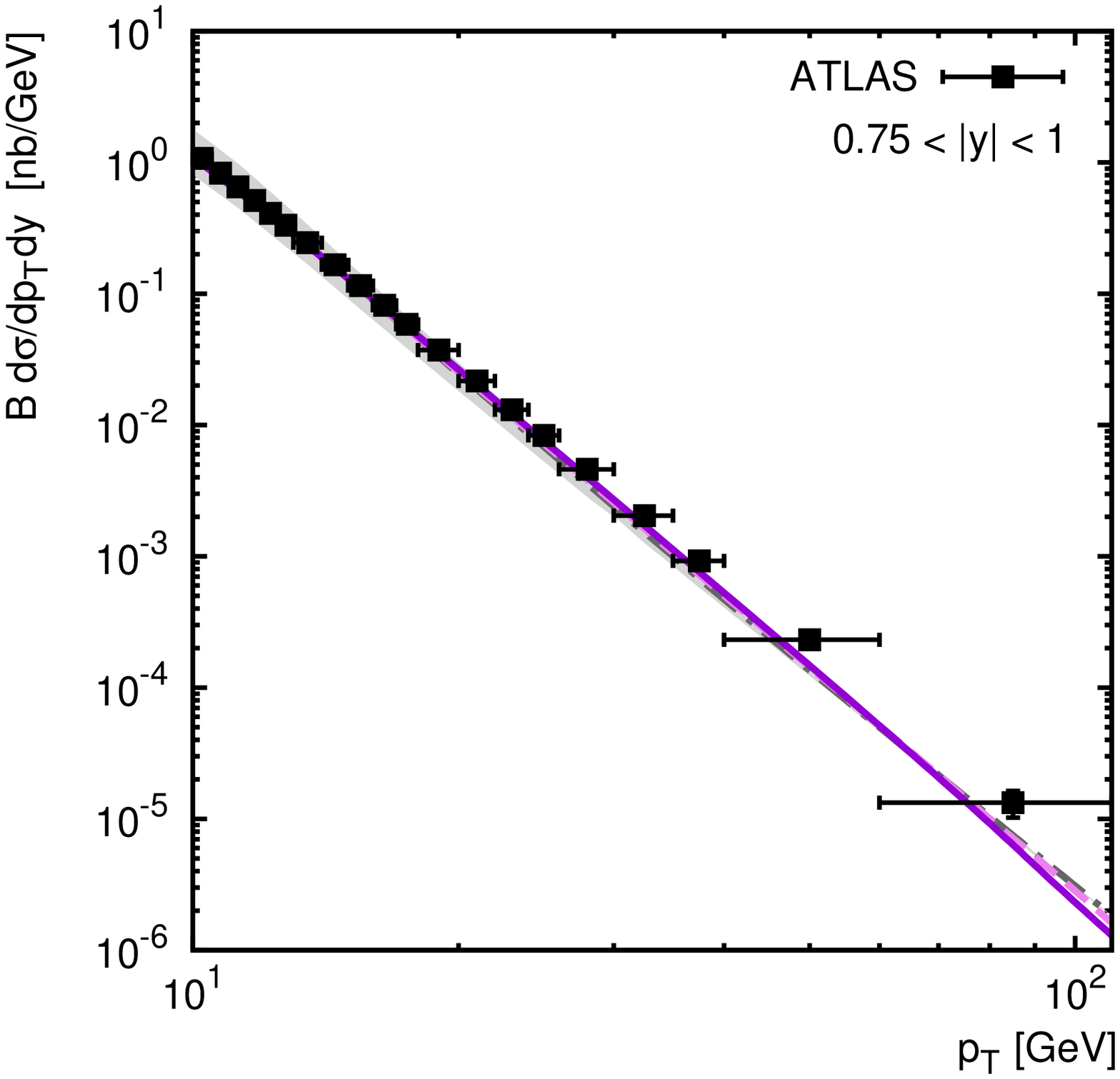, width = 5cm, angle = 270}
\epsfig{figure=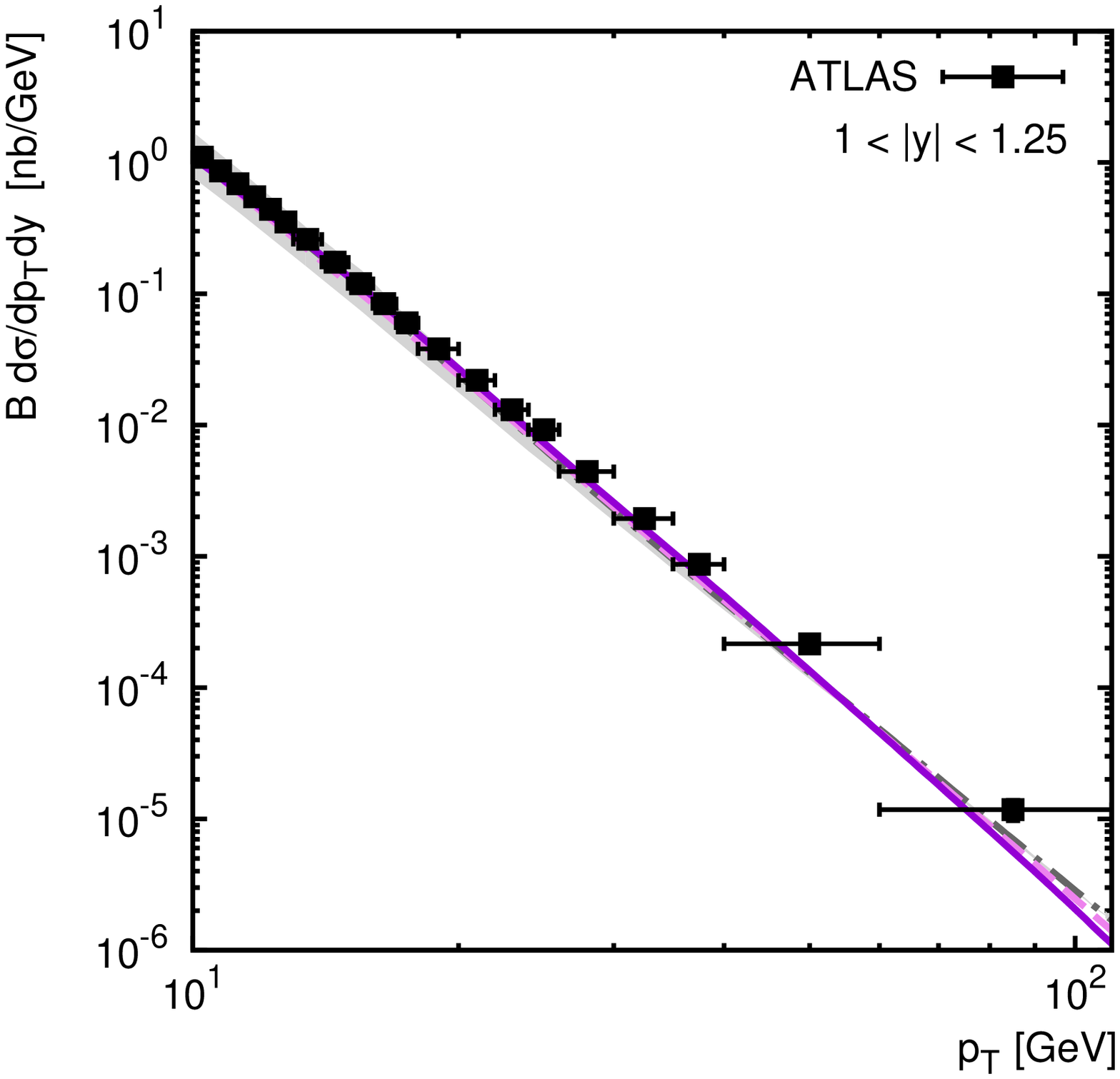, width = 5cm, angle = 270}
\epsfig{figure=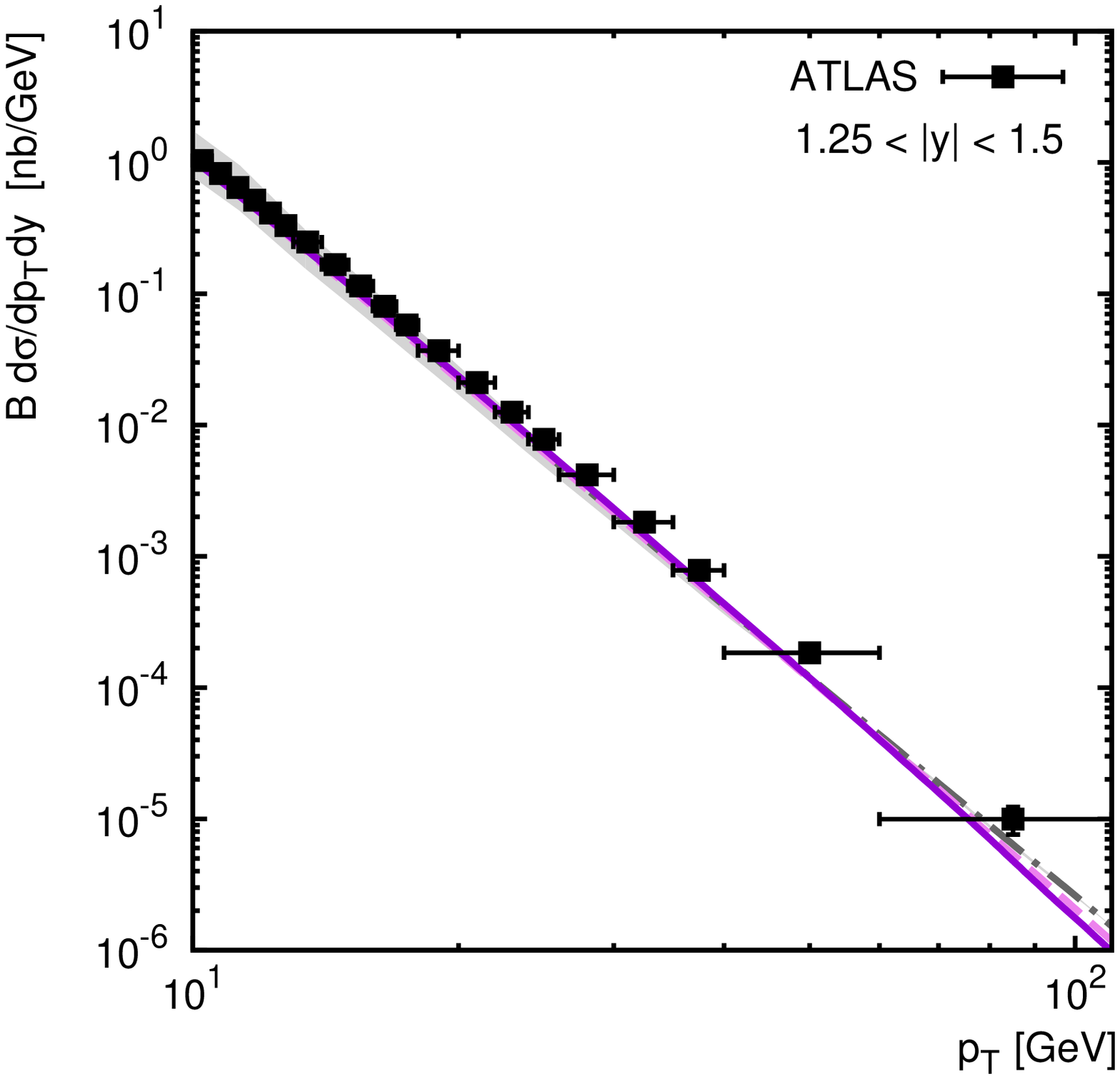, width = 5cm, angle = 270}
\epsfig{figure=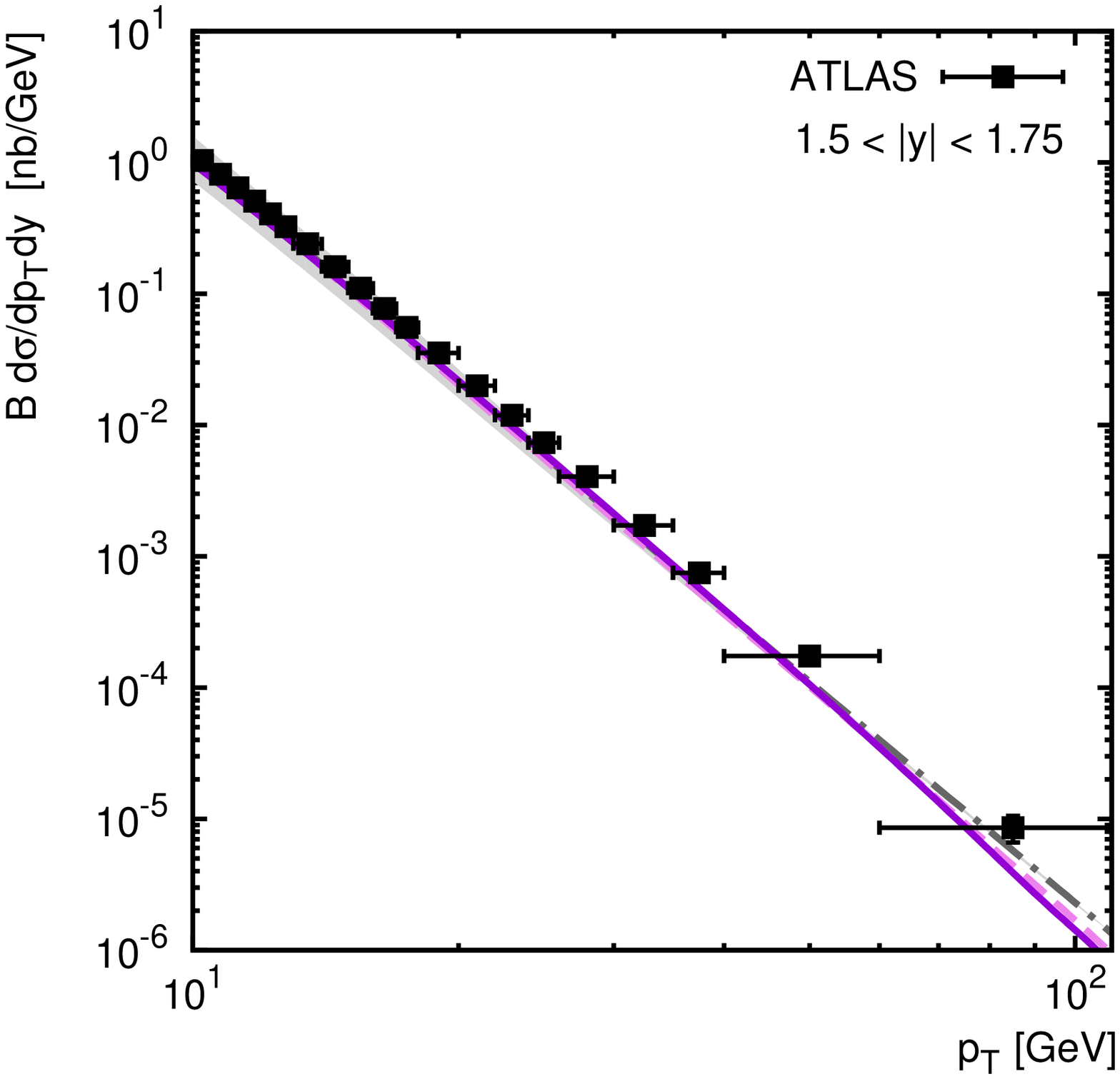, width = 5cm, angle = 270}
\epsfig{figure=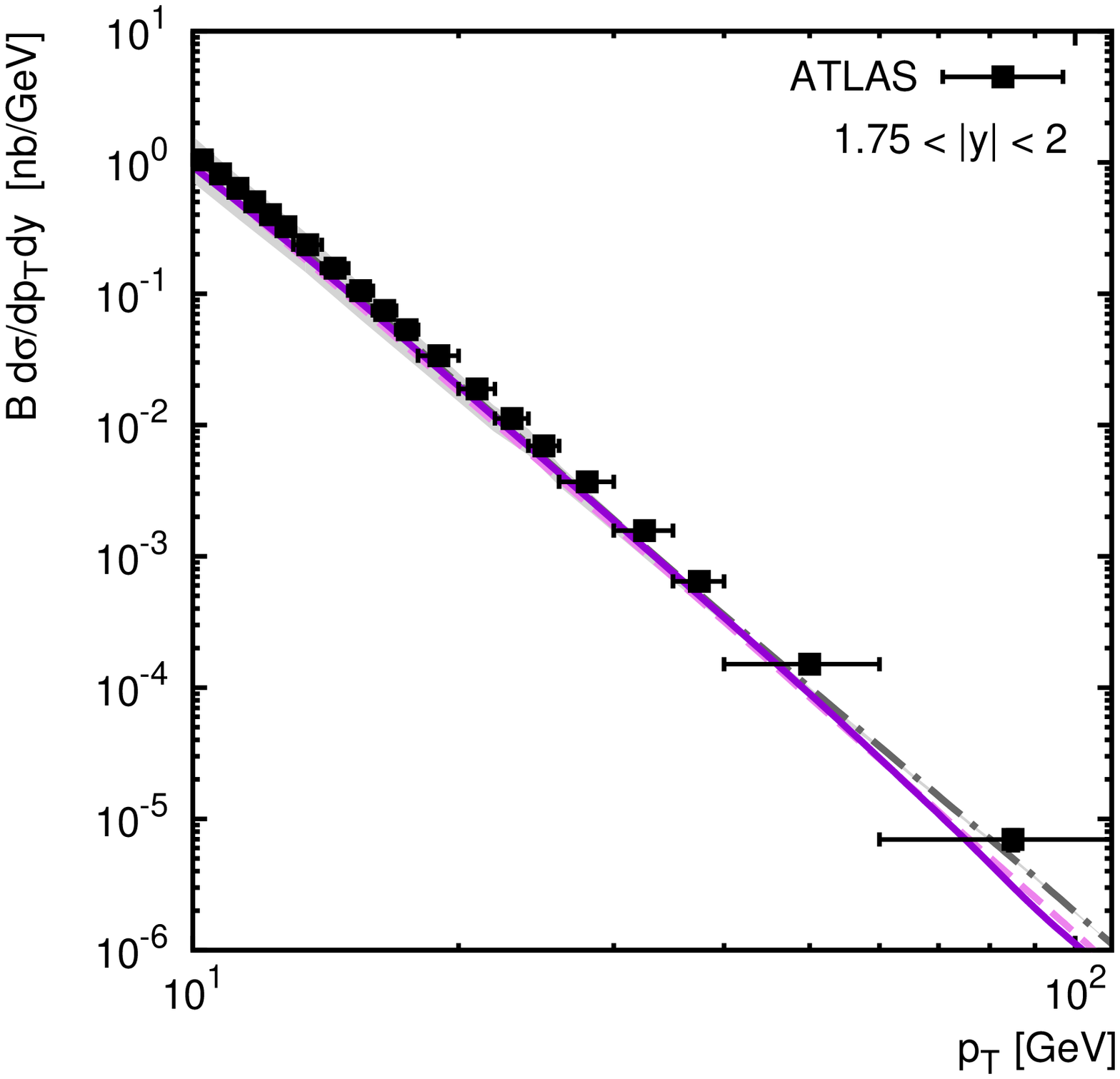, width = 5cm, angle = 270}
\caption{The double differential cross sections of prompt $J/\psi$ meson production in $pp$ collisions
at $\sqrt s = 8$~TeV. Notation of all curves is the same as in Fig.~2.
The experimental data are from ATLAS\cite{23}.}
\label{fig4}
\end{center}
\end{figure}

\begin{figure}
\begin{center}
\epsfig{figure=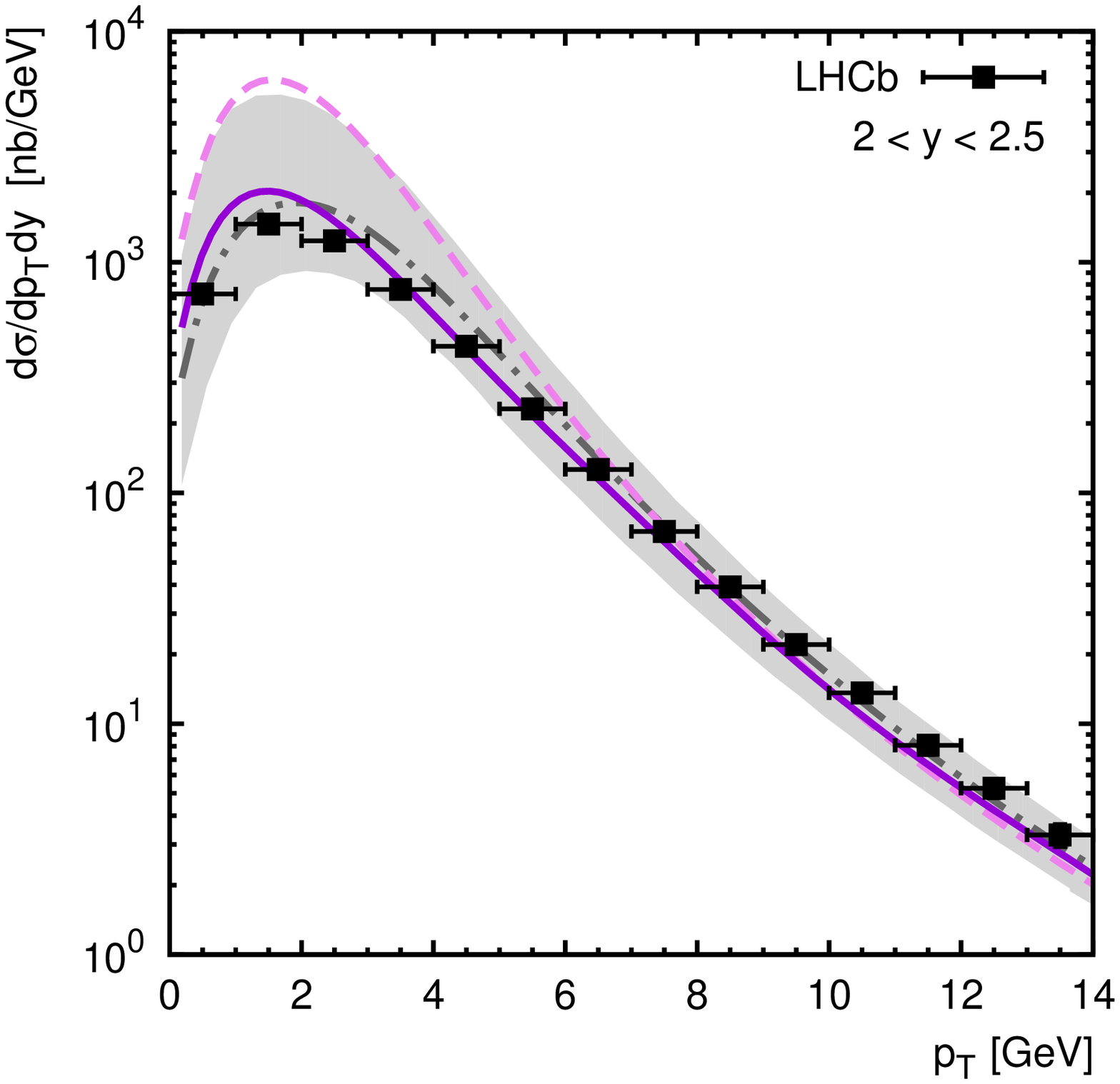, width = 5cm, angle = 270} 
\epsfig{figure=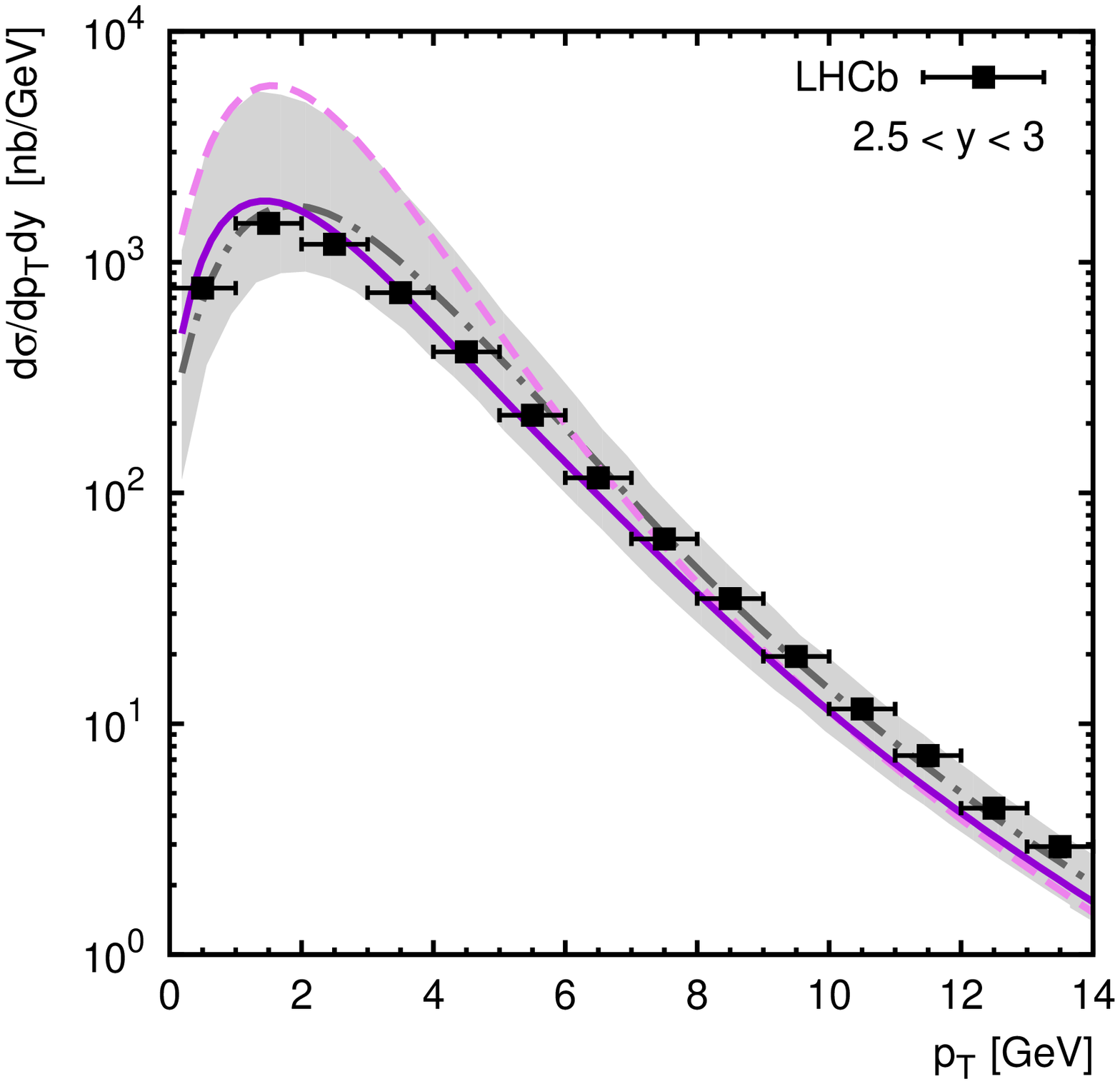, width = 5cm, angle = 270} 
\epsfig{figure=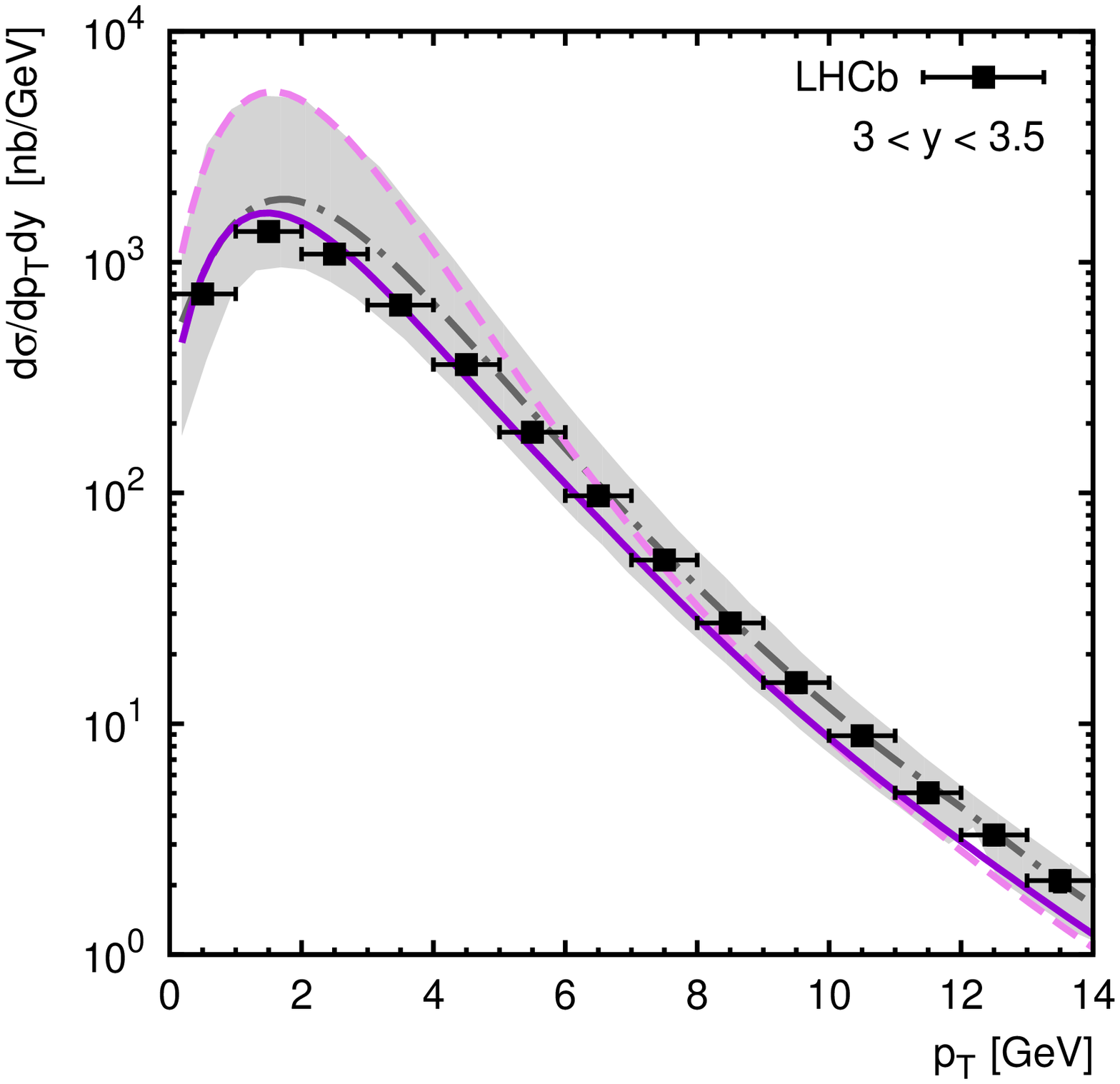, width = 5cm, angle = 270}
\epsfig{figure=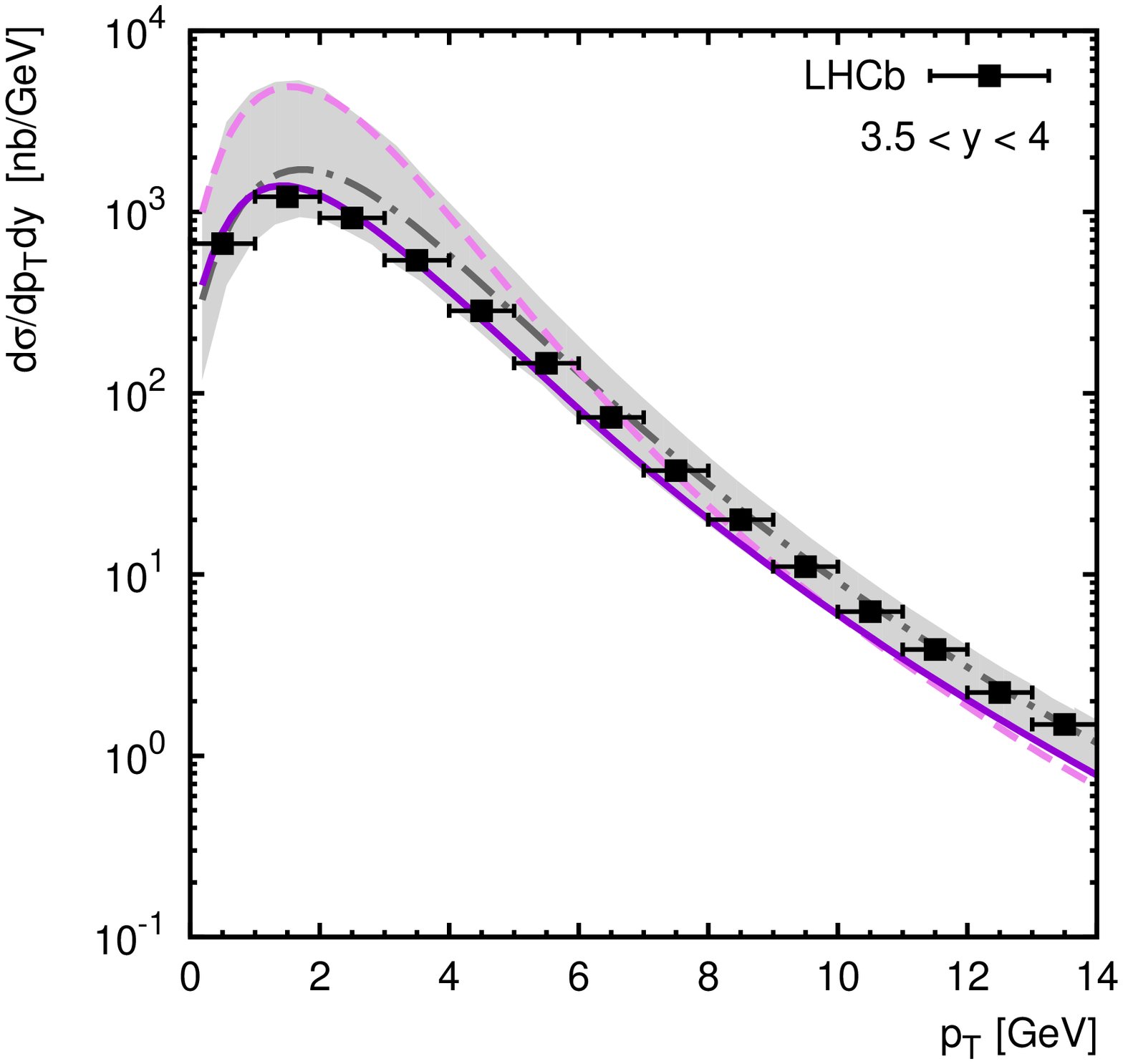, width = 5cm, angle = 270}
\epsfig{figure=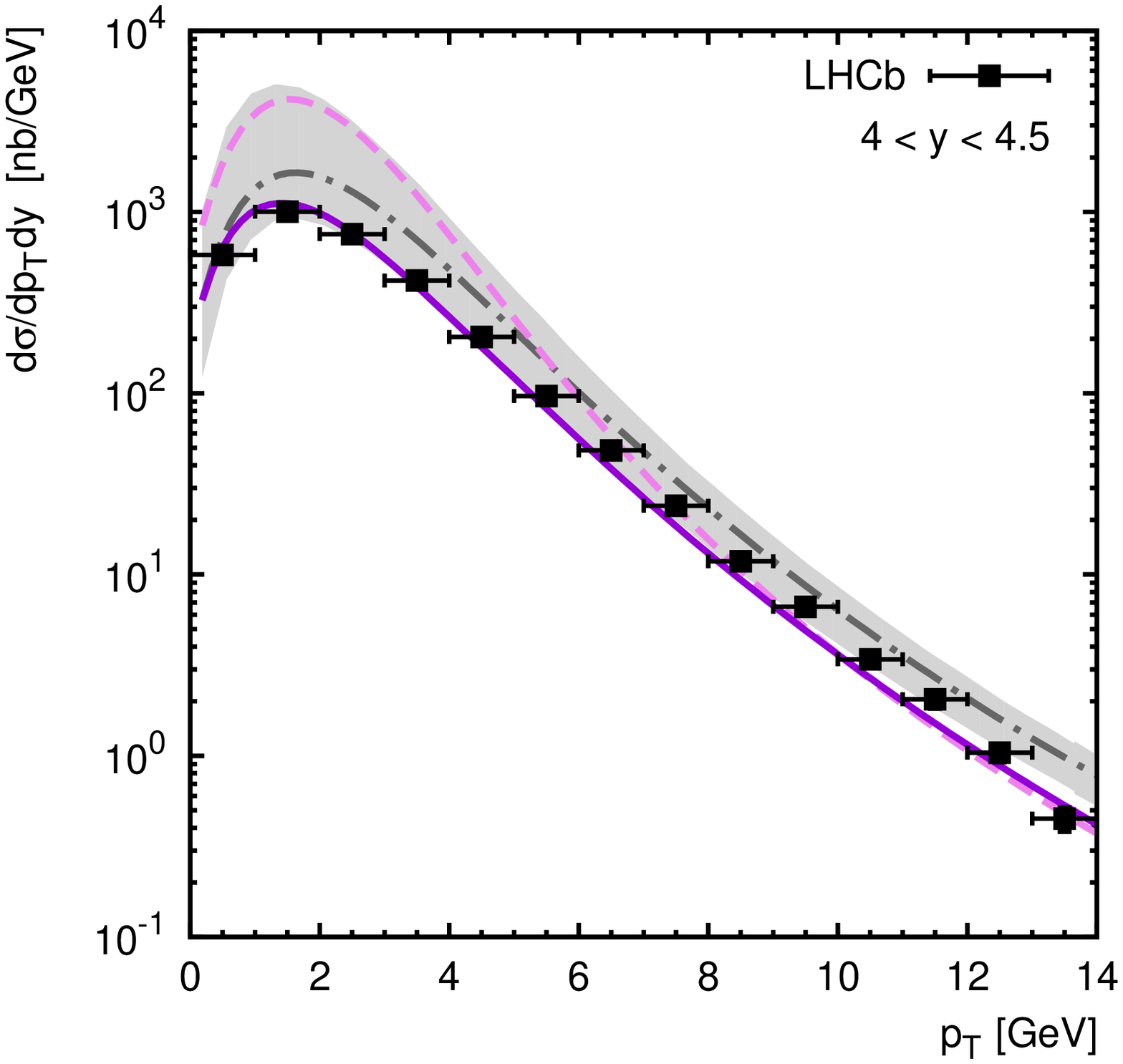, width = 5cm, angle = 270}
\caption{The double differential cross sections of prompt $J/\psi$ meson production in $pp$ collisions
at $\sqrt s = 8$~TeV. Notation of all curves is the same as in Fig.~2.
The experimental data are from LHCb\cite{25}.}
\label{fig5}
\end{center}
\end{figure}

\begin{figure}
\begin{center}
\epsfig{figure=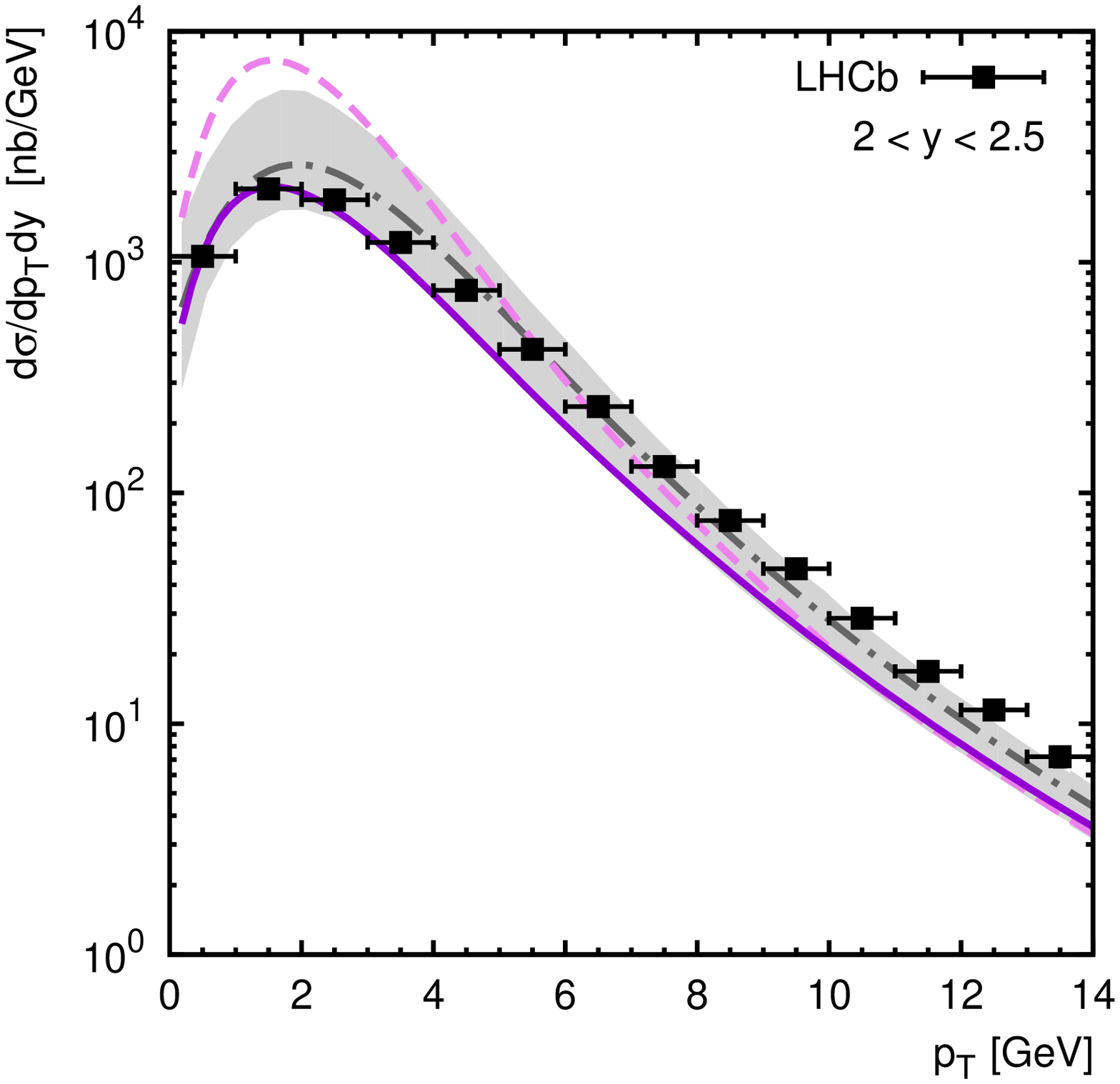, width = 5cm, angle = 270} 
\epsfig{figure=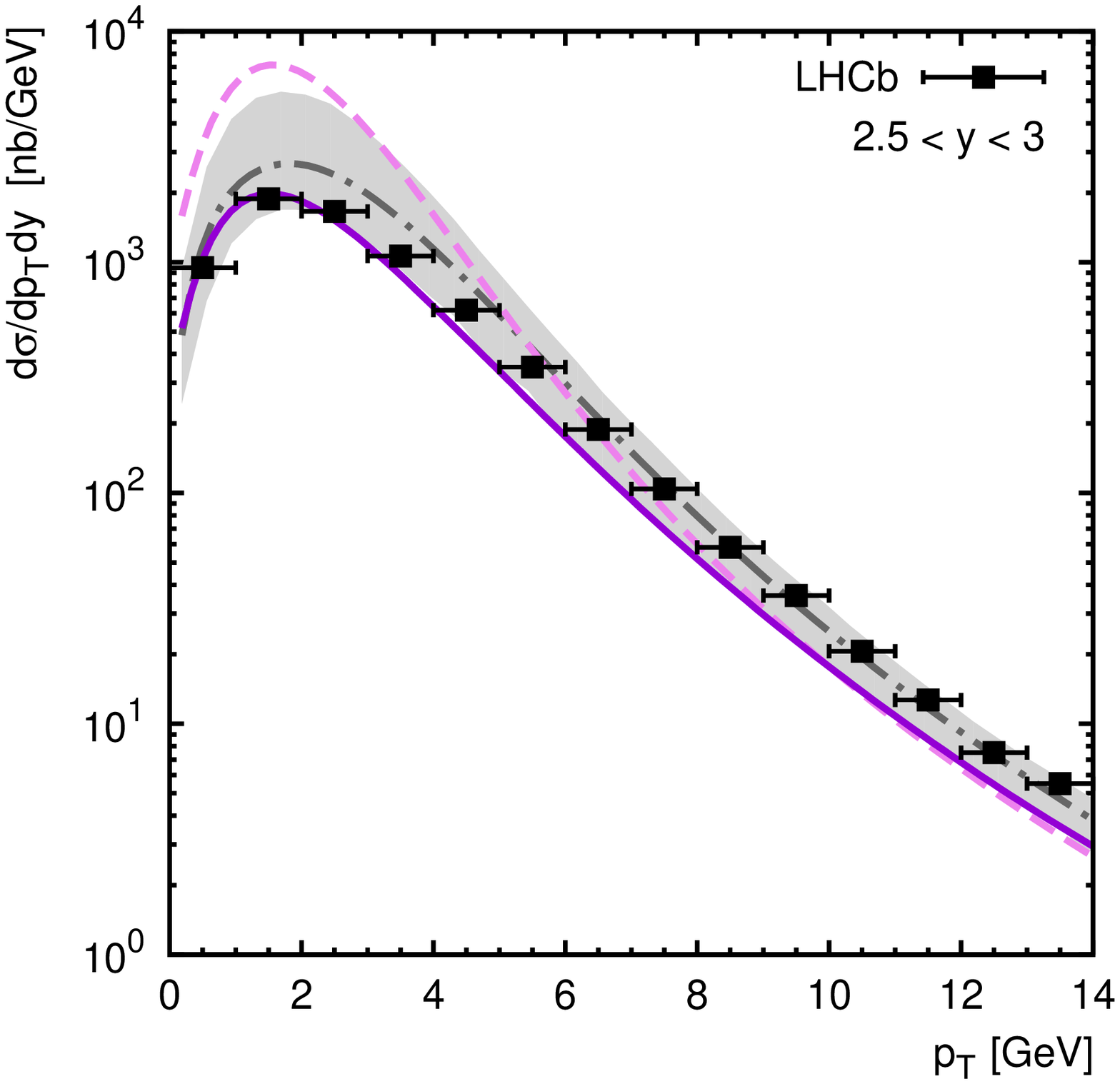, width = 5cm, angle = 270} 
\epsfig{figure=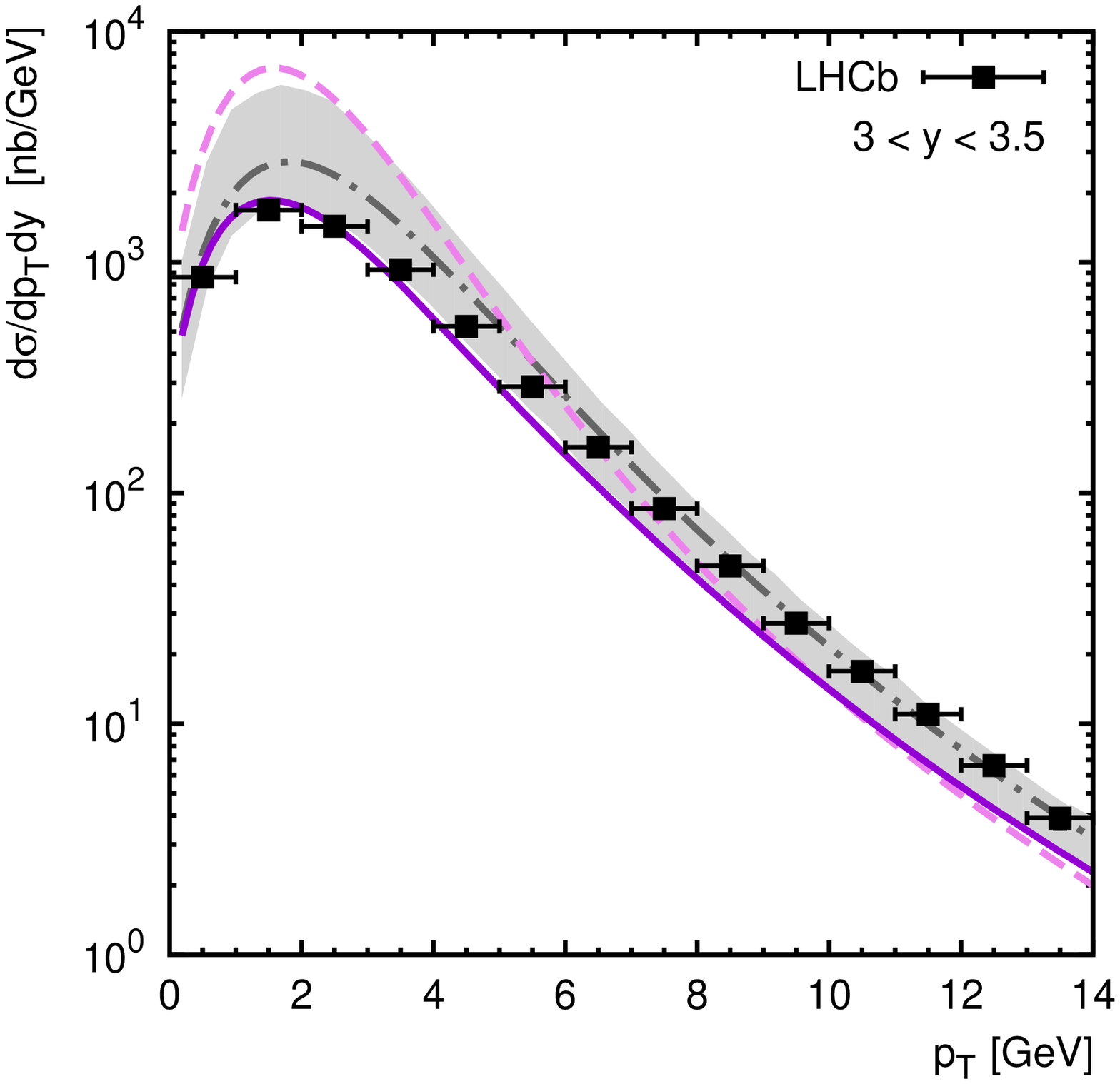, width = 5cm, angle = 270}
\epsfig{figure=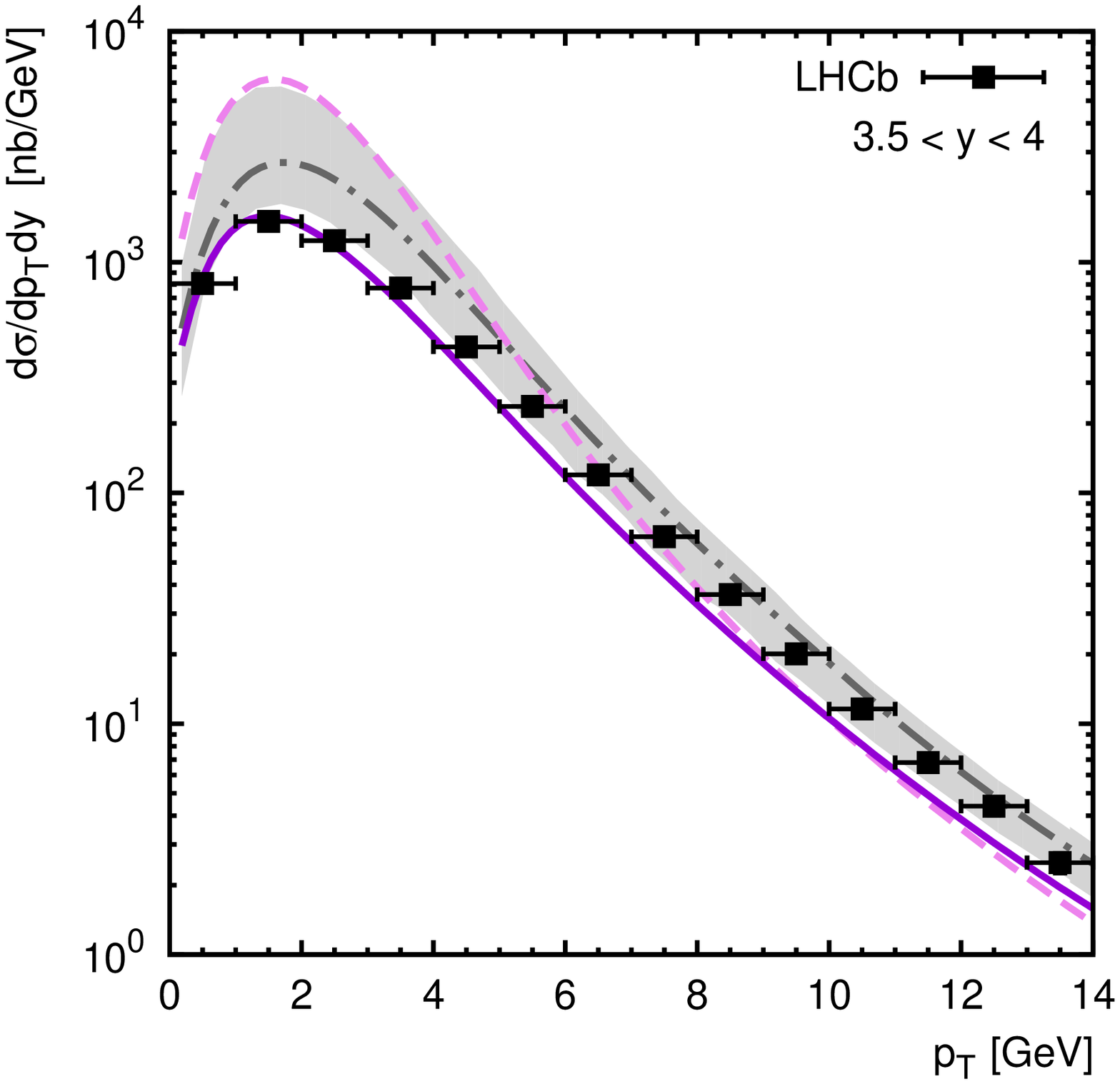, width = 5cm, angle = 270}
\epsfig{figure=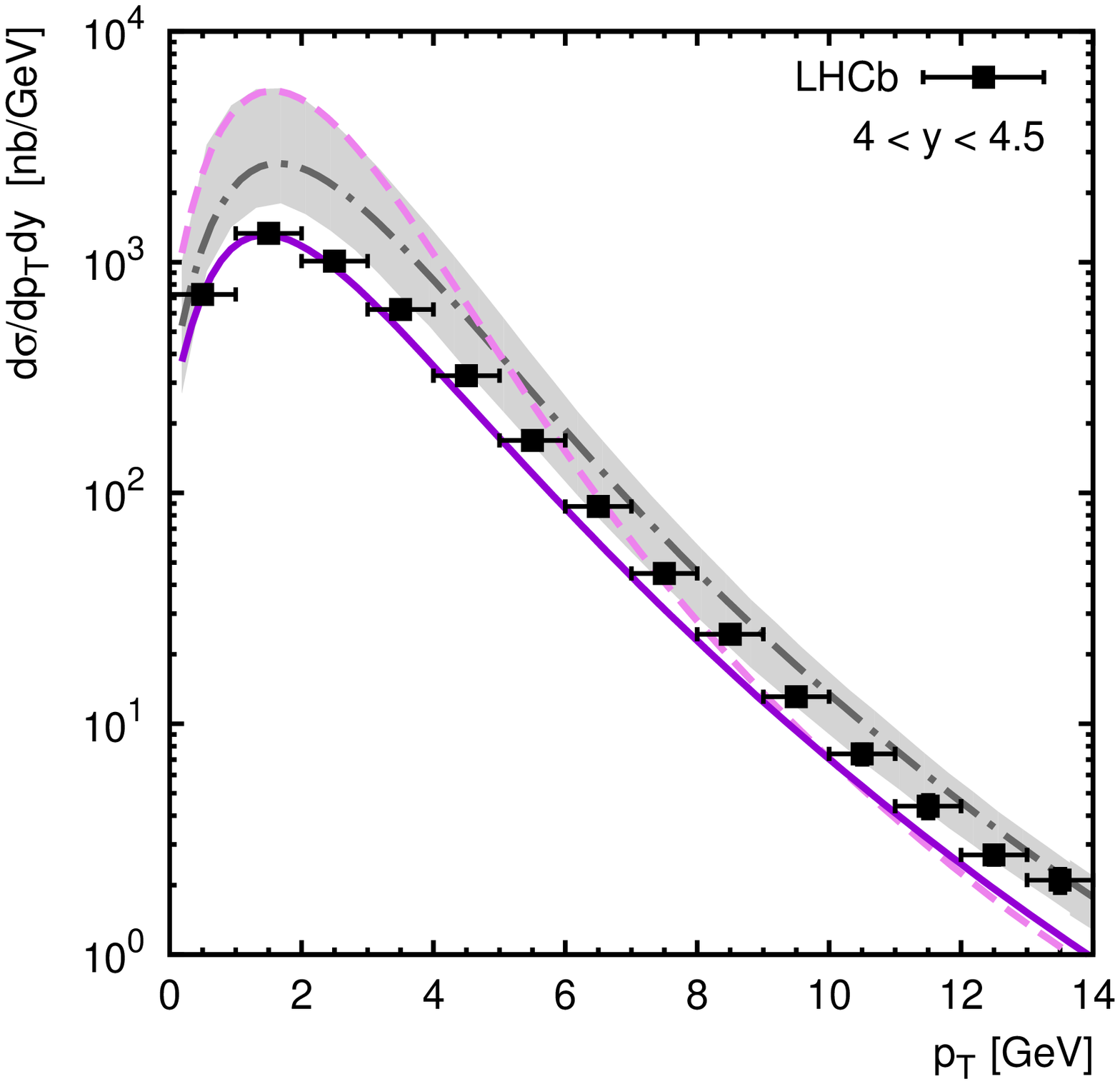, width = 5cm, angle = 270}
\caption{The double differential cross sections of prompt $J/\psi$ meson production in $pp$ collisions
at $\sqrt s = 13$~TeV. Notation of all curves is the same as in Fig.~2.
The experimental data are from LHCb\cite{26}.}
\label{fig6}
\end{center}
\end{figure}

\begin{figure}
\begin{center}
\epsfig{figure=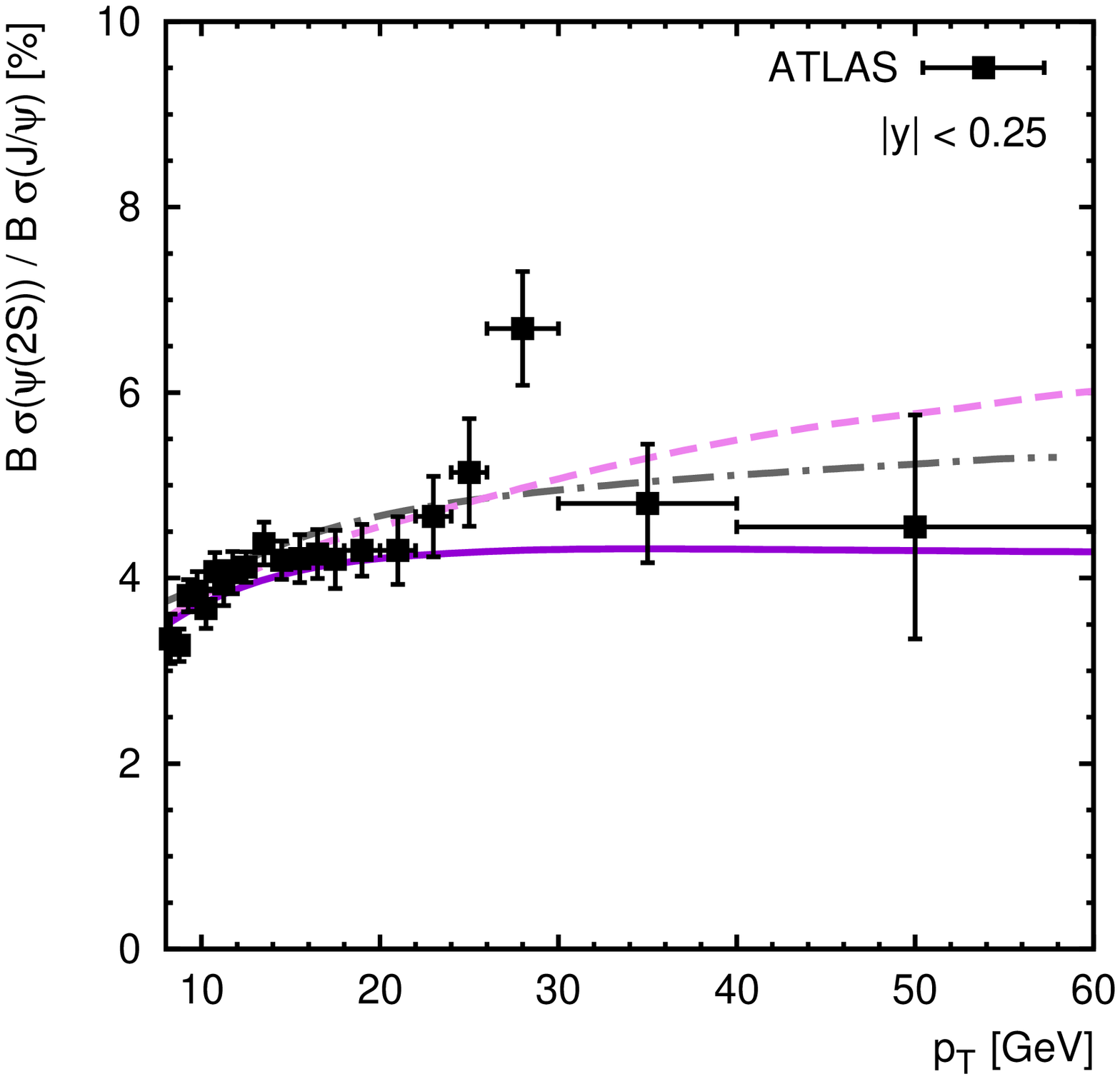, width = 5cm, angle = 270} 
\epsfig{figure=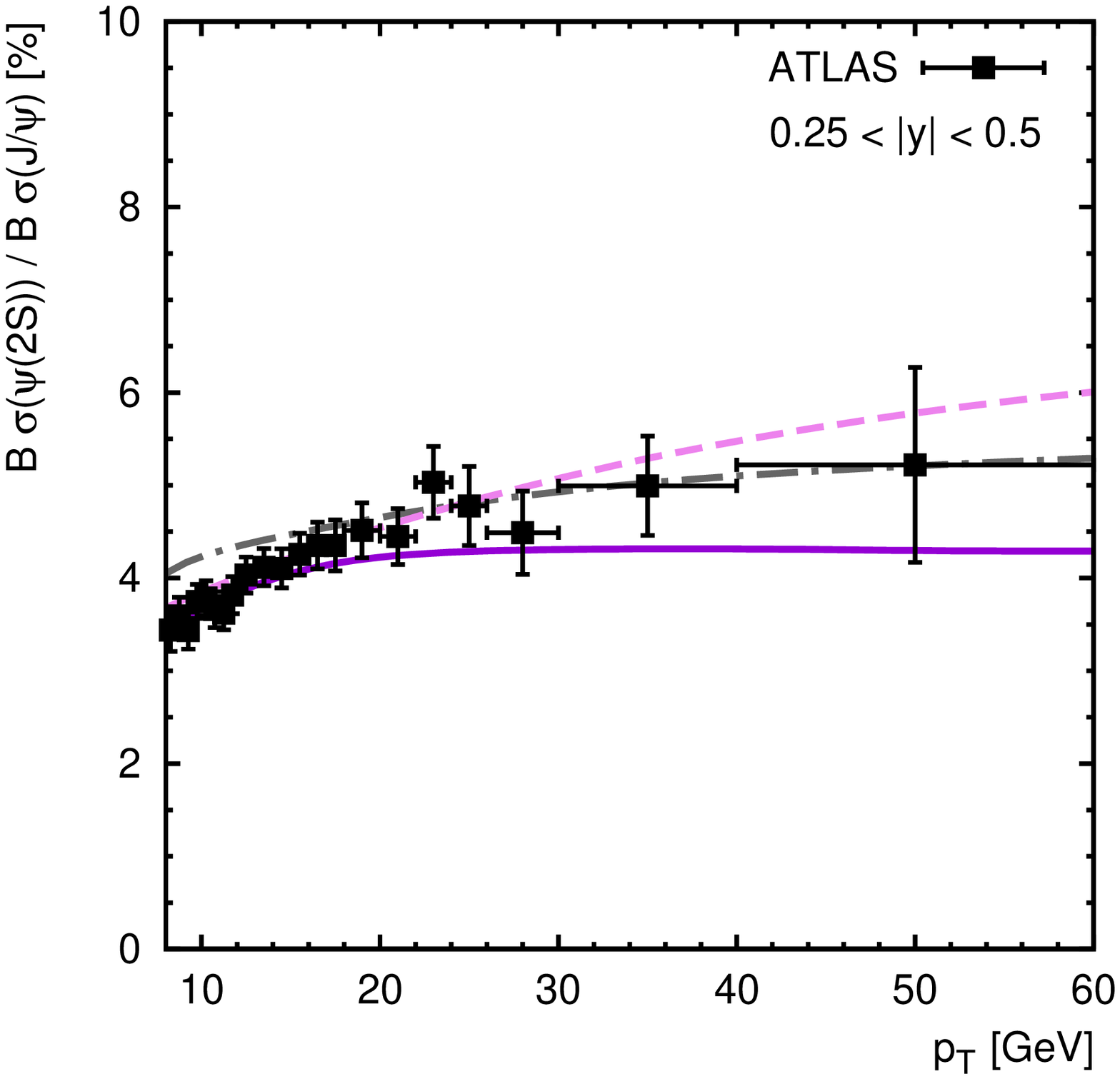, width = 5cm, angle = 270} 
\epsfig{figure=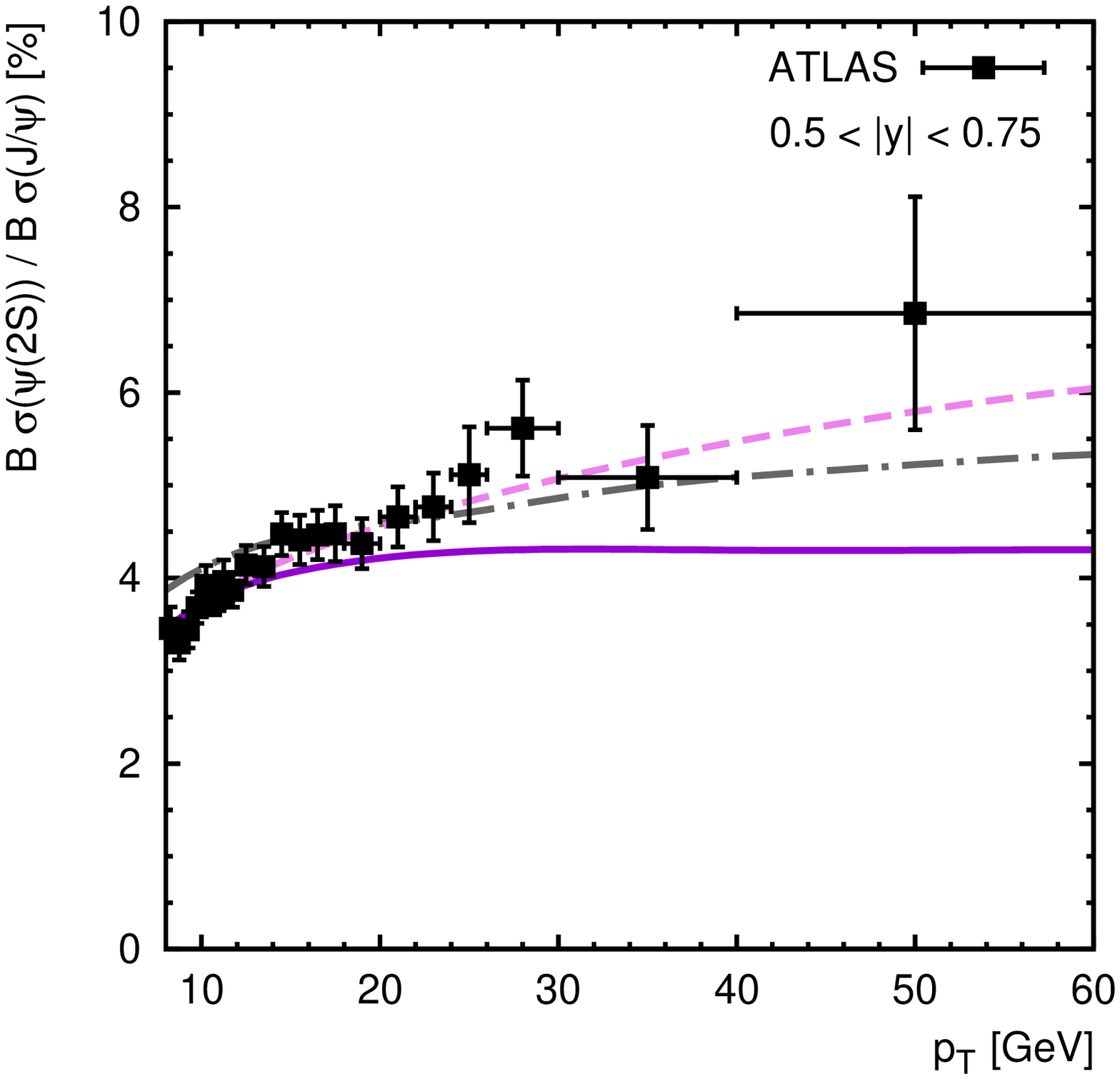, width = 5cm, angle = 270}
\epsfig{figure=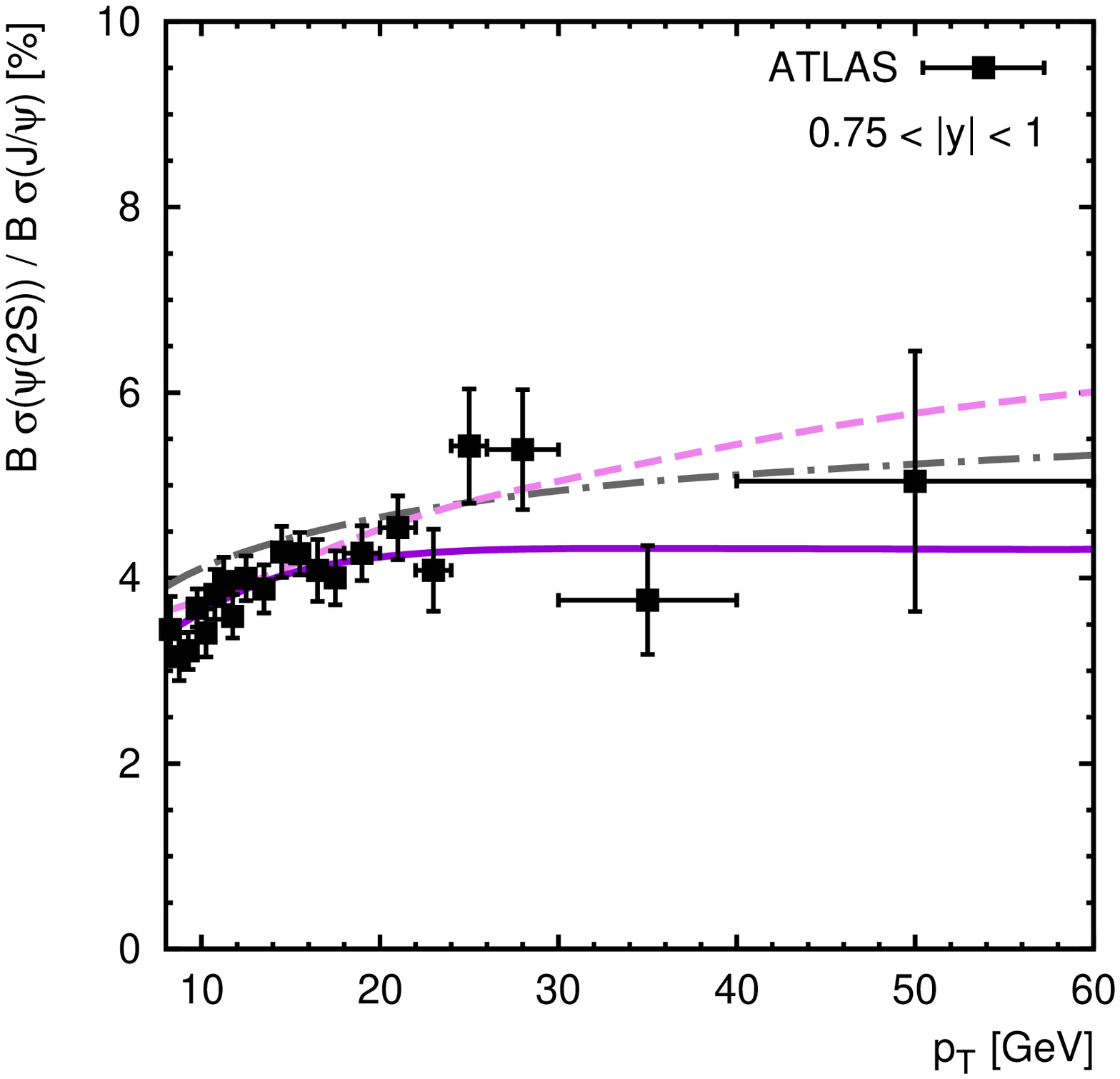, width = 5cm, angle = 270}
\epsfig{figure=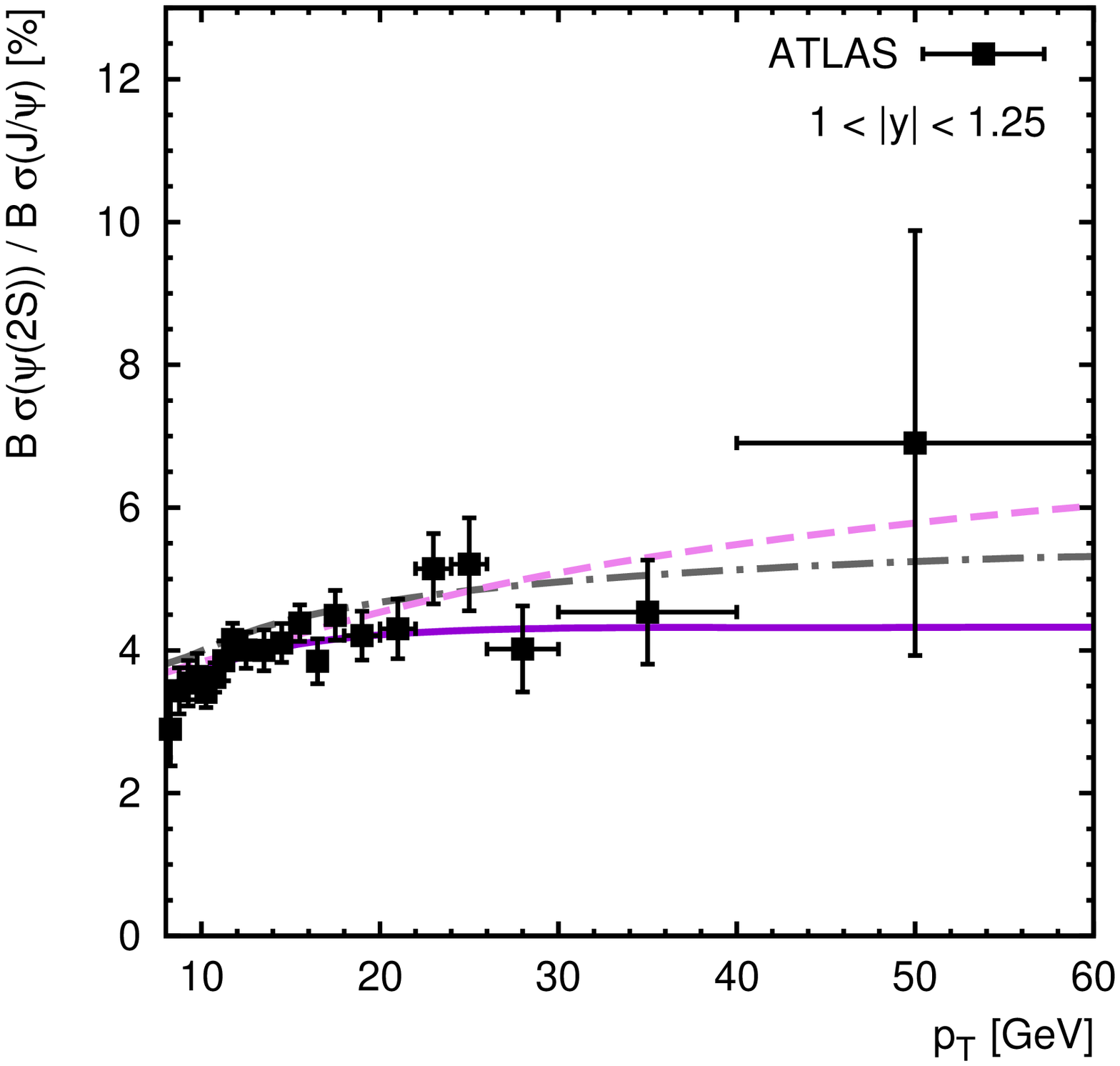, width = 5cm, angle = 270}
\epsfig{figure=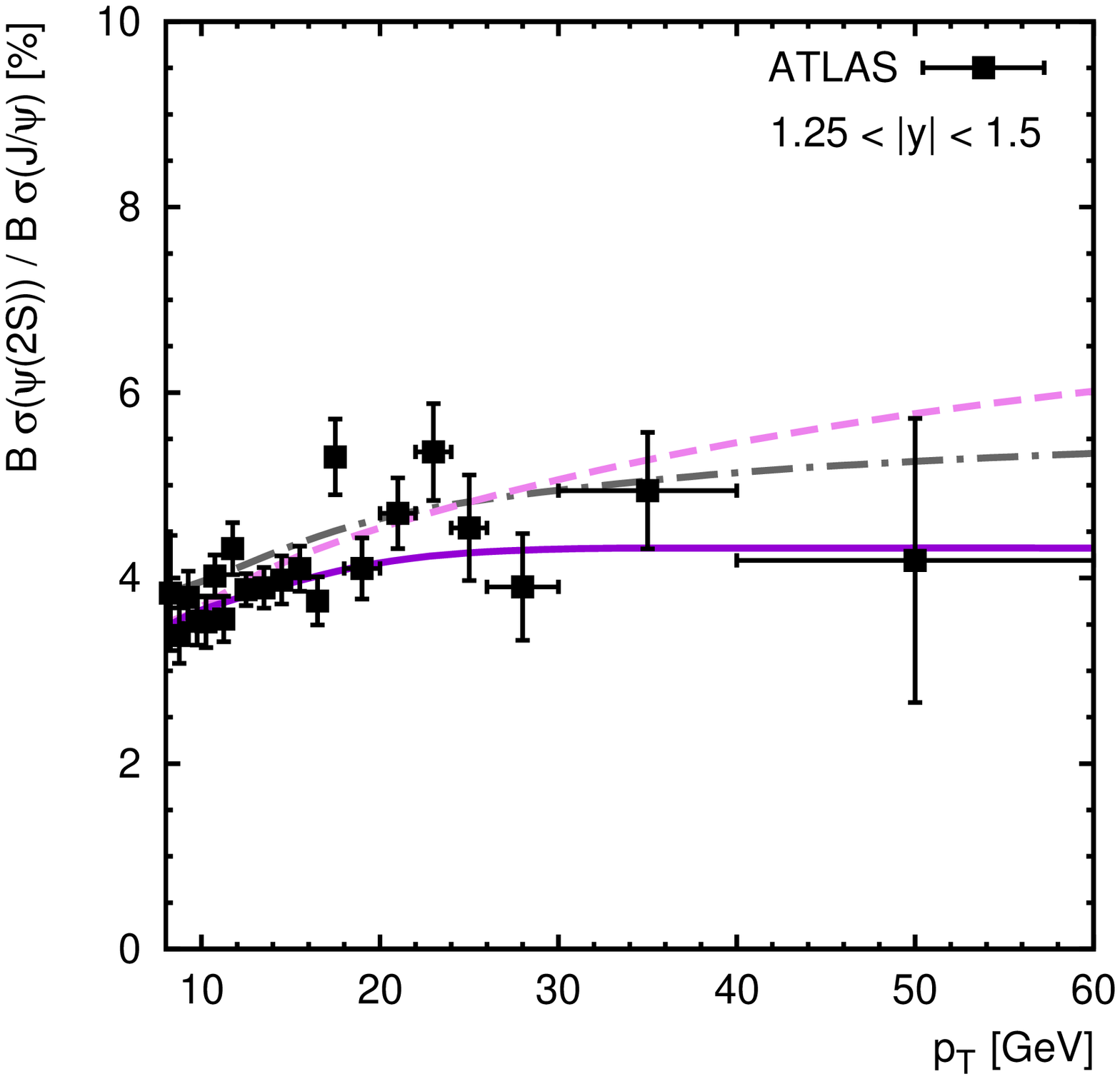, width = 5cm, angle = 270}
\epsfig{figure=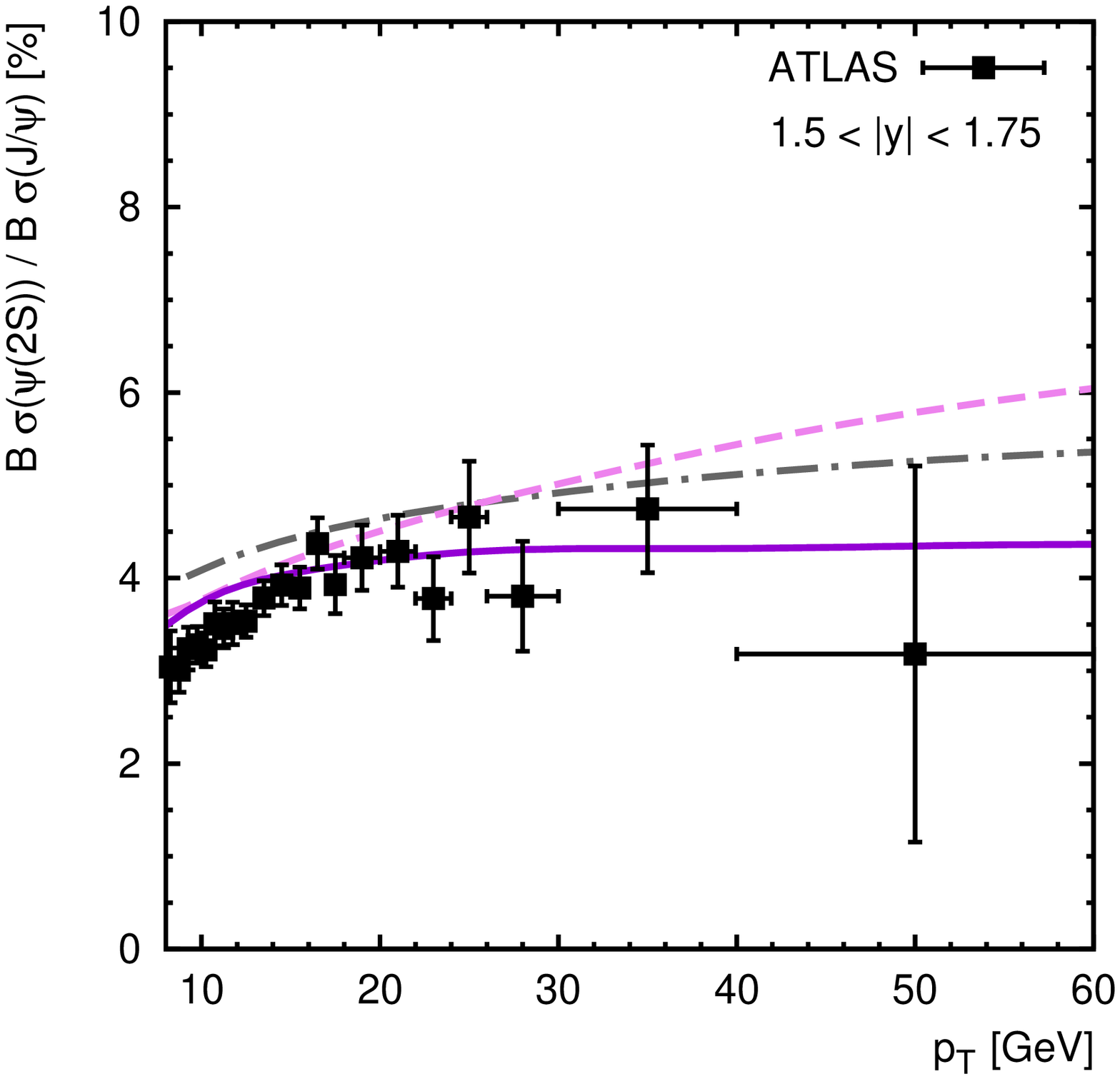, width = 5cm, angle = 270}
\epsfig{figure=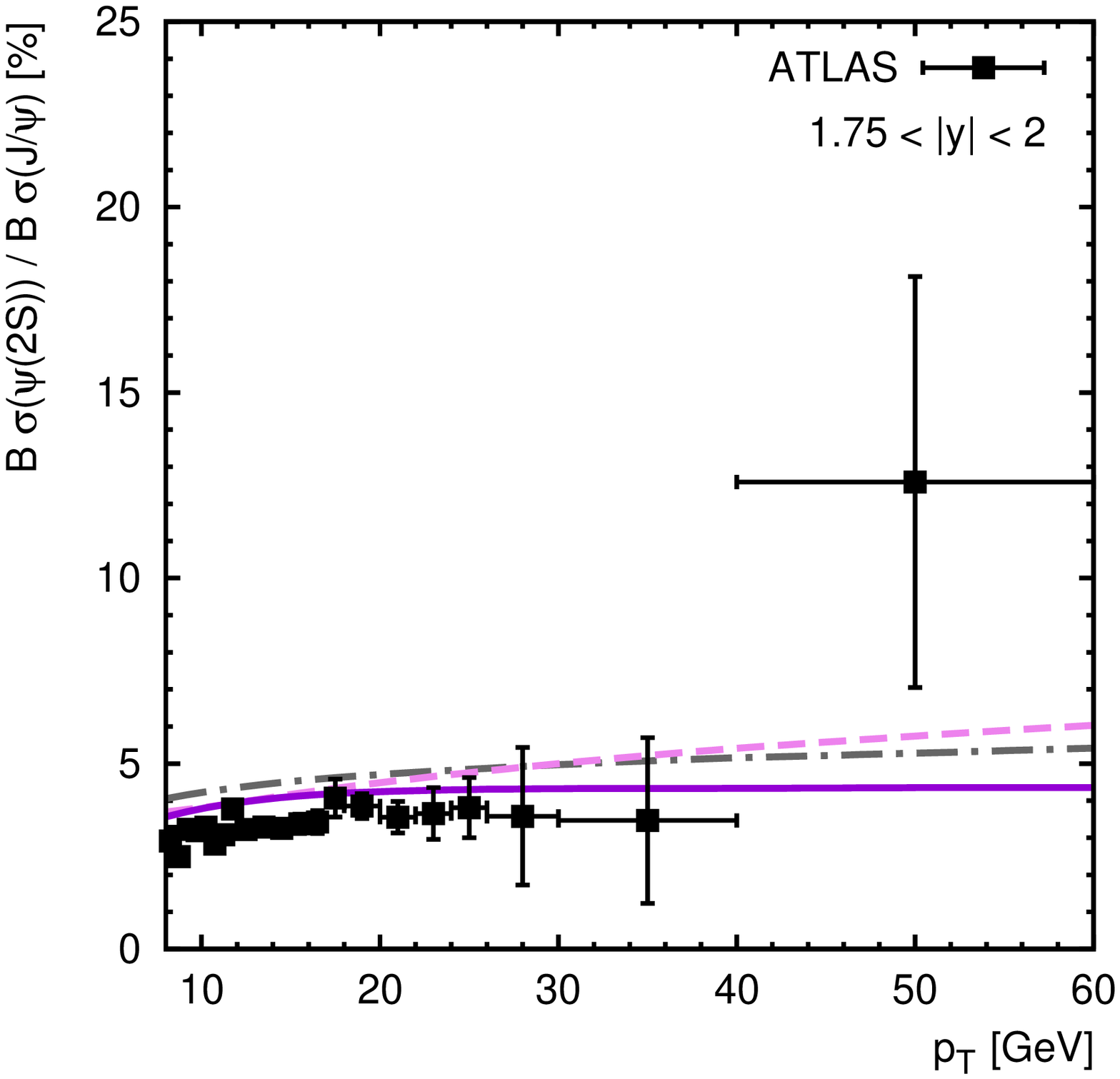, width = 5cm, angle = 270}
\caption{Relative production rate $\sigma(\psi^\prime)/\sigma(J/\psi)$ 
calculated as a function of $J/\psi$ meson transverse momenta 
at $\sqrt s = 7$~TeV. Notation of all curves is the same as in Fig.~2.
The experimental data are from ATLAS\cite{23}.}
\label{fig7}
\end{center}
\end{figure}

\begin{figure}
\begin{center}
\epsfig{figure=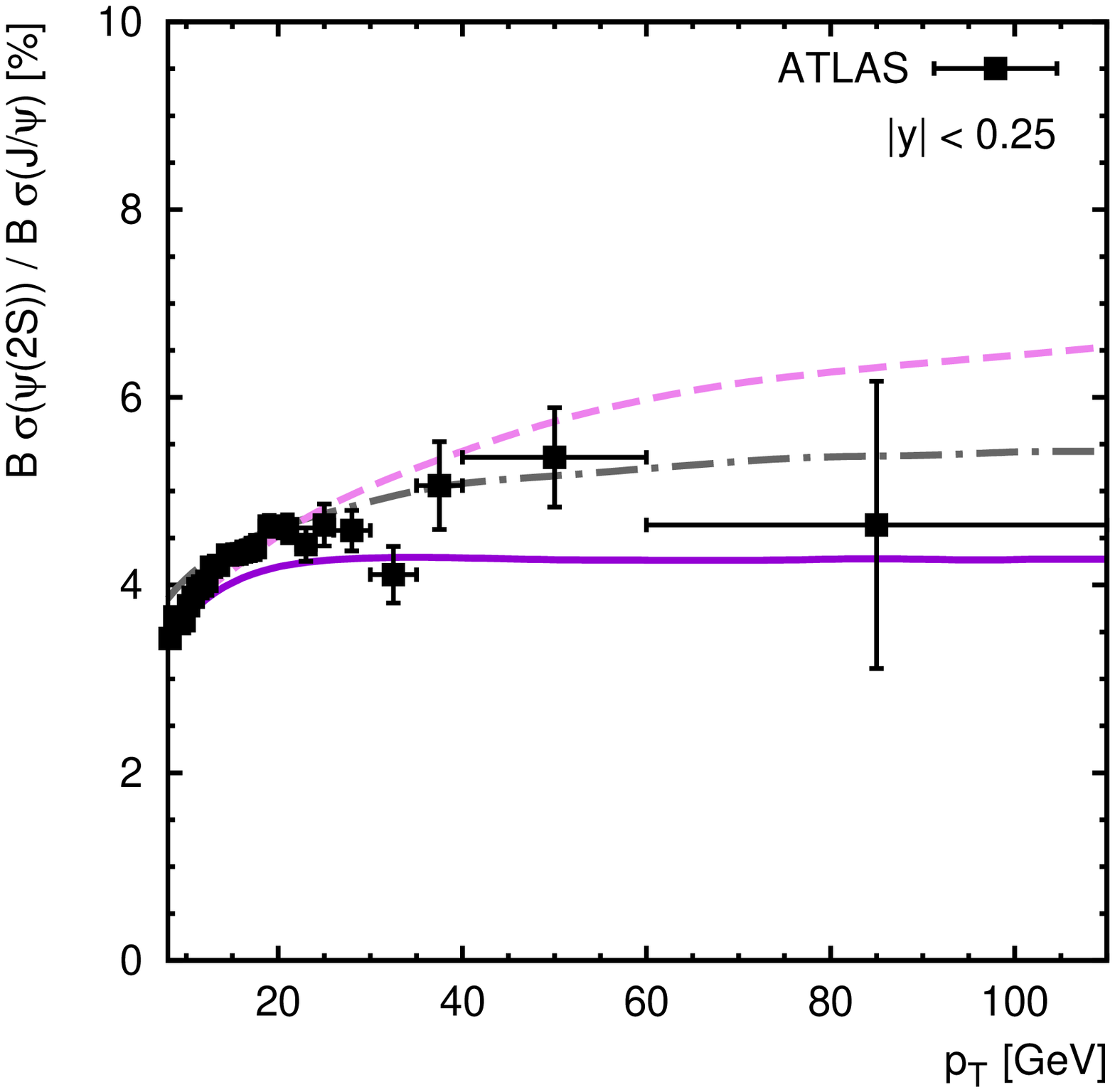, width = 5cm, angle = 270} 
\epsfig{figure=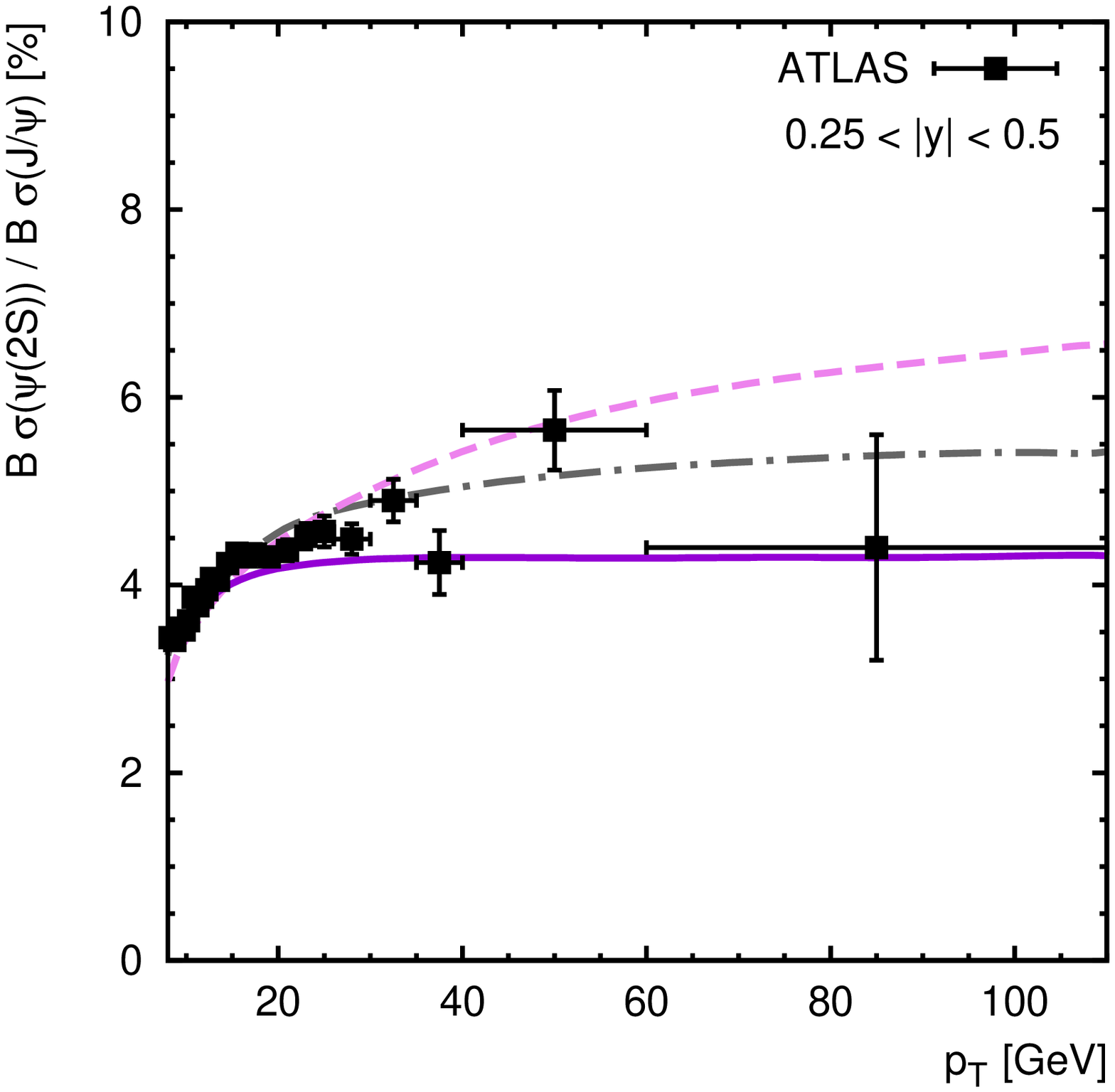, width = 5cm, angle = 270} 
\epsfig{figure=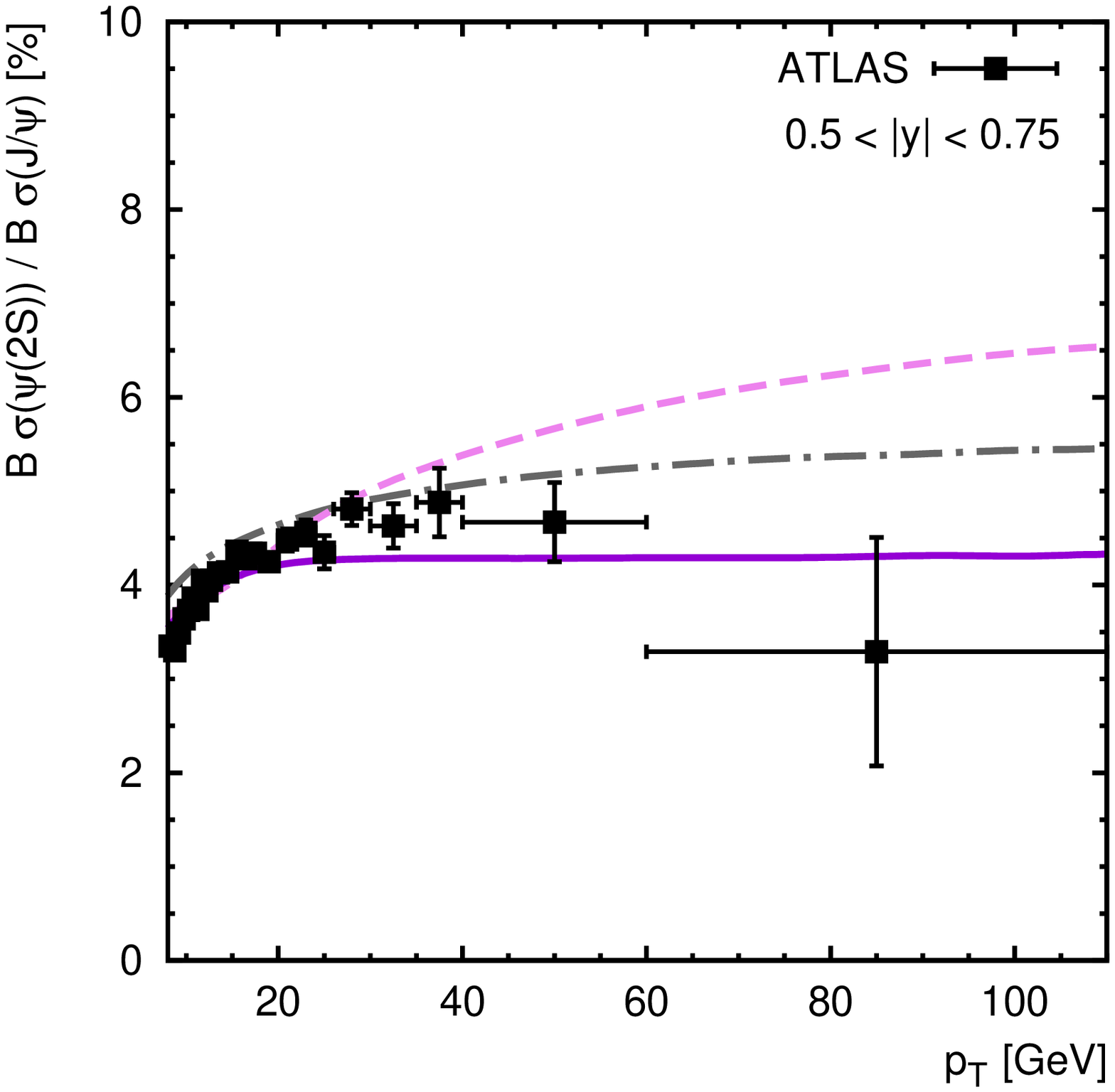, width = 5cm, angle = 270}
\epsfig{figure=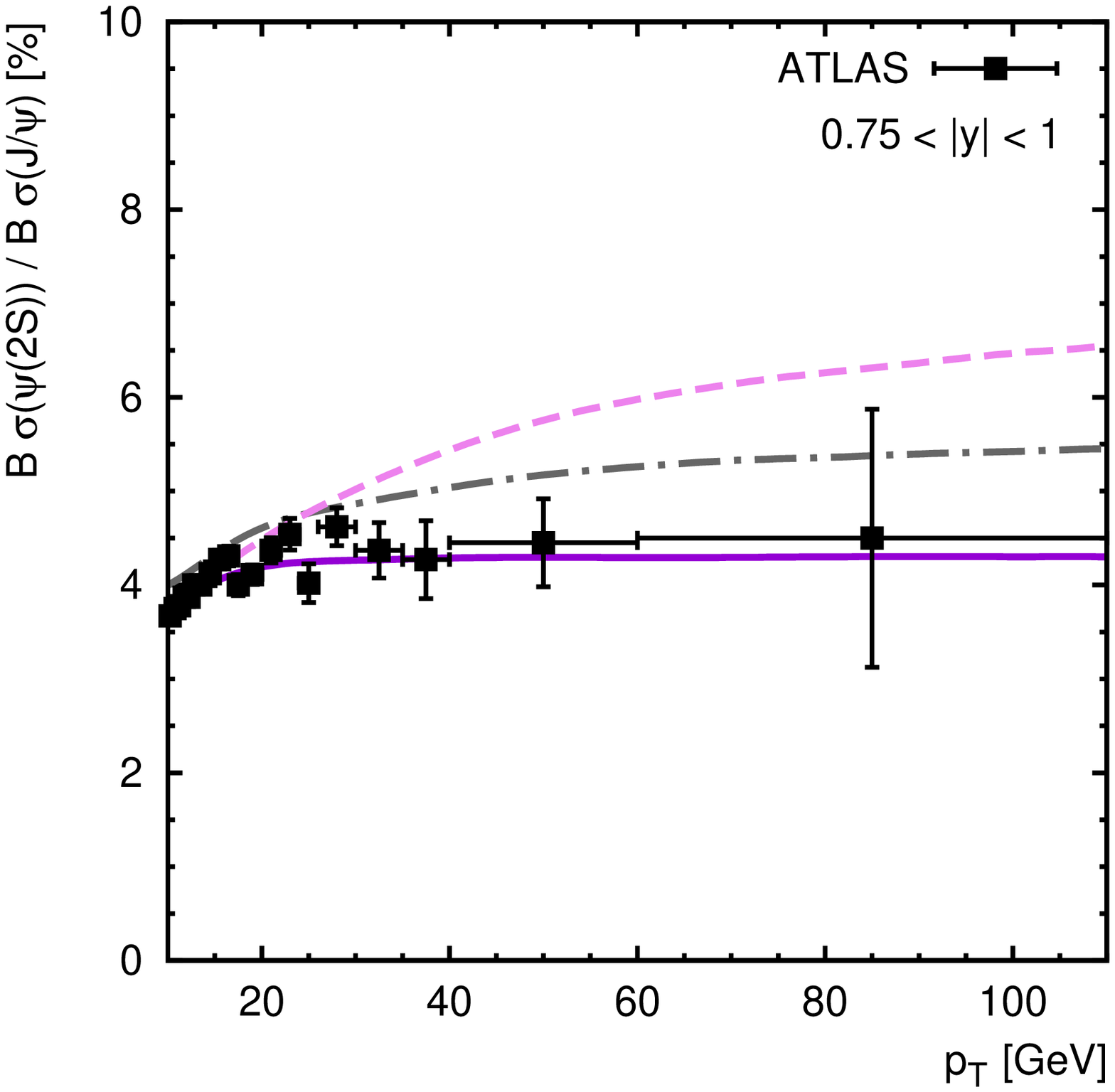, width = 5cm, angle = 270}
\epsfig{figure=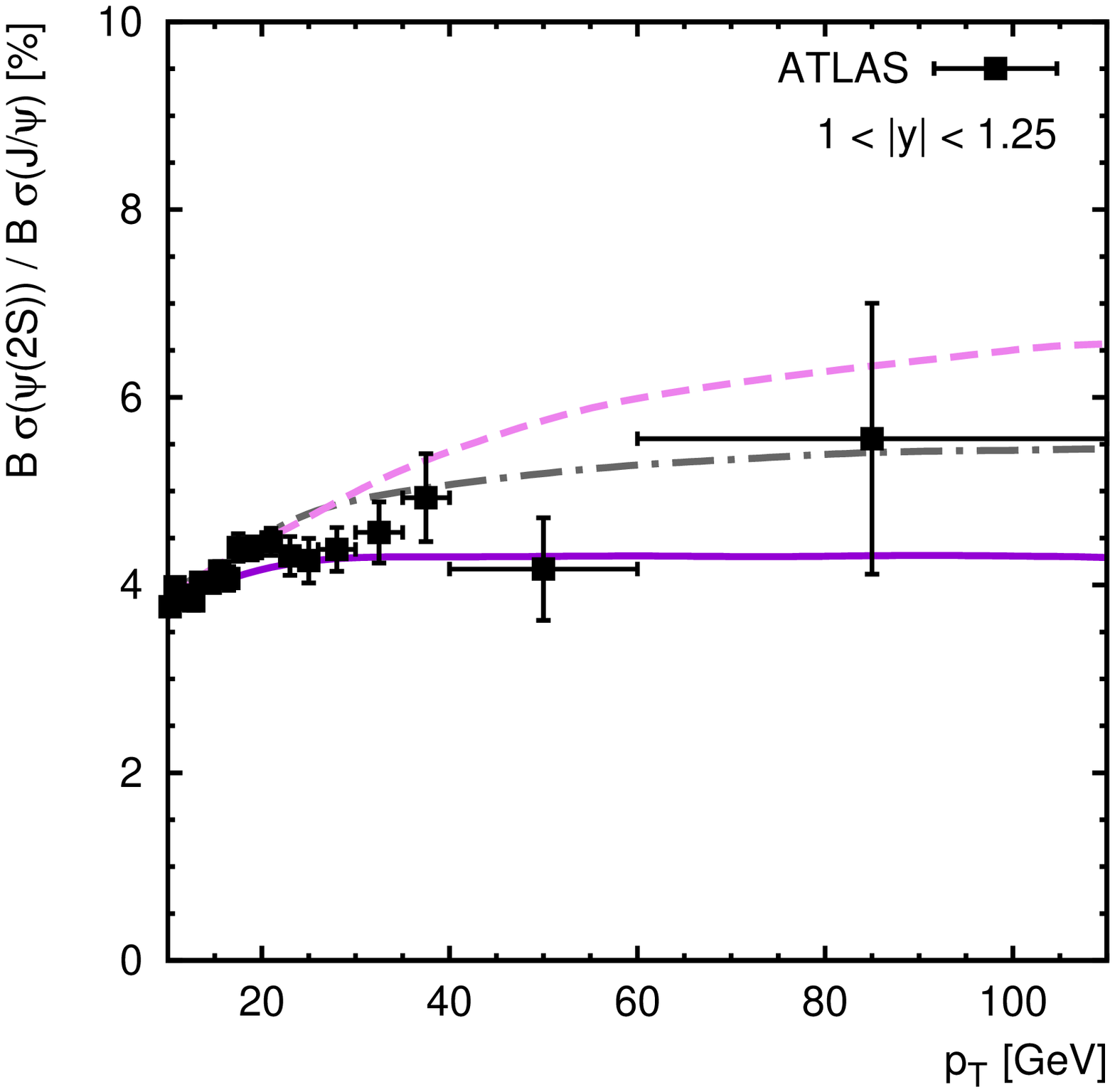, width = 5cm, angle = 270}
\epsfig{figure=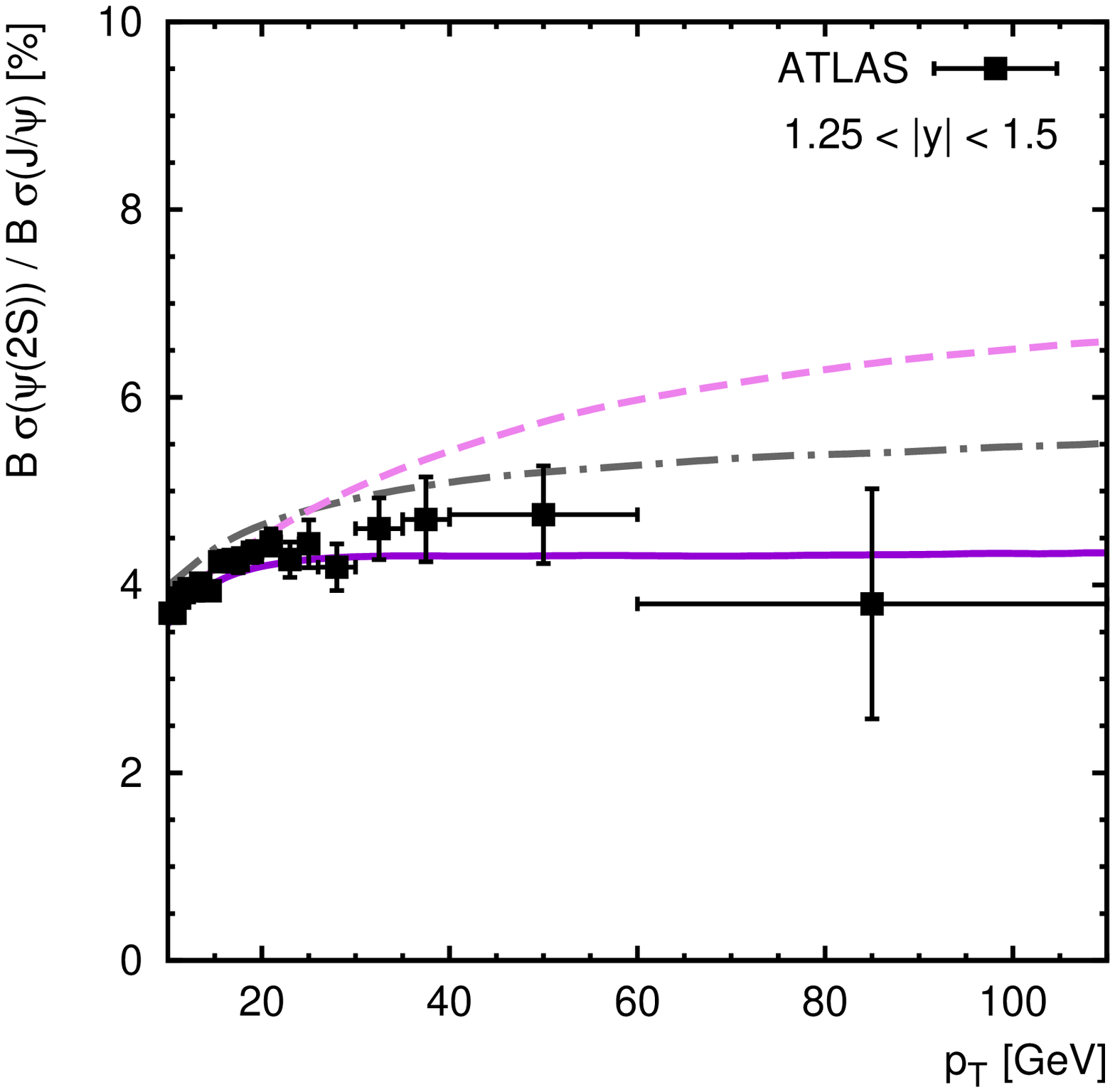, width = 5cm, angle = 270}
\epsfig{figure=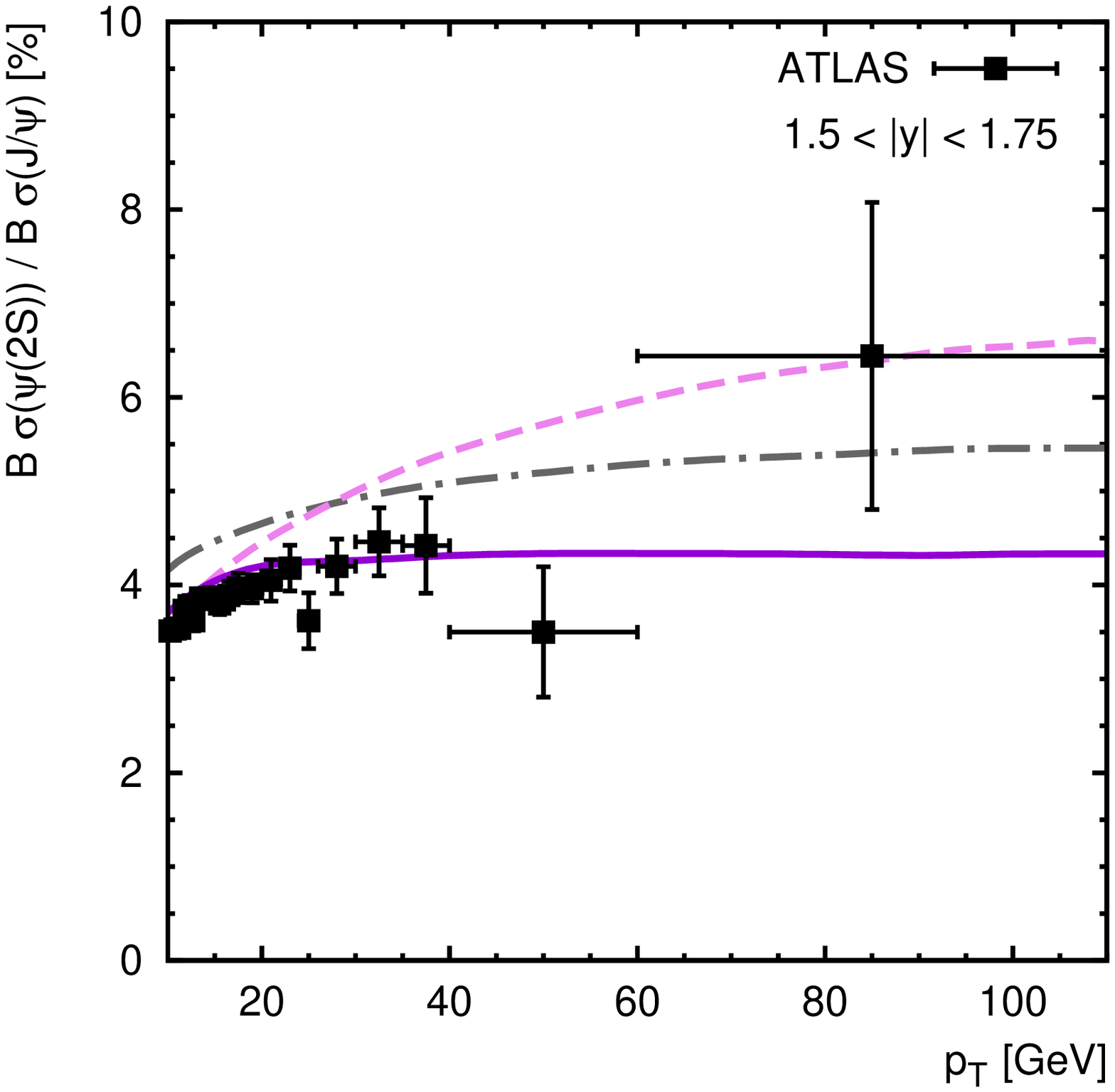, width = 5cm, angle = 270}
\epsfig{figure=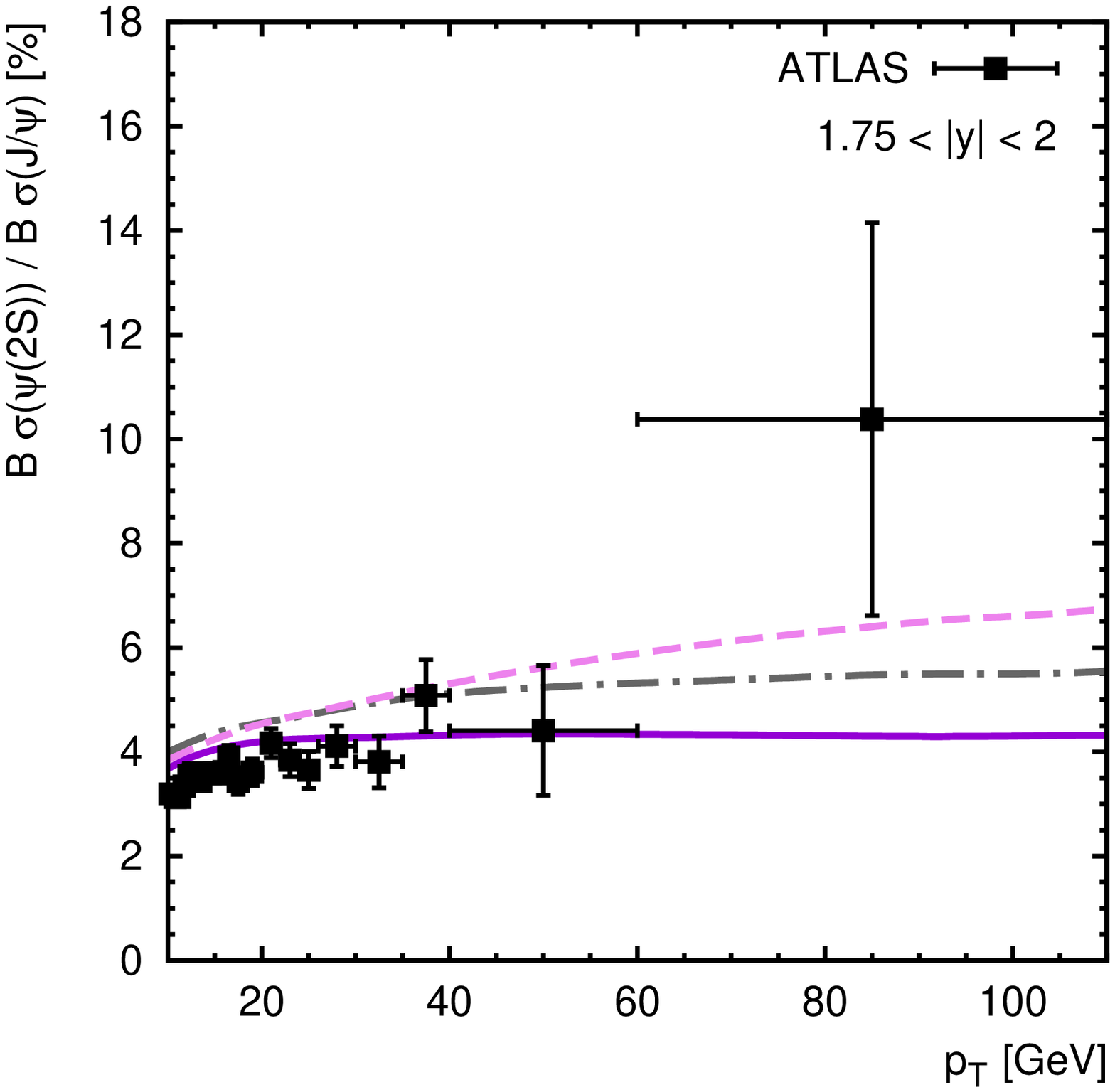, width = 5cm, angle = 270}
\caption{Relative production rate $\sigma(\psi^\prime)/\sigma(J/\psi)$ 
calculated as a function of $J/\psi$ meson transverse momenta 
at $\sqrt s = 8$~TeV. Notation of all curves is the same as in Fig.~2.
The experimental data are from ATLAS\cite{23}.}
\label{fig8}
\end{center}
\end{figure}

\begin{figure}
\begin{center}
\epsfig{figure=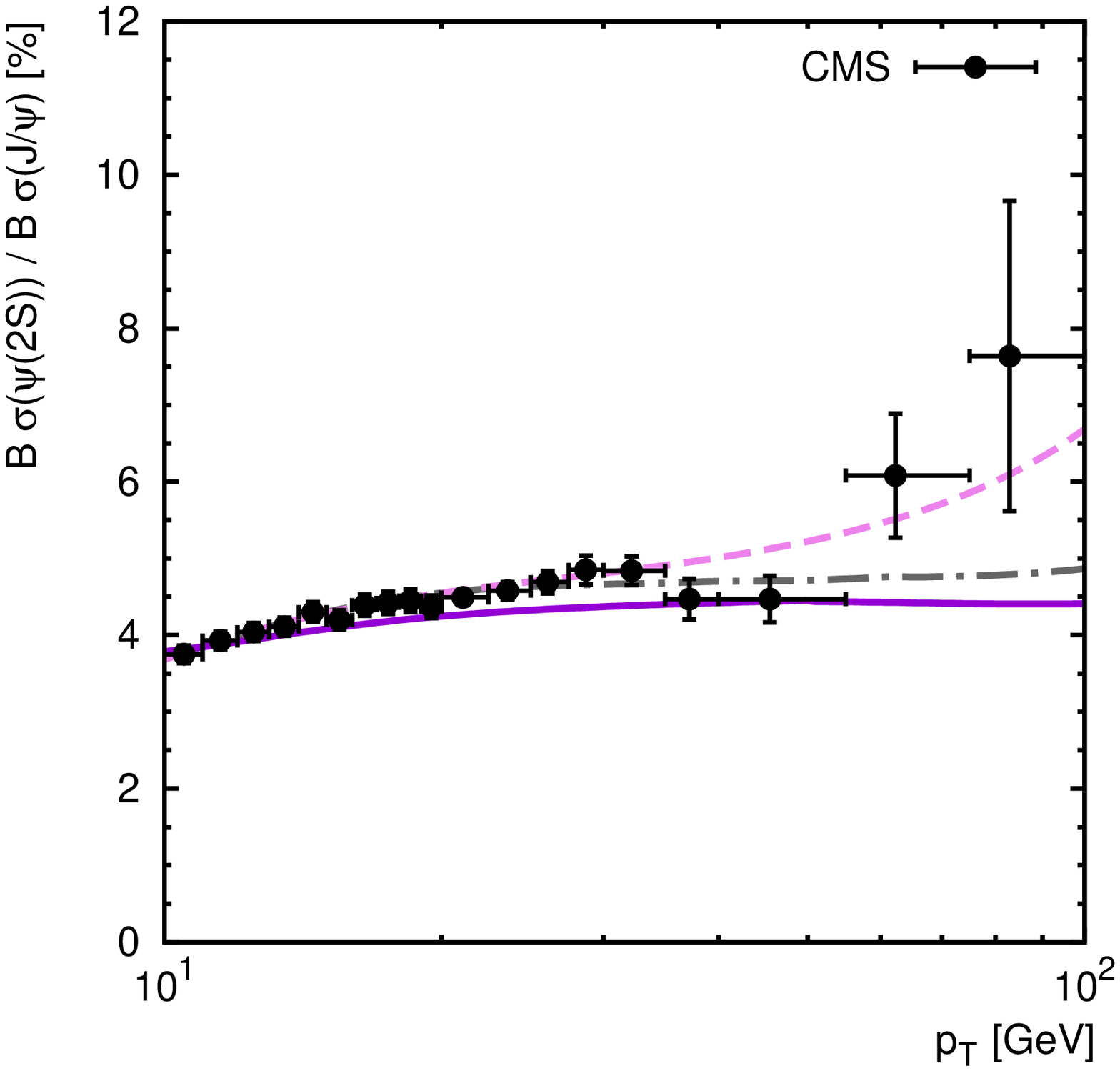, width = 5cm, angle = 270}
\epsfig{figure=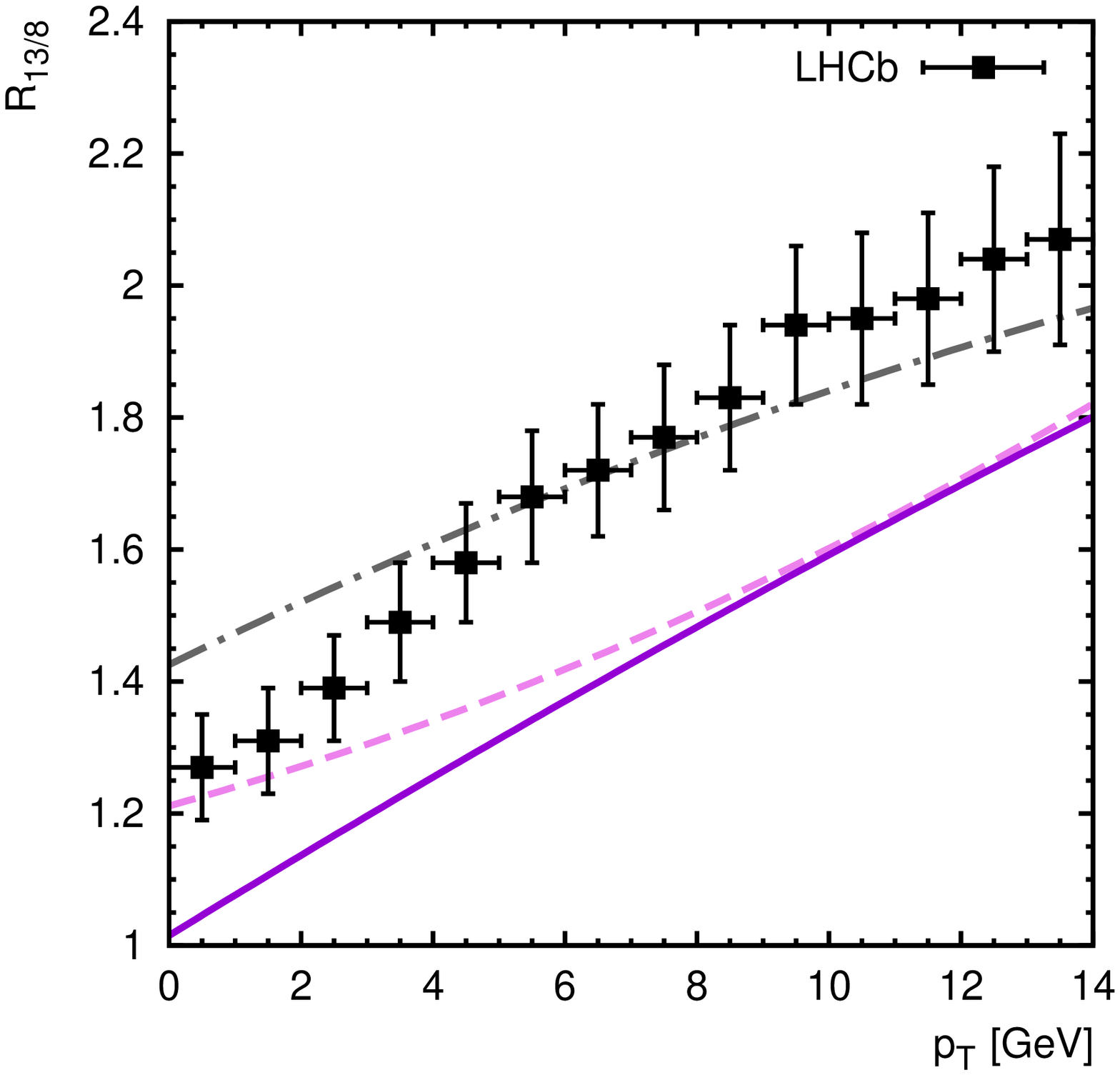, width = 5cm, angle = 270}
\caption{Left panel: relative production rate $\sigma(\psi^\prime)/\sigma(J/\psi)$ 
calculated as a function of $J/\psi$ meson transverse momenta 
at $\sqrt s = 7$~TeV. Right panel: relative ratio $R_{13/8}$ of the $J/\psi$ meson production cross 
sections calculated at $\sqrt s = 13$~TeV and $\sqrt s = 8$~TeV. 
Notation of all curves is the same as in Fig.~2.
The experimental data are from CMS\cite{24} and LHCb\cite{26}.}
\label{fig9}
\end{center}
\end{figure}

\begin{figure}
\begin{center}
\epsfig{figure=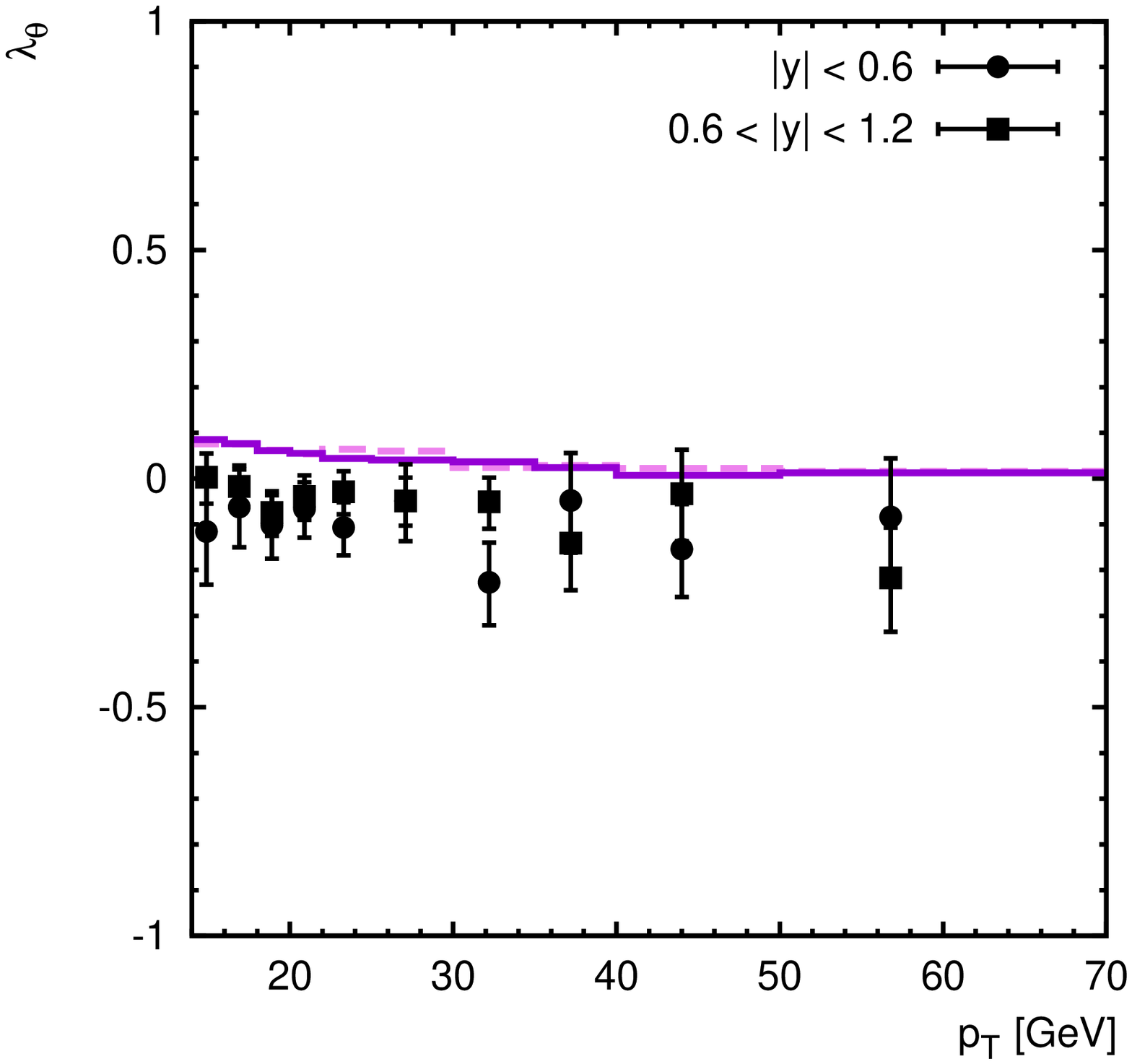, width = 5cm, angle = 270} 
\epsfig{figure=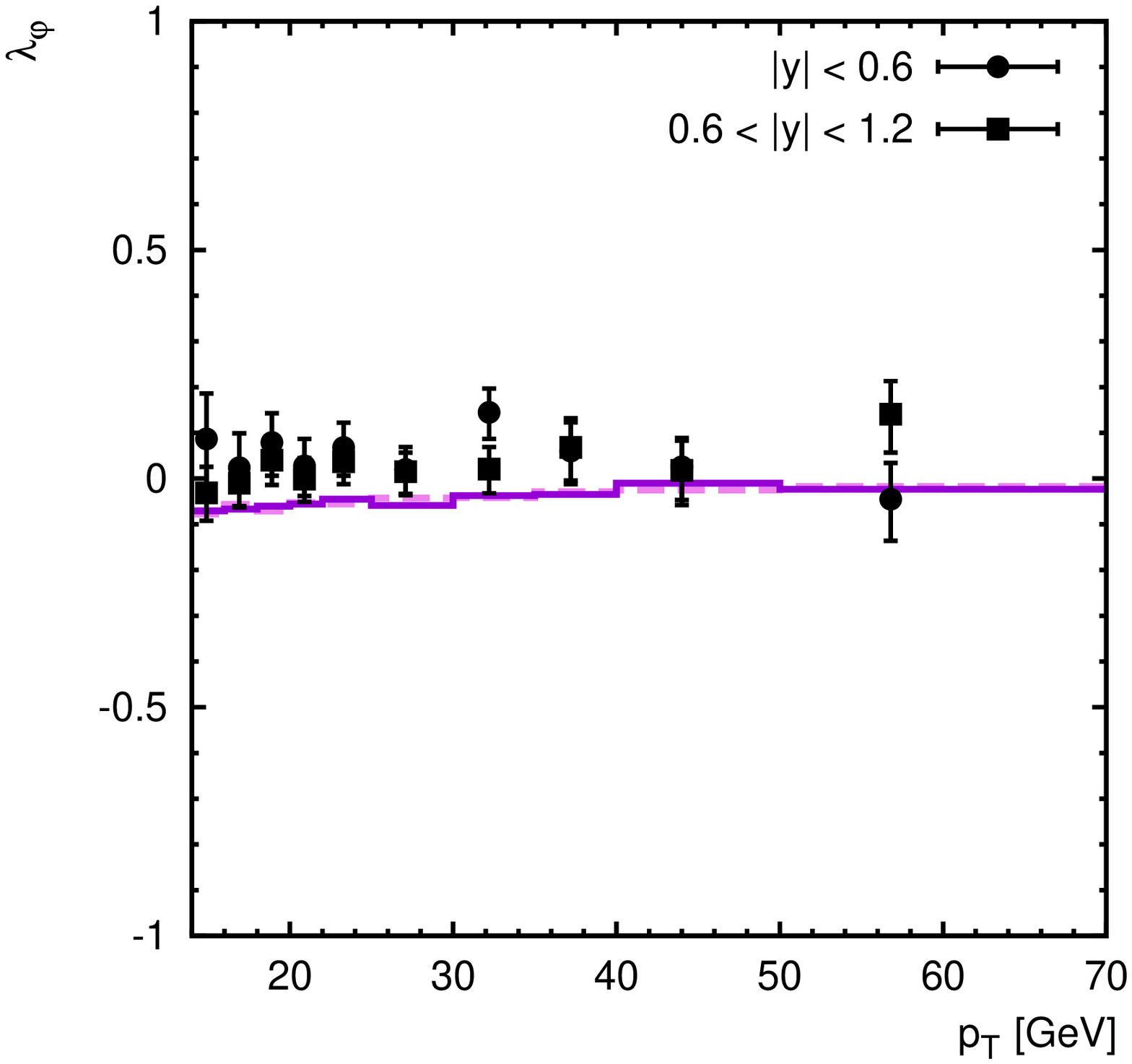, width = 5cm, angle = 270} 
\epsfig{figure=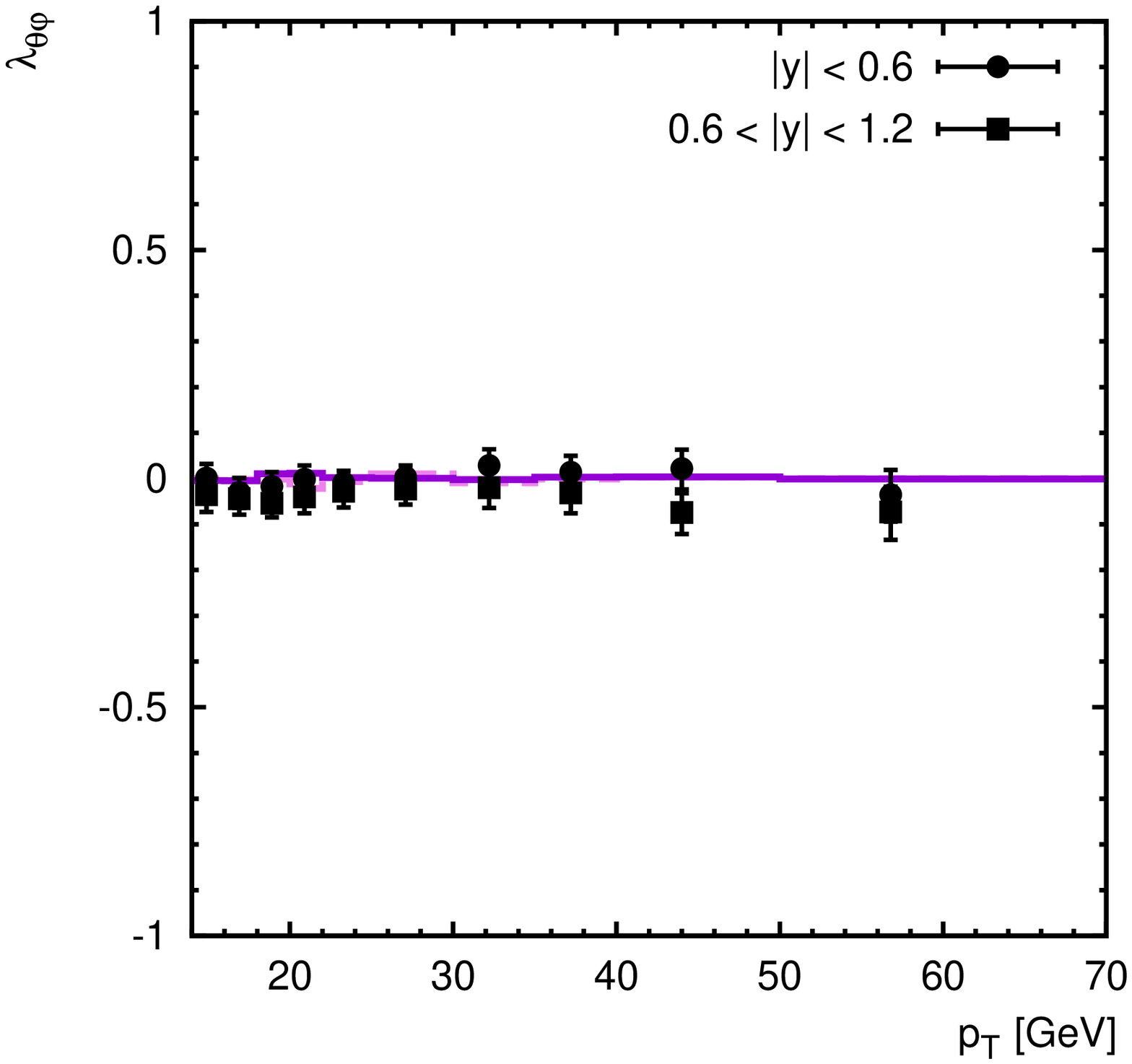, width = 5cm, angle = 270}
\epsfig{figure=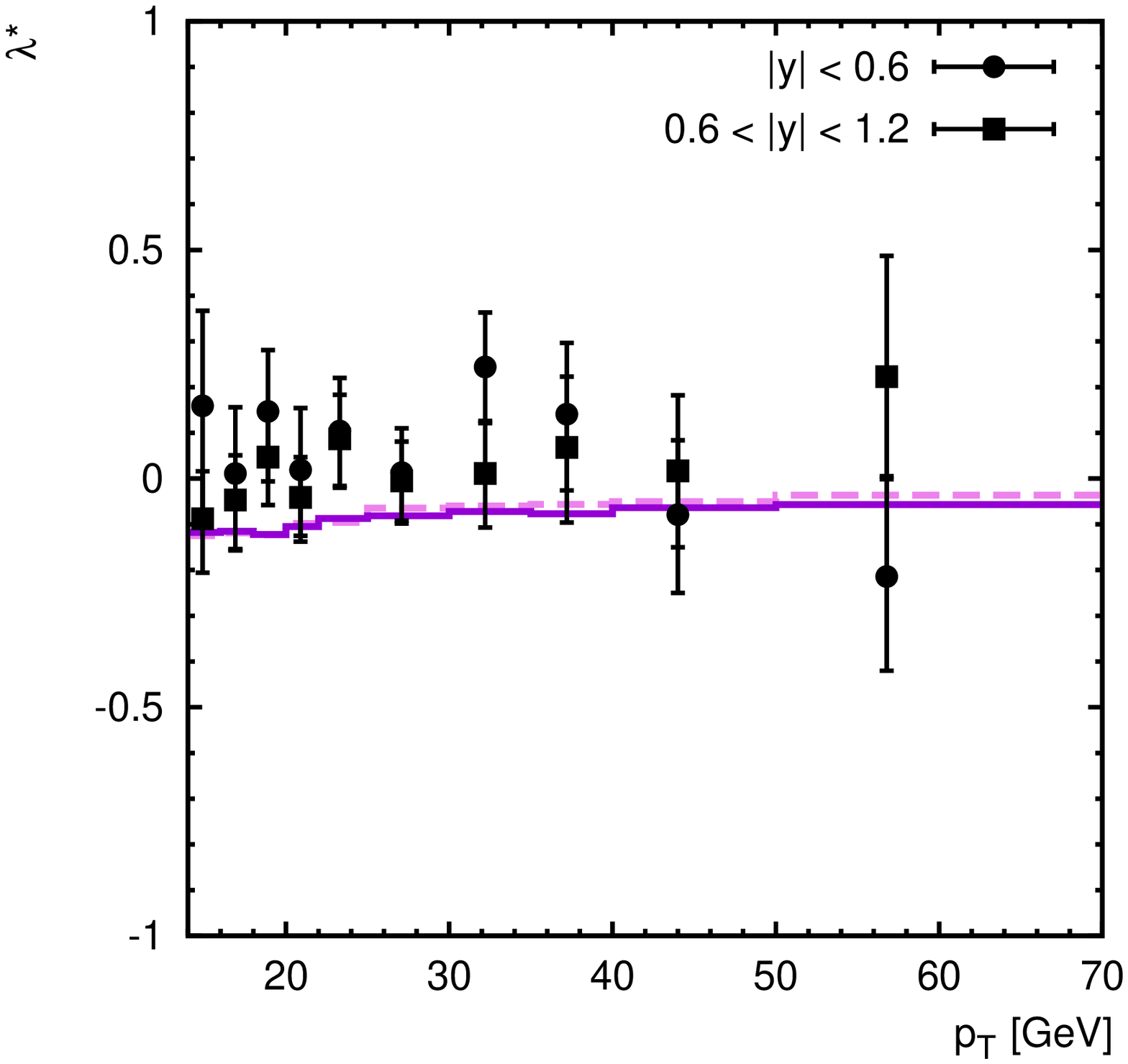, width = 5cm, angle = 270}
\caption{Polarization parameters 
$\lambda_\theta$, $\lambda_\phi$, $\lambda_{\theta \phi}$
and $\lambda^*$ of prompt $J/\psi$ mesons 
calculated as a function of their transverse momentum in the Collins-Soper frame.
The solid and dashed histograms correspond
to the predictions obtained at $|y| < 0.6$ and $0.6 < |y| < 1.2$, respectively. 
The KMR gluon distribution is used.
The experimental data are from CMS\cite{21}.}
\label{fig10}
\end{center}
\end{figure}

\begin{figure}
\begin{center}
\epsfig{figure=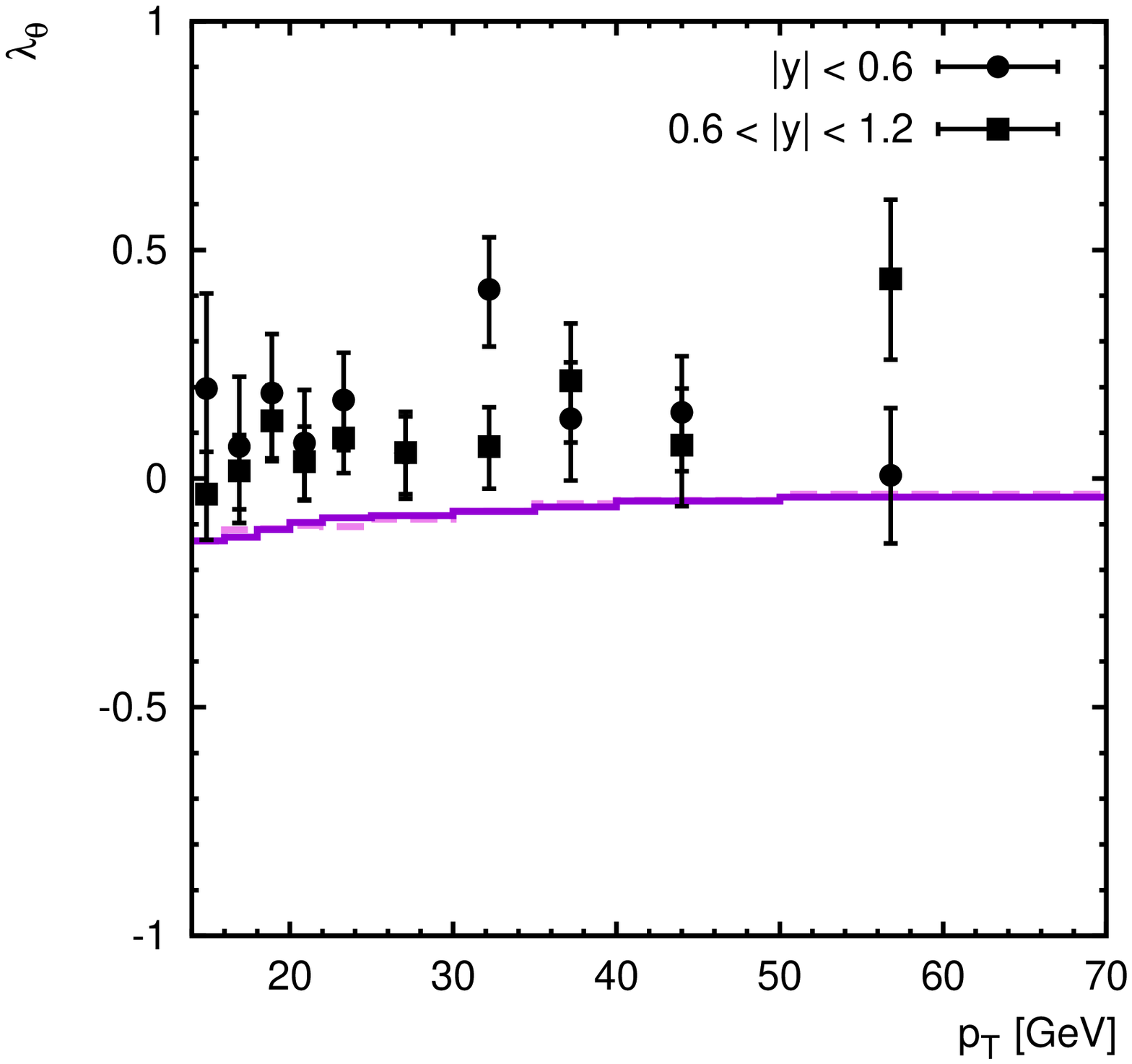, width = 5cm, angle = 270} 
\epsfig{figure=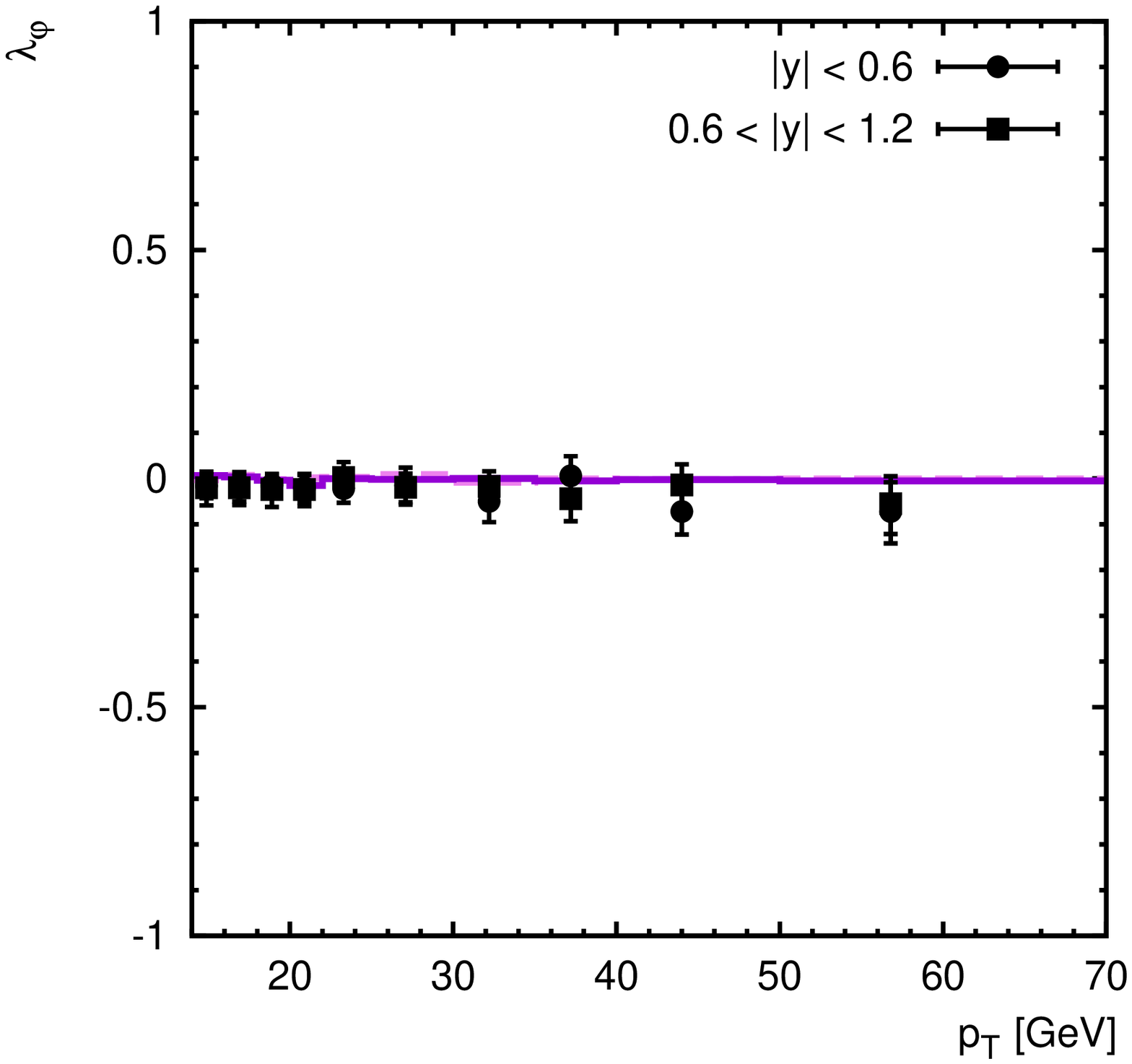, width = 5cm, angle = 270} 
\epsfig{figure=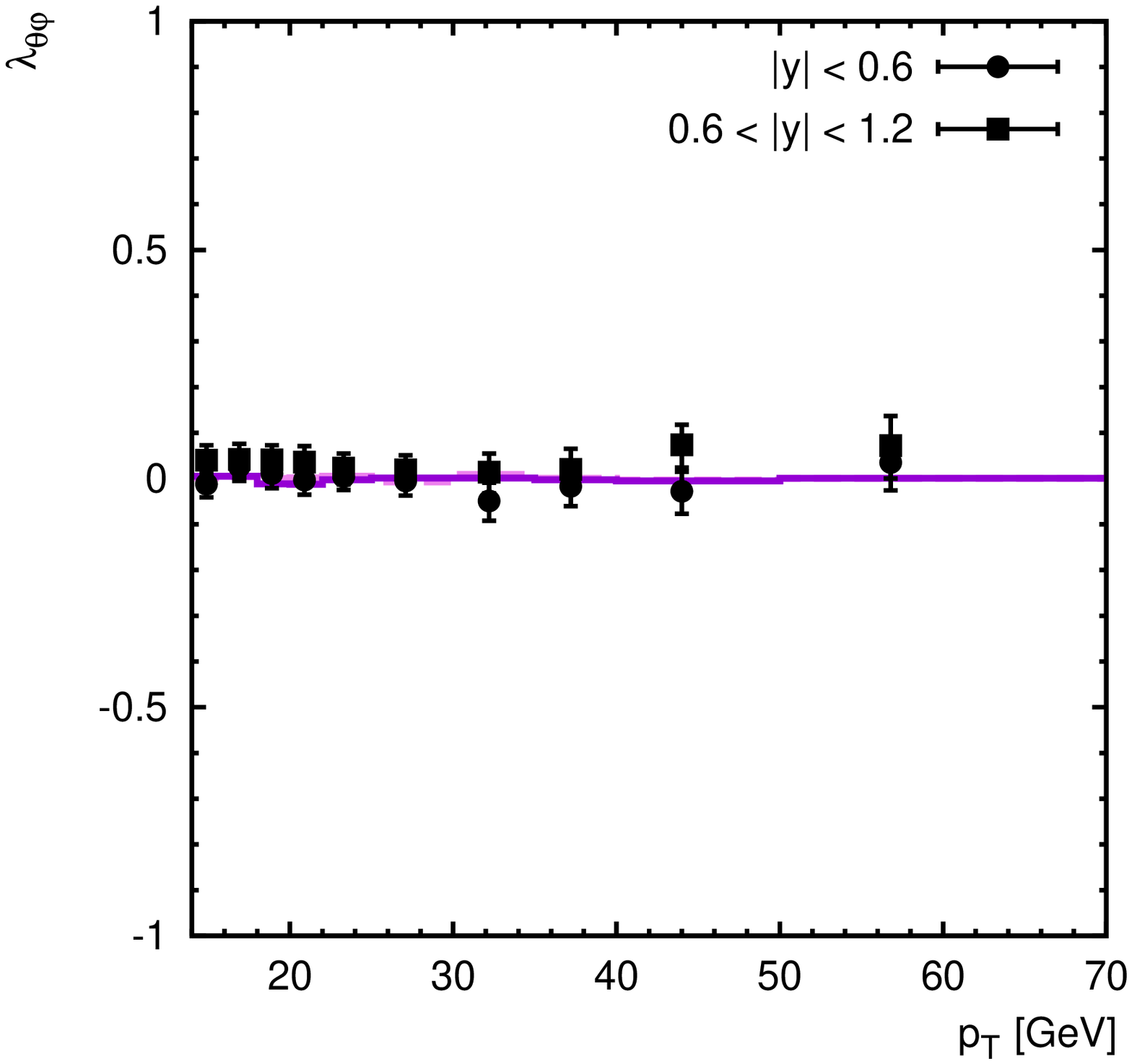, width = 5cm, angle = 270}
\epsfig{figure=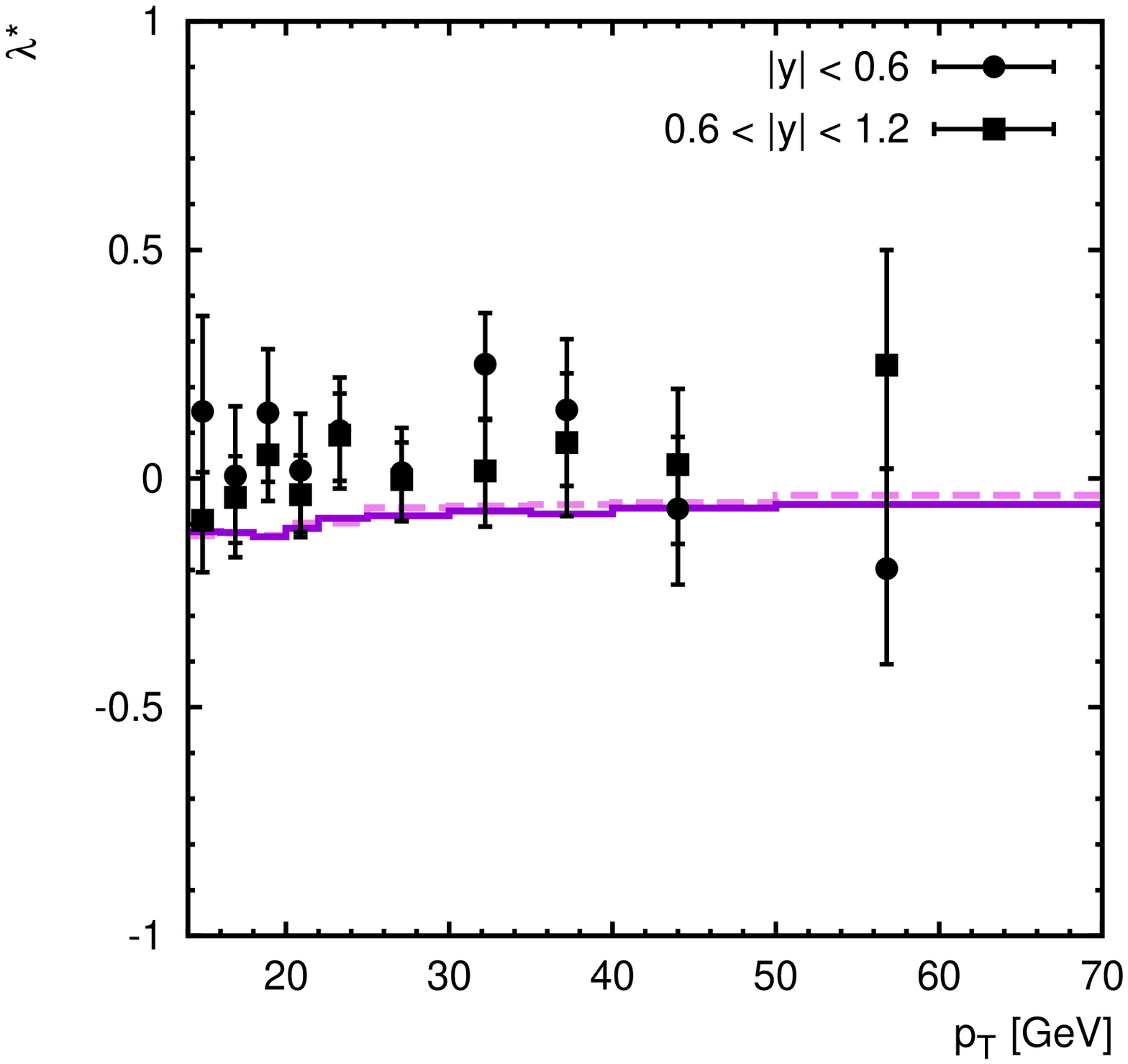, width = 5cm, angle = 270}
\caption{Polarization parameters 
$\lambda_\theta$, $\lambda_\phi$, $\lambda_{\theta \phi}$
and $\lambda^*$ of prompt $J/\psi$ mesons 
calculated as a function of their transverse momentum in the helicity frame.
Notation of all curves is the same as in Fig.~10.
The experimental data are from CMS\cite{21}.}
\label{fig11}
\end{center}
\end{figure}

\begin{figure}
\begin{center}
\epsfig{figure=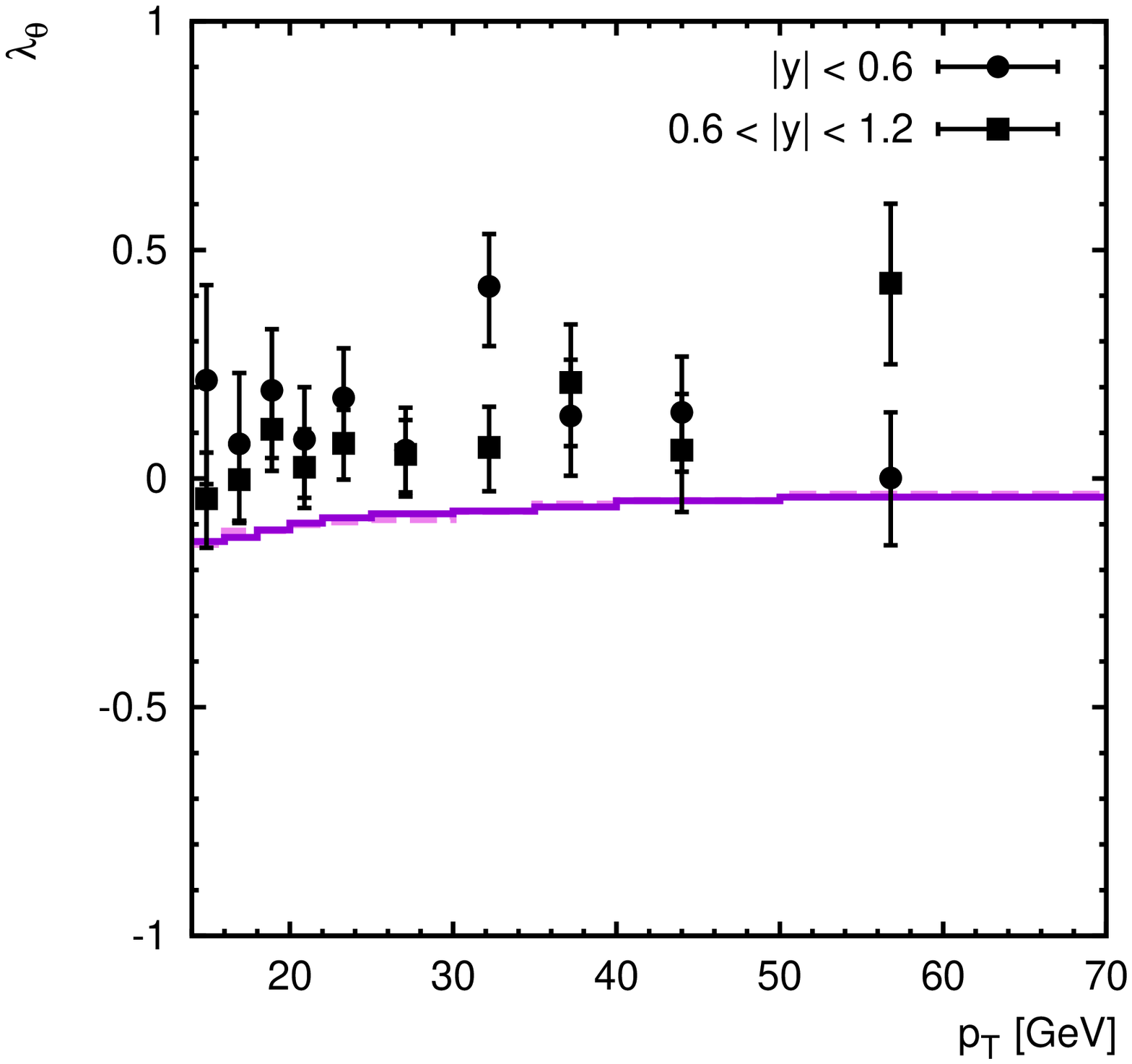, width = 5cm, angle = 270} 
\epsfig{figure=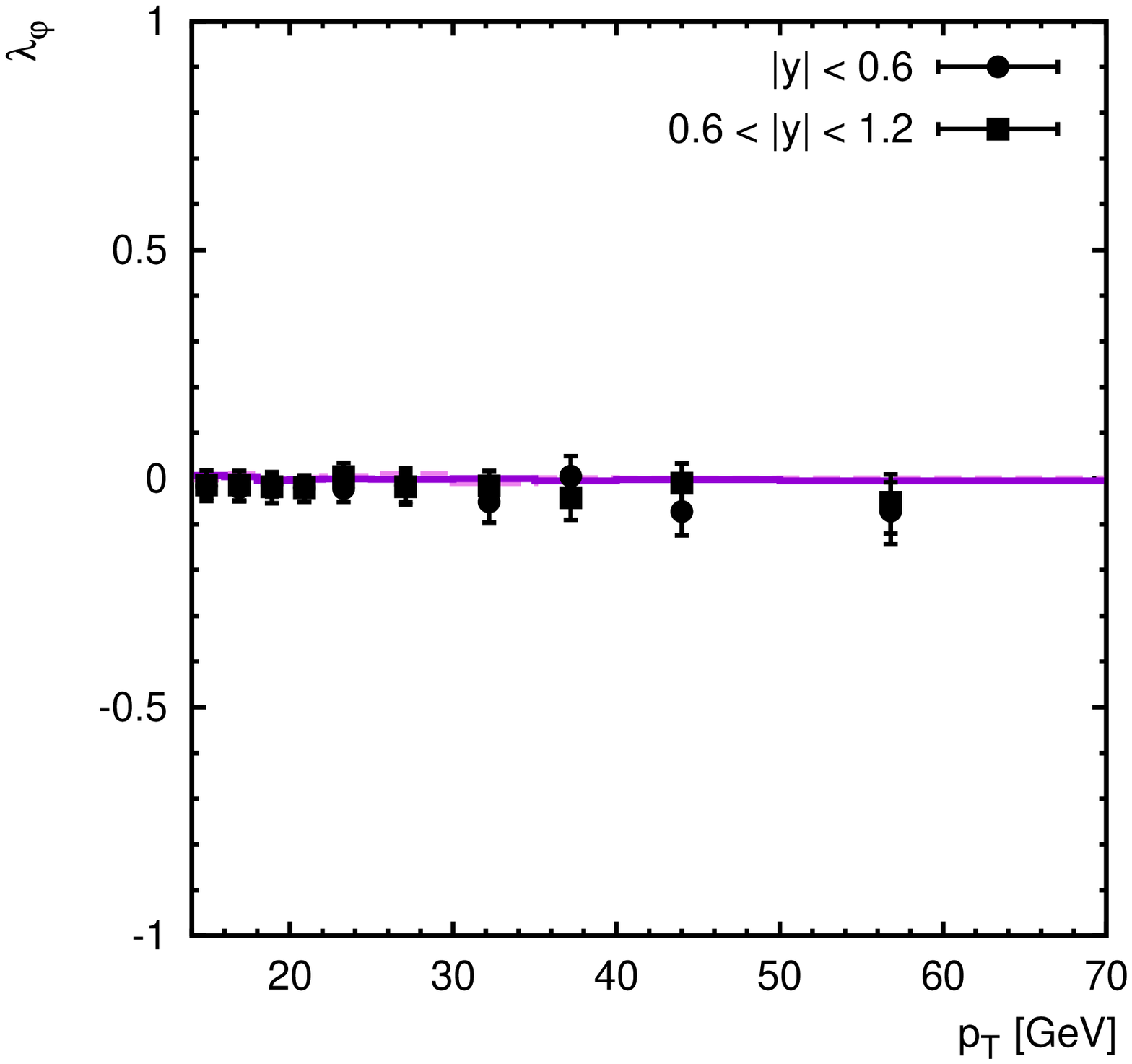, width = 5cm, angle = 270} 
\epsfig{figure=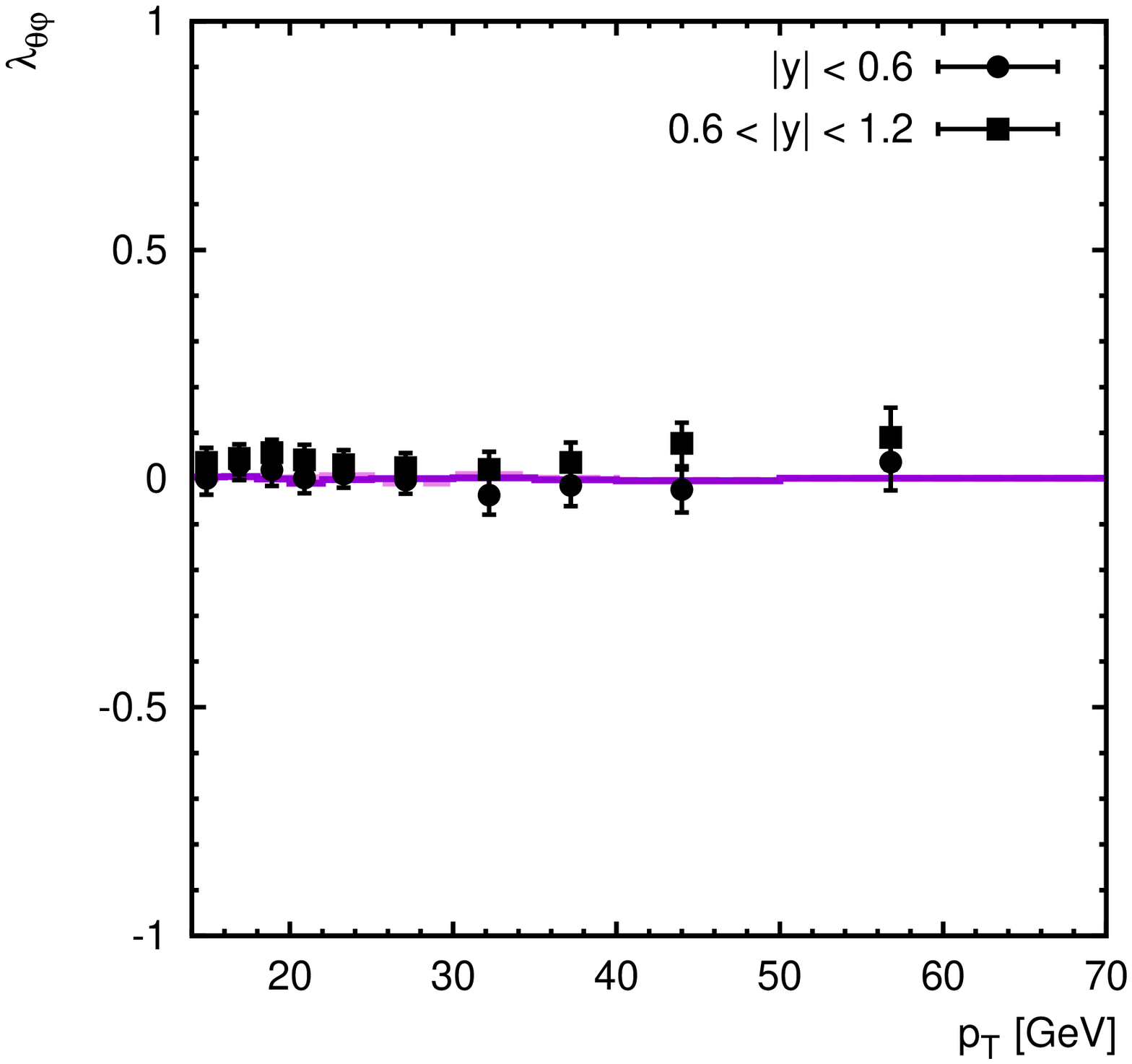, width = 5cm, angle = 270}
\epsfig{figure=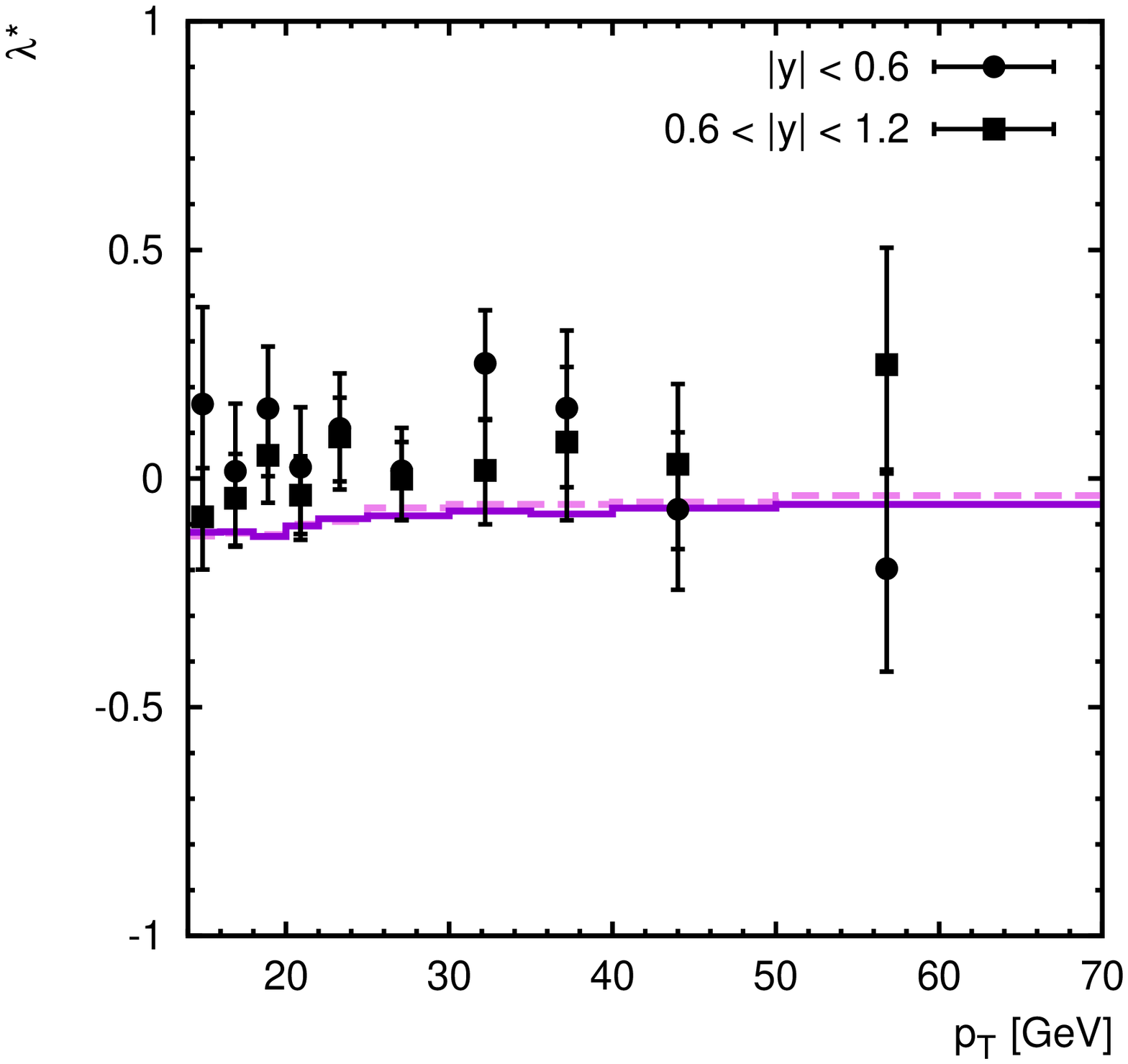, width = 5cm, angle = 270}
\caption{Polarization parameters 
$\lambda_\theta$, $\lambda_\phi$, $\lambda_{\theta \phi}$
and $\lambda^*$ of prompt $J/\psi$ mesons 
calculated as a function of their transverse momentum in the perpendicular helicity frame.
Notation of all curves is the same as in Fig.~10.
The experimental data are from CMS\cite{21}.}
\label{fig12}
\end{center}
\end{figure}

\begin{figure}
\begin{center}
\epsfig{figure=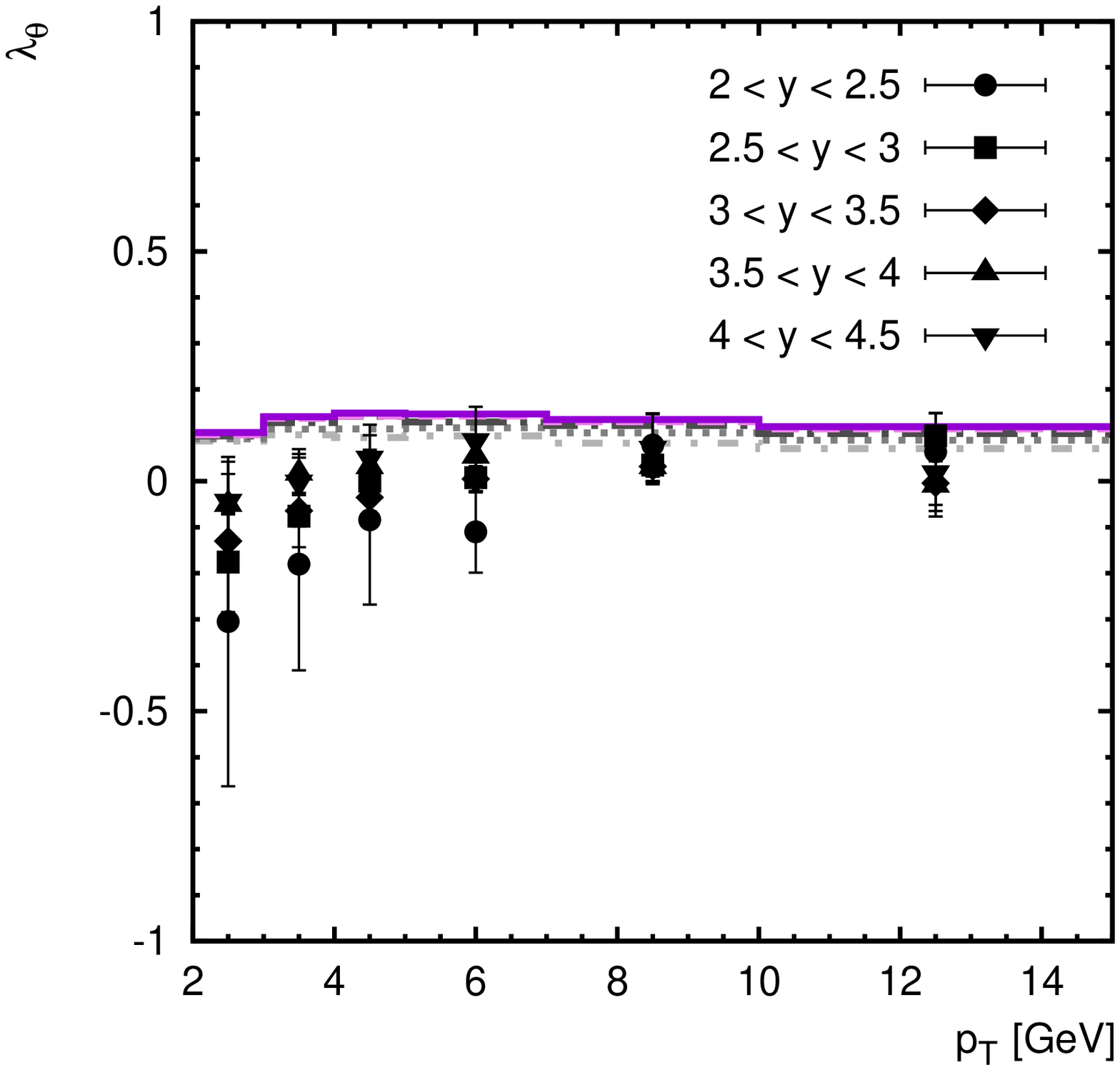, width = 5cm, angle = 270} 
\epsfig{figure=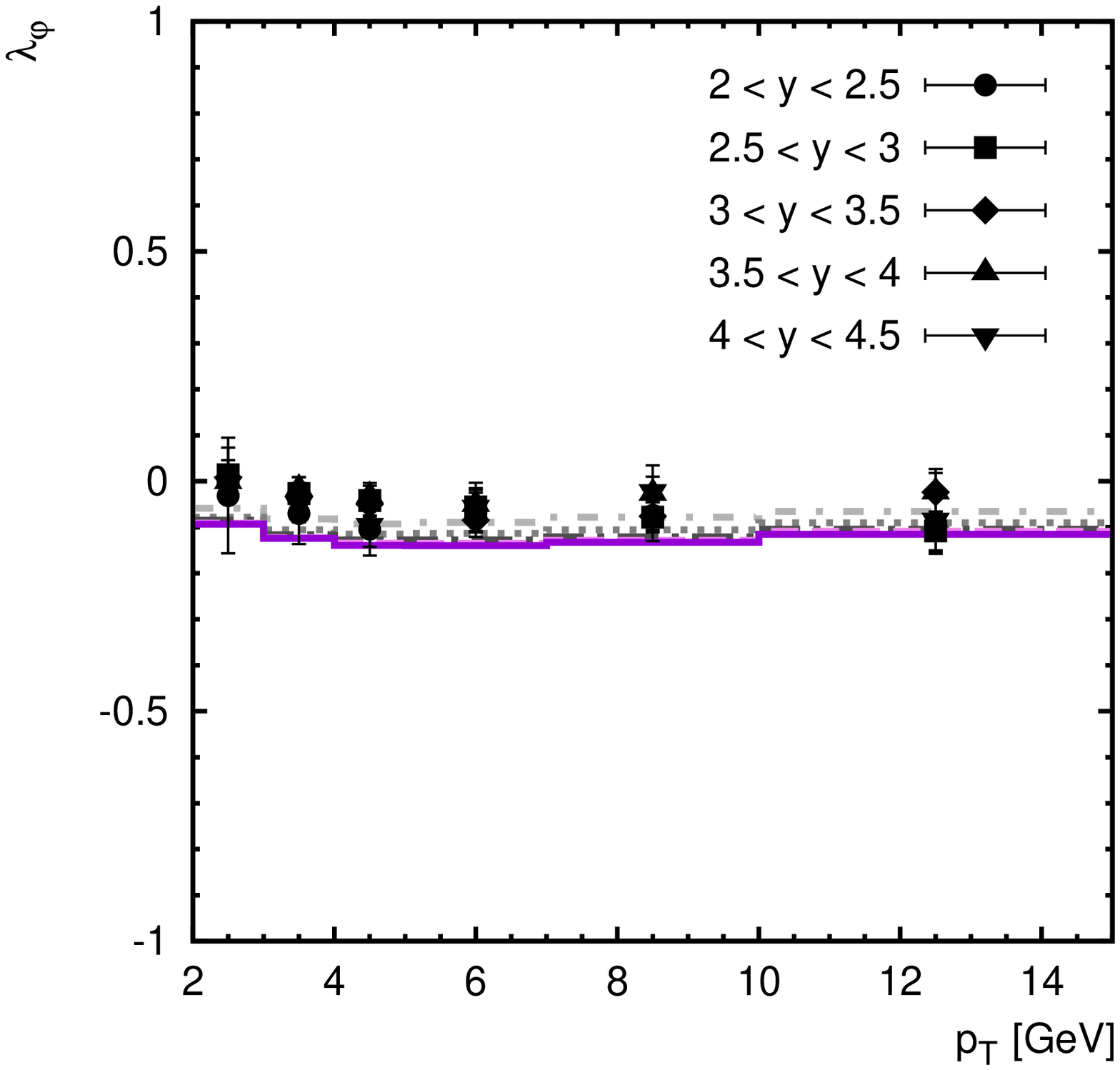, width = 5cm, angle = 270} 
\epsfig{figure=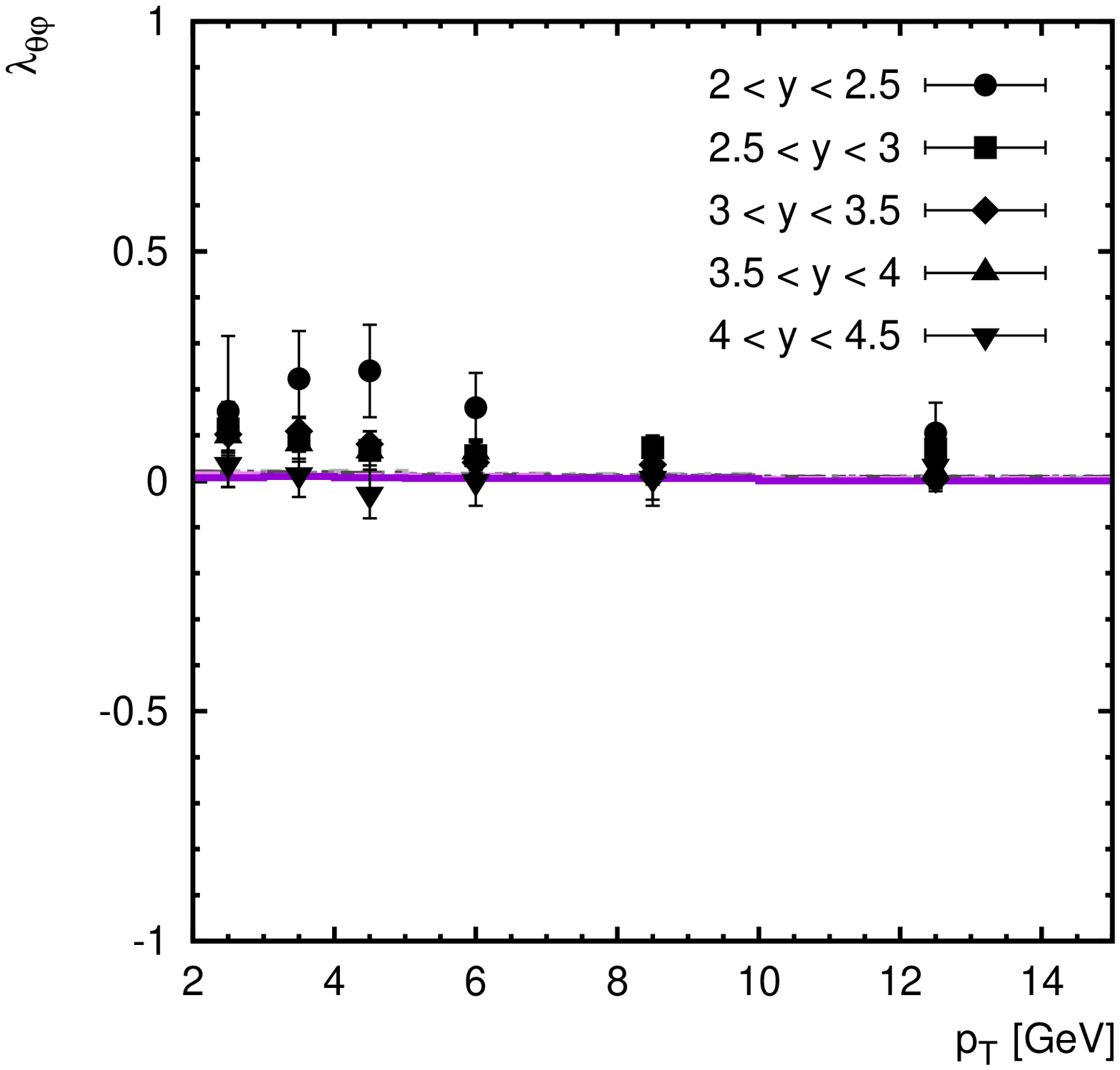, width = 5cm, angle = 270}
\caption{Polarization parameters 
$\lambda_\theta$, $\lambda_\phi$ and $\lambda_{\theta \phi}$
of prompt $J/\psi$ mesons calculated as a function of their transverse momentum in the 
Collins-Soper frame. 
The solid, dashed, dash-dotted, dotted and short dash-dotted histograms correspond
to the predictions obtained at $2 < y < 2.5$, $2.5 < y < 3$,
$3 < y < 3.5$, $3.5 < y < 4$ and $4 < y < 4.5$. 
The experimental data are from LHCb\cite{20}.}
\label{fig13}
\end{center}
\end{figure}

\begin{figure}
\begin{center}
\epsfig{figure=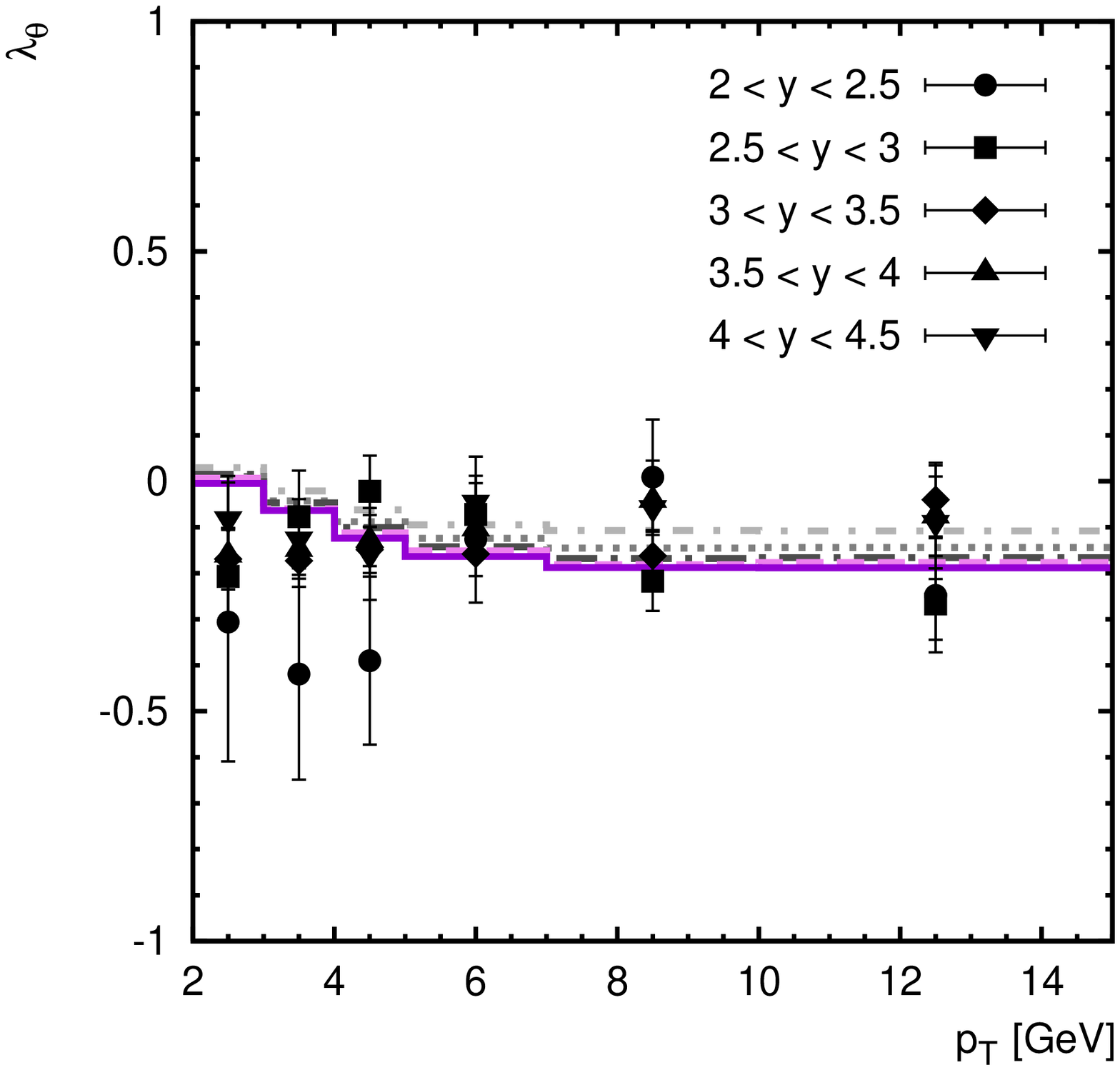, width = 5cm, angle = 270} 
\epsfig{figure=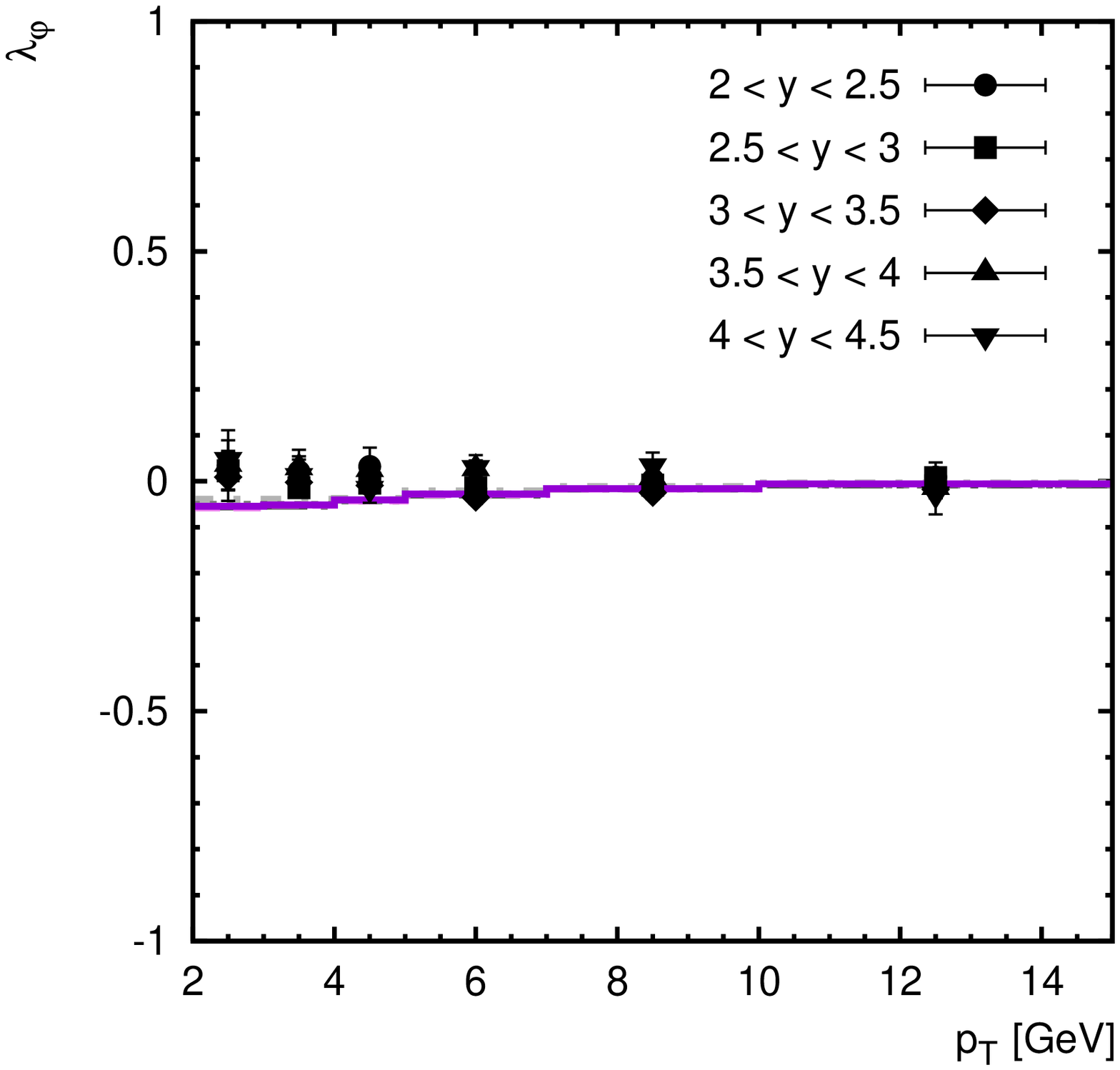, width = 5cm, angle = 270} 
\epsfig{figure=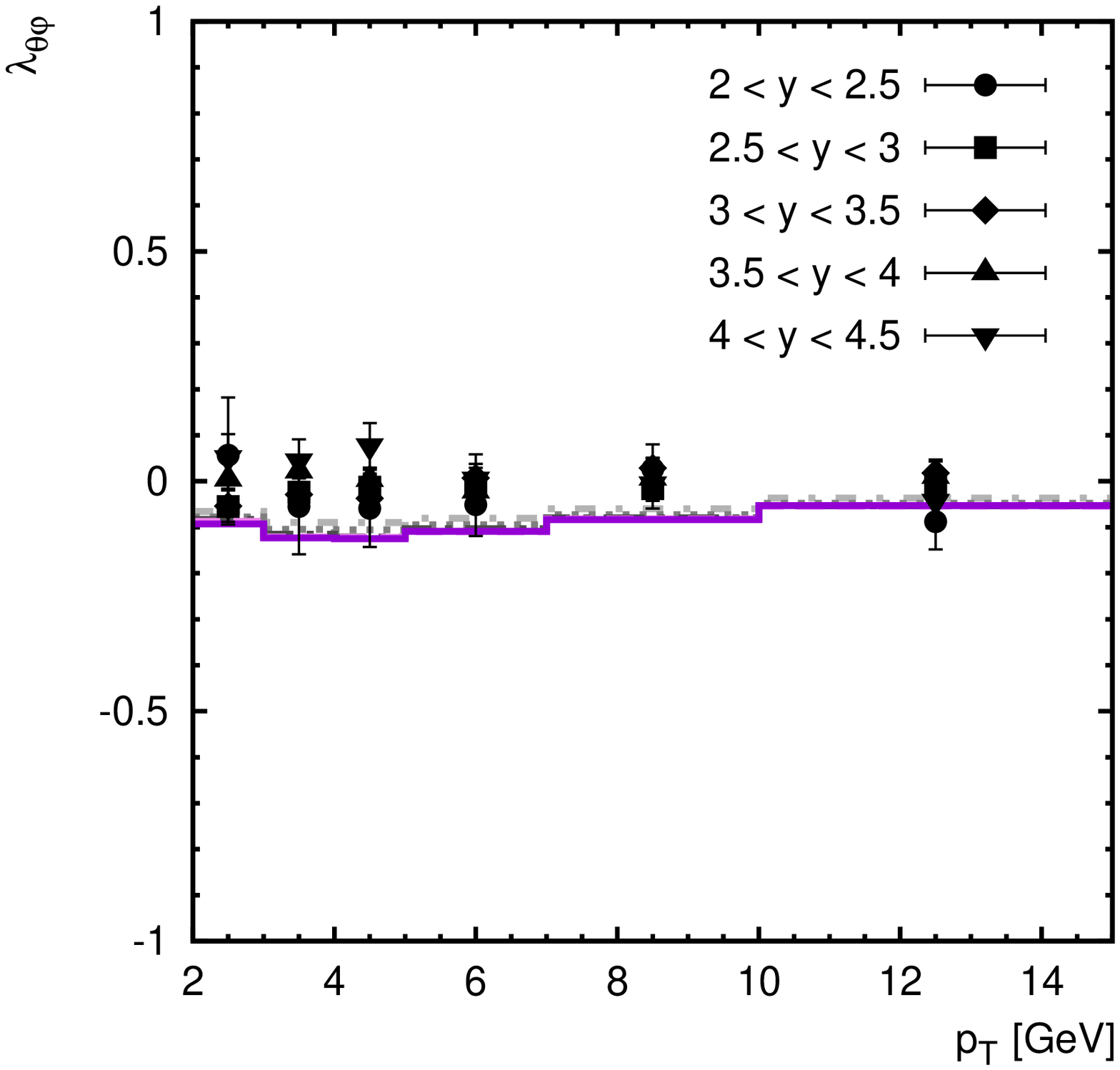, width = 5cm, angle = 270}
\caption{Polarization parameters 
$\lambda_\theta$, $\lambda_\phi$ and $\lambda_{\theta \phi}$
of prompt $J/\psi$ mesons calculated as a function of their transverse momentum in the 
helicity frame. Notation of all curves is the same as in Fig.~13. 
The experimental data are from LHCb\cite{20}.}
\label{fig14}
\end{center}
\end{figure}

\end{document}